\documentclass{aa}
\usepackage{amsmath}
\usepackage{amssymb}
\usepackage[varg]{txfonts}
\usepackage{epsfig,graphicx}
\usepackage[colorlinks=true, linkcolor=blue, citecolor=blue, urlcolor=blue]{hyperref}  
\usepackage{natbib}
\bibpunct{(}{)}{;}{a}{}{,} 
\newcommand{\ser}{S\'ersic}
\usepackage{color}

\defcitealias{2018A&A...609A..37C}{C18}
\begin{document}

\title{Dust emission profiles of DustPedia\thanks{
DustPedia is a project funded by the EU under the heading 
`Exploitation of space science and exploration data'. 
It has the primary goal of exploiting existing data in the 
\textit{Herschel} Space Observatory and Planck Telescope databases. 
}  galaxies}

\titlerunning{Dust emission profiles}

\author{
A.~V.~Mosenkov\inst{1,2,3}
\and M.~Baes\inst{1} 
\and S.~Bianchi\inst{4}  
\and V.~Casasola\inst{4,5} 
\and L.~P.~Cassar\`a\inst{6} 
\and C.~J.~R.~Clark\inst{7} 
\and J.~Davies\inst{7}  
\and I.~De~Looze\inst{1,8} 
\and P.~De~Vis\inst{9} 
\and J.~Fritz\inst{10} 
\and M.~Galametz\inst{11} 
\and F.~Galliano\inst{11} 
\and A.~P.~Jones\inst{9} 
\and S.~Lianou\inst{11}  	
\and S.~C.~Madden\inst{11} 
\and A.~Nersesian\inst{1,6,12} 
\and M.~W.~L.~Smith\inst{7} 
\and A.~Tr\v cka\inst{1}  
\and S.~Verstocken\inst{1}  
\and S.~Viaene\inst{1,13} 
\and M.~Vika\inst{6} 
\and E.~Xilouris\inst{6}}  
\institute{Sterrenkundig Observatorium, Department of Physics and Astronomy, Universiteit Gent Krijgslaan 281 S9, B-9000 Gent, Belgium \\
\email{mosenkovAV@gmail.com}
\and
Central Astronomical Observatory of RAS, Pulkovskoye Chaussee 65/1, 196140, St. Petersburg, Russia
\and
St. Petersburg State University, Universitetskij Pr. 28, 198504, St. Petersburg, Stary Peterhof, Russia
\and
INAF-Osservatorio Astrofisico di Arcetri, Largo E. Fermi 5, I-50125, Florence, Italy
\and
INAF-Istituto di Radioastronomia, Via Piero Gobetti 101, 40129, Bologna, Italy
\and
National Observatory of Athens, Institute for Astronomy, Astrophysics, Space Applications and Remote Sensing, 
Ioannou Metaxa \\ and Vasileos Pavlou GR-15236, Athens, Greece
\and
School of Physics and Astronomy, Cardiff University, The Parade, Cardiff CF24 3AA, UK
\and
Department of Physics and Astronomy, University College London, Gower Street, London WC1E 6BT, UK
\and
Institut d'Astrophysique Spatiale, CNRS, Univ. Paris-Sud, Universit\'e Paris-Saclay, B\^{a}t. 121, 91405, Orsay Cedex, France
\and
Instituto de Radioastronom\'\i a y Astrof\'\i sica, UNAM, Campus Morelia, A.P. 3-72, C.P. 58089, Mexico
\and
Laboratoire AIM, CEA/DSM - CNRS - Universit\'e Paris Diderot, IRFU/Service d'Astrophysique, CEA Saclay, 91191, Gif-sur-Yvette, France
\and
Department of Astrophysics, Astronomy \& Mechanics, Faculty of Physics, University of Athens, Panepistimiopolis, GR15784, Zografos, Athens, Greece
\and
Centre for Astrophysics Research, University of Hertfordshire, College Lane, Hatfield, AL10 9AB, UK
}
\date{}
\abstract{
Most radiative transfer models assume that dust in spiral galaxies is distributed exponentially. In this paper our goal is to verify this assumption by analysing the two-dimensional large-scale distribution of dust in galaxies from the DustPedia sample. For this purpose, we make use of \textit{Herschel} imaging in five bands, from 100 to \,500$\,\mu$m, in which the cold dust constituent is primarily traced and makes up the bulk of the dust mass in spiral galaxies.  
For a subsample of 320 disc galaxies, we successfully perform a simultaneous fitting with a single \ser\ model of the \textit{Herschel} images in all five bands using the multiband modelling code \textsc{galfitm}.
We report that the \ser\ index $n$, which characterises the shape of the \ser\ profile,
lies systematically below 1 in all \textit{Herschel} bands and is almost constant with wavelength. The average value at 250$\,\mu$m is $0.67\pm0.37$ (187 galaxies are fitted with $n^{250}\leq0.75$, 87 galaxies have $0.75<n^{250}\leq1.25$, and 46 -- with $n^{250}>1.25$).
Most observed profiles exhibit a depletion in the inner region (at $r<0.3-0.4$ of the optical radius $r_{25}$) and are more or less exponential in the outer part. We also find breaks in the dust emission profiles at longer distances ($0.5-0.6)\,r_{25}$ which are associated with the breaks in the optical and near-infrared. 
We assume that the observed deficit of dust emission in the inner galaxy region is related to the depression in the radial profile of the H{\sc i} surface density in the same region because the atomic gas reaches high enough surface densities there to be transformed into molecular gas. If a galaxy has a triggered star formation in the inner region (for example, because of a strong bar instability, which transfers the gas inwards to the centre, or a pseudobulge formation), no depletion or even an excess of dust emission in the centre is observed.
} 


\maketitle

\section{Introduction}
\label{sec:intro}

Dust is one of the several constituents of galaxies which holds a major role in the interstellar medium (ISM). First, it acts as a catalyst in the transformation of the atomic hydrogen into molecular hydrogen from which stars form \citep{1995ApJ...443..152W}. Thus, the dust is an important element in the chemical evolution of the ISM. Second, it allows the gas to cool and condense to form new stars \citep{1978ApJS...36..595D,1986ApJ...302..363D,1997ARA&A..35..179H}. Third, it absorbs some fraction of the emitted starlight and re-emits it in the infrared (IR) domain, where the thermal emission by dust grains dominates in the spectral energy distribution (SED) of galaxies between $\approx10$ and 1000\,$\mu$m \citep{2002MNRAS.335L..41P,2016A&A...586A..13V}. Dust is thus an effective tracer of the star formation activity. Because of the important role of dust in the ISM and its tight link to the other components of galaxies, the study of dust emission is mandatory to have a better understanding of all of the processes at play.

Spiral galaxies have cold and warm dust, as first suggested by \citet{1984ApJ...278L..67D}, and confirmed by
the Infrared Space Observatory (ISO, \citealt{1996A&A...315L..27K}) (see review of \citealt{2005SSRv..119..313S}). The cold dust component ($T_\mathrm{d} = 10-20$~K) is associated with molecular and atomic hydrogen clouds, which are heated mostly by the general interstellar radiation field (ISRF, \citealt{1989A&ARv...1...49C,2015MNRAS.448..135B}).

Since most of the total interstellar dust mass of a galaxy resides in its cold phase (see review of \citealt{2018ARA&A..56..673G} and references therein), the cold dust component can therefore be used to estimate the total dust distribution over the galaxy. 

Radiative transfer (RT) modelling serves as a good tool to study the dust distribution by the observed dust attenuation in edge-on galaxies. To retrieve structural parameters of the dust component, optical and near-infrared (NIR) observations are used to create monochromatic (i.e. based on one photometric band, see e.g. \citealt{1997A&A...325..135X,1998A&A...331..894X,1999A&A...344..868X,2007A&A...471..765B,2012MNRAS.427.2797D,2013A&A...550A..74D}) or oligochromatic (based on several bands, see e.g. \citealt{2010A&A...518L..39B,2014MNRAS.441..869D,2015IAUS..309..309D,2016A&A...592A..71M,2017MNRAS.464...48P,2018A&A...616A.120M}) models. In the optical, the dust attenuation in the edge-on view is severe and can be observed in many edge-on galaxies as a prominent dust lane which corresponds to an absorbing dust disc. The dust surface density of this disc is usually represented by a double-exponential model, and from the fitting its main free parameters (the total dust mass, the dust disc scalelength and scaleheight) can be recovered. Despite the fact that the model of a double exponential disc is plausible for most galaxies, few attempts were made to fit the dust component with a different dust model, for example, a dust ring \citep{2012MNRAS.427.2797D,2015IAUS..309..309D,2014ApJ...795..136S,2015A&A...579A.103V}. However, a model of a non-exponential dust disc has not ever been adopted in RT modelling.

In face-on galaxies RT modelling has been also successfully used \citep[see e.g.][]{2014A&A...571A..69D,2017A&A...599A..64V}.

Apart from RT fitting of the observed dust attenuation in edge-on galaxies, a more straightforward way to study dust distribution in galaxies is by mapping the radial distribution of dust emission in the far-infrared (FIR).
First studies of dust discs in galaxies were performed with the Infrared Astronomical Satellite (\textit{IRAS}, \citealt{1984ApJ...278L...1N,1990BAAS...22Q1325M,2003AJ....126.1607S,2007A&A...462..507L}) and the ISO, and later with the \textit{Spitzer Space Telescope} \citep{2004ApJS..154....1W}. 

Often, to study the radial distribution of dust emission in galaxies, a 1D azimuthally averaged or 2D profile is used. Below we briefly list some studies in which the large-scale distribution of dust was traced.

In \citet{1998A&A...335..807A}, 8 nearby galaxies observed with the ISO at 200~$\mu$m were studied. Based on these observations, the exponential scalelengths of the dust emission were estimated and compared with those determined in the IRAS  bands and the optical $B$ band. 

\citet{2009ApJ...701.1965M} carried out a detailed analysis of the radial distributions of stars, gas, and dust in the \textit{Spitzer} Infrared Nearby Galaxies Survey (SINGS, \citealt{2003PASP..115..928K}). They found that the dust density profiles are exponential. However, in S0/a and Sab galaxies the dust profiles usually present a central depletion.

In recent years, the \textit{Herschel} Space Observatory \citep{2010A&A...518L...1P} offered a unique opportunity to study the FIR/submm spectral domain (from 55 to $672\,\mu$m, thus covering wavelengths beyond the peak of thermal dust emission). Thanks to the \textit{Herschel} telescope,  we are able to map the distribution of the cold dust component in a large number of nearby objects in an unprecedented way.

In \citet{2015A&A...576A..33H}, radial surface brightness (SB) profiles of 61 galaxies from the KINGFISH (Key Insights into Nearby Galaxies: Far-Infrared Survey with \textit{Herschel}, \citealt{2011PASP..123.1347K}) sample were fitted with an exponential function in order to compare stellar and cool-dust disc scalelengths, as measured at 3.6~$\mu$m and 250~$\mu$m.

In \citet{2016MNRAS.462..331S}, combining FIR emission of 110 galaxies from the \textit{Herschel} Reference Survey \citep[HRS,][]{2010PASP..122..261B}, an exponential distribution for the dust was recovered. However, the average SB profile at 250~$\mu$m for a subsample of 45 large galaxies clearly shows evidence for a shallower slope at large radii than in the inner part (a broken- or double-exponential model was adopted in this case). Combining signal has the advantage that very large radii of the SB profile can be probed. On the other hand, some individual galaxy features are obviously washed out, with a general smooth profile created.

In most of the studies mentioned above the azimuthally averaged (1D) distribution of the galaxy SB was used to present the radial distribution of dust emission in galaxies. However, as was first shown by \citet{1995ApJ...448..563B}, while being simple to implement, the 1D approach has some serious drawbacks. As noticed by \citet{2010AJ....139.2097P}, `for compact galaxies, 1D profile fitting cannot properly correct for image smearing by the point spread function (PSF) because 1D profile convolution is not mathematically equivalent to convolution in a 2D image'. 2D fitting is more beneficial than 1D profile analysis when a multi-component decomposition is required \citep[see e.g.][]{1998ASPC..145..108S,2002AJ....124..266P,2010AJ....139.2097P,2004ApJS..153..411D,2015ApJ...799..226E}. Furthermore, the radial distribution of dust is usually a priori approximated by an exponential law. Often, the inner part of the galaxy, which exhibits some deviation from this unified function, is not taken into account when fitting the radial profile. To this day, no 2D fitting with a non-exponential disc model (e.g. with a general \ser\ law, \citealt{1968adga.book.....S}) in the FIR/submm has yet been done.   

The aim of this paper is to study the 2D distribution of dust for a large sample of galaxies using the \ser\ law where the shape of the profile is controlled by the \ser\ index $n$ (see Sect.~\ref{sec:decomp}). The advantage of using this law is that setting $n=0.5$ gives a Gaussian, for $n=1$ it converts into an exponential law, and for $n=4$ we get a de Vaucouleurs profile. Therefore, in this work we verify whether the dust distribution in disc galaxies is truly exponential as was assumed earlier, mostly based on analysing 1D profiles. We apply the \ser\ law for describing the 2D distribution of dust emission in exactly the same way as it is done for the stellar emission. To the best of our knowledge, this is done here for the first time.

In this paper we focus our attention on the spatial distribution of cold dust in galaxies by exploiting five \textit{Herschel} bands, from 100~$\mu$m to 500~$\mu$m.
It is the distribution of the emission of the diffuse dust disc which we investigate in this paper. In addition, we trace old stellar populations using the NIR imaging (3.4~$\mu$m) available due to the \textit{WISE} space observatory \citep{2010AJ....140.1868W}. We make use of the DustPedia sample, which aims at characterising the dust in the Local Universe using a series of robust and homogeneous modelling and analysis tools. We apply a unified approach for fitting the \ser\ model to both the NIR and FIR/submm sets of observations. From this, we are able to draw conclusions about the shape of the radial distribution (given by the \ser\ index) and the effective radius of the dust component, as well as about the change of these parameters with wavelength. 

This paper is structured as follows. In Sect.~\ref{sec:sample}, we briefly describe the DustPedia sample and define a subsample of galaxies, on which we focus our analysis. We also provide a description of the data preparation we use in our study. The procedure for performing the fitting of galaxy images is presented in Sect.~\ref{sec:decomp}. We provide the results of our fitting in Sect.~\ref{sec:results}. The discussion of the obtained results is presented in Sect.~\ref{sec:discussion}. Finally, our conclusions are summarised in Sect.~\ref{sec:conclusions}. In the Appendix we provide some details and results of an additional analysis which is important for our study.

Throughout this paper, we assume regression in the form $y=k\,x+b$ with the Spearman rank-order correlation coefficient $\rho$ (lying between -1 and +1).

\section{The sample and data}
\label{sec:sample}

The DustPedia sample, which we use in our study, was originally selected to fulfil the following criteria. It includes galaxies with a radial velocity $v < 3000$\,km\,s$^{-1}$. The selected galaxies are large, with an optical diameter larger than $1\arcmin$. They are detected at 3.4\,$\mu$m (signal-to-noise ratio $>5\sigma$) and have \textit{Herschel} Photoconductor Array Camera and Spectrometer (PACS, \citealt{2010A&A...518L...2P}) or Spectral and Photometric Imaging Receiver (SPIRE, \citealt{2010A&A...518L...3G}) observations. The total number of the selected galaxies is 875 and, at the moment, it is the most representative sample of galaxies built upon the \textit{Herschel} legacy; it spans a wide range of galaxy masses and morphologies.

The imagery and photometry for the DustPedia sample was specially prepared by \citet[][hereafter \citetalias{2018A&A...609A..37C}]{2018A&A...609A..37C} and can be freely retrieved from the DustPedia database\footnote{\url{http://dustpedia.astro.noa.gr/}}.
Below, we use the DustPedia sample and its imagery to select galaxies for our subsequent analysis.

To study the cold dust distribution in galaxies, we make use of the \textit{Herschel} observations which have the best spatial resolution in the FIR/submm to date. We consider five \textit{Herschel} wavebands: PACS\,100$\,\mu$m, PACS\,160$\,\mu$m, SPIRE\,250$\,\mu$m, SPIRE\,350$\,\mu$m, and SPIRE\,500$\,\mu$m. We do not exploit PACS\,70 data as they are available for only 244 galaxies of the DustPedia sample. 
On top of that, in this band emission from stochastically heated warm dust can be substantial and, therefore, could potentially bias our current study.

In order to select galaxies for a \ser\ modelling of their \textit{Herschel} images, we impose the following criteria.
If a galaxy has a measured flux density lower than $3\sigma$ in any \textit{Herschel} band, then this galaxy is removed from the subsequent consideration. Also, we rejected galaxies which have a major flag associated with the \textit{Herschel} flux due to contamination, artefacts, or insufficient sky coverage in this band (flags `C', `A', `N', respectively, given in the aperture photometry table from \citetalias{2018A&A...609A..37C}). This results in a sample of 344 galaxies. After a visual inspection of the selected galaxies, we additionally rejected five galaxies (IC\,3521, IC\,3611, NGC\,2992, PGC\,166077, and UGC\,7249) which appeared to be too small to be fitted. 
The final subsample thus consists of 339 DustPedia galaxies. Our sample comprises only one ``early-type'' galaxy, NGC\,2974, which has a morphological type code in HyperLeda \citep{2014A&A...570A..13M} of $-4.3\pm1.2$. However, more careful investigation of this galaxy reveals a spiral structure: \citet{1995MNRAS.274.1107N} classified this galaxy as Sa. Therefore, we can consider our subsample as consisting entirely of disc galaxies.

\textit{Herschel} imagery for all PACS\,100--SPIRE\,500$\,\mu$m wavebands is taken completely from the DustPedia archive (see details in \citealt{2017PASP..129d4102D} and \citetalias{2018A&A...609A..37C} on how the images were reduced), including the flux density maps and the error-maps. The maps have pixel sizes of $3\arcsec$ (PACS\,100), $4\arcsec$ (PACS\,160), $6\arcsec$ (SPIRE\,250), $8\arcsec$ (SPIRE\,350), and $12\arcsec$ (SPIRE\,500), with a corresponding FWHM of $10\arcsec$, $13\arcsec$, $18\arcsec$, $25\arcsec$, and $36\arcsec$. For PACS\,100 and PACS\,160, we retrieved the extended PSF kernels provided in \citet{2016A&A...591A.117B} which were created by combining Vesta (central part of the PSFs) and Mars (wings) dedicated observations. For SPIRE\,250, SPIRE\,350, and SPIRE\,500, we adopted the kernels which were used in \citet{2016MNRAS.462..331S} (created by means of SPIRE calibration observations of Neptune). These extended PSFs are mandatory to have correct fits of the SB profiles.

Complementary NIR data were also used to trace the old stellar population \citep[see e.g.][]{1984ApJS...54..127E,1995ApJ...447...82R,1993PASP..105..651G}. The \textit{WISE} observatory scanned the whole sky in the \textit{W1} band (3.4$\,\mu$m), and so we make use of its observations for the entire sample of 875 DustPedia galaxies. We do this to obtain some important general characteristics of these galaxies (e.g. galaxy inclination, see the Appendix~\ref{sec:add_analysis}), as well as some galaxy structural properties. These quantities are used in the current study and in the papers to follow.

Since no error-maps are provided for the \textit{WISE} data in the DustPedia database, we applied the same scripts used by \citetalias{2018A&A...609A..37C} to retrieve the \textit{WISE\,W1} intensity maps and the related pixel-uncertainty maps from the NASA/IPAC Infrared Science Archive (IRSA\footnote{\url{http://irsa.ipac.caltech.edu/frontpage/}}). These maps store the 1$\sigma$ uncertainty per pixel corresponding to the co-added intensity values in the AllWISE data release Image Atlas. The final \textit{WISE} cutouts retain the standard AllWISE Image Atlas pixel size of $1.375\arcsec$ and the FWHM of $6.1\arcsec$. The PSF image for \textit{WISE\,W1} is taken from Section IV.4.c.viii of the \textit{WISE} All-Sky Data Release Explanatory Supplement\footnote{\url{http:/wise2.ipac.caltech.edu/docs/release/allsky/expsup}}, which is constructed from co-added mosaics of the deep Ecliptic Pole observations, and rescaled to match the pixel size and resolution of the \textit{WISE} cutouts.

In order to prepare galaxy images in the \textit{WISE\,W1} and \textit{Herschel} bands, we used the following steps which are described in full detail in the Appendix~\ref{sec:data_preparation}. For each galaxy, its image in each band was rebinned to PACS\,100~$\mu$m (except for \textit{WISE\,W1}), background subtracted, and cropped to encompass the whole galaxy body and some free space. We also masked out contaminating objects (stars and galaxies), which should not be taken into account in the process of fitting. The extended PSF in each band was also specially treated to match the corresponding cutout.

Once these routines were run, the final images are ready to be processed with our special fitting tool. The detailed description of our fitting technique is given in the next section.

\section{Single S\'ersic modelling technique}
\label{sec:decomp}

\begin{figure*}
\centering
\includegraphics[width=18cm, angle=0, clip=]{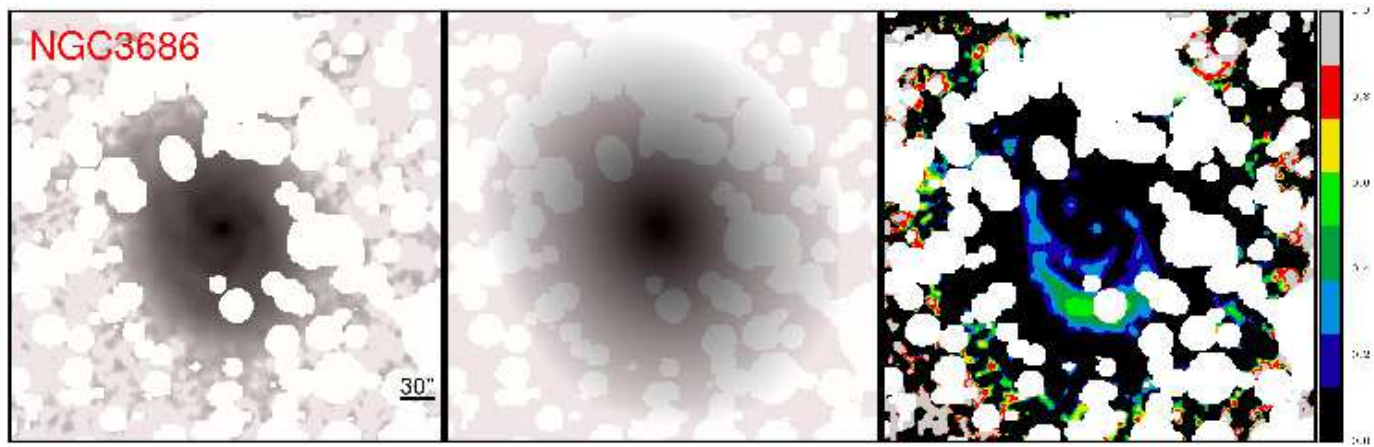}
\caption{The results of the fitting for NGC\,3686 in the \textit{WISE\,W1} band: the \textit{WISE\,W1} image (left panel), the best fitting image (middle panel),
and the residual image (right panel) which indicates the relative deviation between
the fit and the image in absolute values, i.e. $| data-model |/model$. The right color bar shows the scaling of the residual image. The masked objects are highlighted by the white colour (the left and right panels) or are transparent areas (the middle panel). The galaxy and model images are given in a logarithmic scale. All images cover a field-of-view of $6.5\arcmin\times6.5\arcmin$.}
\label{NGC3686_decomp}
\end{figure*}

\begin{figure}
\centering
\includegraphics[height=7.5cm, angle=0, clip=]{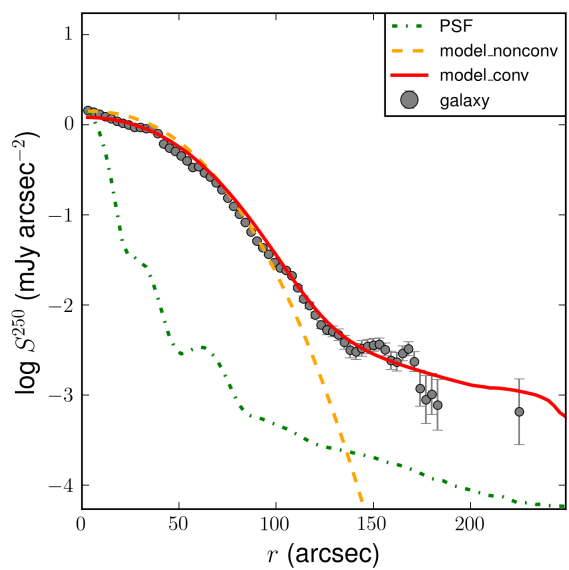}
\caption{The azimuthally averaged profile for NGC\,3686 (grey circles), its convolved (red solid line) and unconvolved (orange dashed line) model, and the PSF kernel (green dotdash line) in the SPIRE\,250$\,\mu$m band.}
\label{NGC3686_decomp_GM_azim}
\end{figure}

In order to quantitatively describe the SB distribution in galaxies in a chosen waveband (which traces the stellar density distribution of a stellar population or highlights a mixture of different populations with intrinsic dust attenuation), two-dimensional \citep[2D,][]{1995ApJ...448..563B,1996A&AS..118..557D,1994A&A...285..723E,2002MNRAS.333..400C} profiles are usually fitted with an analytic function or a superposition of functions which represent different structural components of the galaxy. This is the so-called parametric method, which, if used appropriately, gives reliable results on the galaxy structure, and, at the same time, can be considered as a useful indicator of galaxy morphology.

A simple way to fit the light radial profile of a galaxy is by using a \ser\ model \citep{1968adga.book.....S}, with several free parameters: the effective radius $r_\mathrm{e}$ (the radius which encompasses half of the total luminosity of the system), the \ser\ index $n$ (which controls the degree of curvature of the profile), and the effective SB $\mu_\mathrm{e}$ (in units of Jy\,arcsec$^{-2}$) at the radius $r_\mathrm{e}$ (equivalently, the total luminosity of the model or its central SB $\mu_0$ can be used):

\begin{equation}
 \mu(r) = \mu_0 + 
 \frac{2.5\, \nu_\mathrm{n}}{\ln{10}} \, 
 \left(\frac{r}{r_\mathrm{e}}\right)^{1/n}\,,
\label{form_ser}
\end{equation}
where $\nu_\mathrm{n} \simeq \ln10\,(0.868 \, n-0.142)$ \citep{1993MNRAS.265.1013C}. To distinguish the designations of the same \ser\ parameters fitted in different bands, we use an upper index for denoting the passband (or wavelength), e.g. for \textit{WISE\,W1} the \ser\ index is written as $n^{3.4}$.

Since the \ser\ law has several free parameters, which control the view of the profile, a large variety of the observed galaxy profiles can be approximated by this function very well \citep{1993MNRAS.265.1013C,1999AJ....118.1230C,2001MNRAS.326.1517D,2005ApJS..157..218C,2005AJ....130..475A,2017A&A...608A.142V}. Single \ser\ fitting was applied to large samples of galaxies to study how their parameters change with morphological type and other general properties of galaxies \citep[see e.g.][]{2007ApJS..172..615H,2010MNRAS.408.1313L,2011MNRAS.411.2439H,2012MNRAS.421.1007K,2013MNRAS.430..330H,2017A&A...598A..32M}.

The \ser\ index shows how far the light profile is from being exponential.
$n = 1$ shows that a galaxy consists of a single exponential disc. $n < 2$ is an efficient criterion for selecting disc-dominated galaxies, whereas $n > 2$ is likely to be related to an early-type galaxy or a spiral galaxy with a bright bulge (see e.g. \citealt{2004ApJ...604L...9R} and Sect.~\ref{sec:results}). In other words, the general \ser\ index is sensitive to the presence of a central component (a bulge or a bar). Thus, this unified function serves as a good first step in describing the multicomponent structure of galaxies \citep[see e.g.][]{1995MNRAS.275..874A,1996A&AS..118..557D,2004A&A...415...63M,2005MNRAS.362.1319L,2007A&A...467..541A,2007AJ....133.2846R,2008MNRAS.384..420G,2009MNRAS.393.1531G,2010MNRAS.401..559M}. 

In this paper we apply single \ser\ modelling to describe 2D stellar and dust emission of the selected galaxies. Apart from the parameters in Eq.~\ref{form_ser}, the free parameters of the 2D model include the coordinates of the galaxy centre, its apparent flattening, and position angle. The applicability of this fitting method to the \textit{Herschel} observations and its robustness is discussed in the Appendix~\ref{sec:mock}.

In this study we implement the \textsc{galfit} and \textsc{galfitm} software to fit a \ser\ profile to the galaxy light profile. \textsc{galfit} performs a Levenberg-Marquardt minimisation of the $\chi^2$ residual between a galaxy image and its PSF-convolved model by modifying the free parameters. The final model is found when $\chi^2$ reaches a minimum. The calculation of $\chi^2$ is weighted by a sigma map, which is provided for each galaxy in our sample. This code has been successfully applied to individual galaxies with a diverse morphology \citep{2009ApJ...691..705G,2013ApJ...766...47H,2015MNRAS.447.2287R} and to large samples of galaxies consisting of thousands of objects \citep{2007ApJS..172..615H,2011MNRAS.411.2439H,2015ApJS..219....4S}.

Recently, a new fitting code \textsc{galfitm} \citep{2011ASPC..442..479B,2013MNRAS.430..330H,2013MNRAS.435..623V,2014MNRAS.444.3603V,2015A&A...577A..97V} has been created (as an extension of \textsc{galfit}) to fit a wavelength-dependent model by using multiple images of the same galaxy in a simultaneous and consistent manner. Rather than fitting the parameter values at the wavelength of each band, \textsc{galfitm} fits the coefficients of a smooth function describing the wavelength dependence of each parameter. It has been shown that multiband modelling significantly improves the extraction of information, particularly from bands with low signal-to-noise ratio when combined with higher signal-to-noise images \citep{2013MNRAS.435..623V, 2014MNRAS.444.3603V, 2014MNRAS.441.1340V,2016MNRAS.460.3458K}. Using such a multiband approach, one can obtain a more reliable model based on several observations of the same galaxy at different wavelengths.

For the purpose of this work, we make use of the \textsc{galfit} wrapper \textsc{deca} written in Python (we refer the reader to \citealt{2014AstBu..69...99M} for a complete description of the initial version of the package). We constructed 2D photometric models of the light profiles for each galaxy, selected in Sect.~\ref{sec:sample}, using a general \ser\ model (eq. \ref{form_ser}). In addition, the \textsc{galfitm} code is also incorporated in \textsc{deca} for a simultaneous fitting of galaxy images in different bands.

\textsc{deca} is especially suitable for a large sample of galaxies, when a modelling of each individual galaxy by hand is problematic. It can work in a fully automated regime if we use a unified \ser\ model for all galaxies selected. \ser\ fitting assumes that each galaxy has a light distribution which can be satisfactorily described by a \ser\ function. Of course, since galaxies are quite complex systems, using this single component can be a rather coarse approximation of the observed 2D profile in some galaxies (for example, in grand-design galaxies, galaxies with strong bars or ring structures, i.e. where different components have different geometry and/or position and orientation towards the observer). Nevertheless, to trace the general distribution of both stars and dust in a unified way, our single-model approach should be applicable to the statistically large sample of DustPedia galaxies.

\begin{figure*}
\centering
\includegraphics[height=18cm, angle=0, clip=]{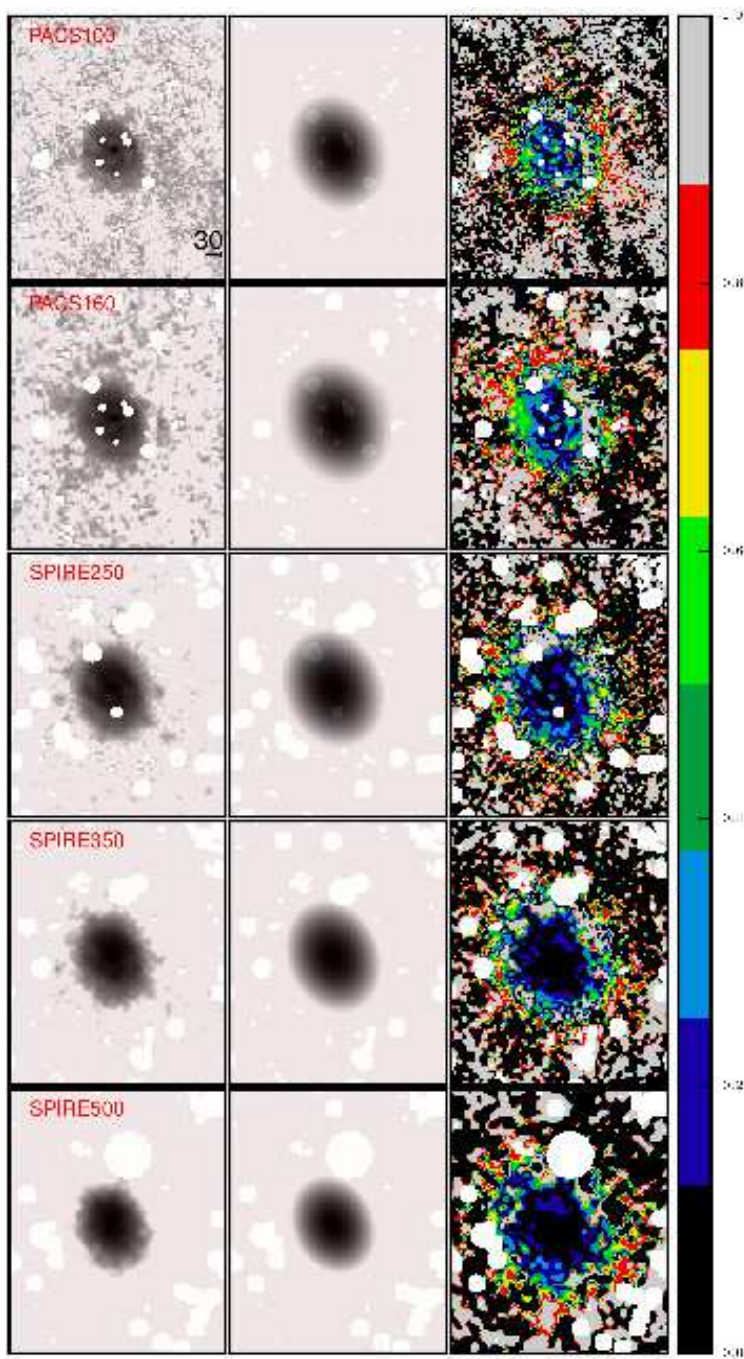}
\caption{The results of the \textsc{galfitm} modelling for NGC\,3686: the \textit{Herschel} images from 100 to 500~$\mu$m (left panel), the models (middle panel), and the relative residual images (right panel, in the same units as presented in Fig.~\ref{NGC3686_decomp}). The righthand color bar shows the scaling of the residual images. Any masked objects are highlighted with a white colour. The galaxy and model images are given in a logarithmic scale. All images cover a field-of-view of $6.75\arcmin\times8.35\arcmin$.}
\label{NGC3686_decomp_GM}
\end{figure*}

For the \textit{Herschel} images, we use first \textsc{galfit} and then \textsc{galfitm}: \textsc{deca} first performs modelling in all five \textit{Herschel} bands separately running \textsc{galfit} and then uses the output results as an input, first-guess model for \textsc{galfitm} five-band fitting. For \textsc{galfit}, we do not put any constraints on the fit parameters, however, for \textsc{galfitm} we perform a multiwavelength modelling where the parameters are replaced with functions of wavelength (Chebyshev polynomials). With these polynomials the user is allowed to control how the structural parameters of the model will vary as a function of wavelength. For the total galaxy magnitude we set the maximum polynomial order (for which the coefficients are free to vary in the fitting process) to be 5 (as the number of bands is 5, i.e., in essence, they are considered as unconstrained). The \ser\ index and $r_\mathrm{e}$ are allowed to vary parabolically with wavelength (this degree of freedom is based on the previous results of gradual increase of the scalelength of the dust emission profile with wavelength, see Introduction). All other parameters (centre position, axis ratio, and position angle) are selected to be constant with wavelength, e.g. the same in all bands.

As \textsc{galfitm} significantly improves the extraction of information, particularly from bands with low signal-to-noise ratio when combined with higher signal-to-noise images, the use of this approach is justified for the \textit{Herschel} observations. In addition, it allows us to link models of the same galaxy in different bands, so that, for example, the fit models would have the same centre, position angle, and axis ratio in all five fitted bands. We found this very convenient when fitting galaxy images in different wavebands.

If there are other sources within the fitting region, they are either masked or, if close enough to the main source to cause significant photometric blending, are included in the fit. In the case of severe foreground Galactic cirrus in some \textit{Herschel} images (as, for instance, for NGC\,6946), the sky component is also fitted simultaneously with the target galaxy.

In Fig.~\ref{NGC3686_decomp} we show an example of a \textsc{galfit} fitting in the \textit{WISE\,W1} band for one of the typical DustPedia galaxies, the Sbc-type galaxy NGC\,3686. An example of the azimuthally averaged profile with the overimposed model in the SPIRE\,250$\,\mu$m band (both created using the IRAF's {\sc ellipse} task\footnote{\url{http://iraf.noao.edu/}}) is shown in Fig.~\ref{NGC3686_decomp_GM_azim}. The profiles in all \textit{Herschel} bands look similar and, therefore, we only show here (and to the end of the paper) the profiles for the SPIRE\,250$\,\mu$m band. We note that for this galaxy the \ser\ index of the dust component $n^{250}=0.52$ (i.e. it has a Gaussian profile), in comparison with $n^{3.4}=1.15$ (an exponential profile) for the SB distribution of the old stellar population. The results of a \textsc{galfitm} fitting in all five \textit{Herschel} bands for the same galaxy are presented in Fig.~\ref{NGC3686_decomp_GM}. 

\section{Results}
\label{sec:results}

\textsc{deca} was applied to the \textit{WISE\,W1} images for all 875 DustPedia galaxies, as well as to the selected subsample of 339 galaxies with available \textit{Herschel} observations in all five bands. Below we present results of the fitting and discuss them in Sect.~\ref{sec:discussion}. 

In the Appendix~\ref{sec:comparison} we provide the robustness of our fitting results by comparing them with the literature. This comparison shows that the model parameters are recovered well using our \textsc{deca} fitting. We also show that 865 galaxies (98.86\% of the whole DustPedia sample) have successful single \ser\ fits in the \textit{WISE\,W1} band. 

Based on the \textsc{galfitm} results for the \textit{Herschel} data, we found that 320 galaxies (94\% out of the selected subsample of 339 galaxies) are large enough to have a plausible \ser\ model. The minimum value of the semimajor axis (measured for all DustPedia galaxies in \citetalias{2018A&A...609A..37C}) for this subsample is 63$\arcsec$, whereas the mean value is $3.2\arcmin\pm1.5\arcmin$. Out of them, we also define a subsample (and call it the reference sample) of 71 `good' galaxies which have a smooth SB distribution in all five bands and a reduced $\chi^2$ which yields $0.8\leq\chi^2\leq1.2$, an arbitrary criterion of the goodness of the fit. To the end of the paper we restrict our analysis to the 320 galaxies (as those for which a single \ser\ fit is reliable in the \textit{WISE\,W1} as well as in all five \textit{Herschel} bands). We also consider the reference sample to make sure that the results for both samples are compatible and our findings are not affected by the effects of resolution and the presence of a complex dust structure which can be observed in well-resolved galaxies.

Because neither \textsc{galfit}, neither \textsc{galfitm} provide errors of the retrieved parameters, we decided to estimate them for a randomly selected, typical spiral galaxy, NGC\,3622, using a genetic algorithm \citep{1989gaso.book.....G}. We applied it ten times and took the scatter in the fitted parameters as their uncertainties. Unfortunately, this operation is very time consuming, therefore we only did it for one galaxy. We found the \ser\ index $n^{250}$ for this galaxy to be $0.79\pm0.22$, which means that the estimated uncertainty is approximately 25\% of the retrieved parameter. For the effective radius we found $r_\mathrm{e}^{250}=18.3\arcsec\pm1.9\arcsec$, i.e its uncertainty is within 10\% of the parameter value.

The tables with the results can be accessed from the DustPedia database and from the VizieR catalogue service\footnote{Will be given for open access once the paper is accepted for publication.}. The final pipeline product includes separate tables with the model results for the \textit{WISE\,W1} band and for the \textit{Herschel} bands. For the \textit{WISE\,W1} band, we also provide some supplementary information described in the Appendix~\ref{sec:add_analysis}, e.g. the estimated galaxy inclinations and stellar masses.

Additionally, in the Appendix~\ref{sec:mock} we describe galaxy simulations which we performed to test the fitting technique described in Sect.~\ref{sec:decomp} and for comparing them with real observations. In our simulations we essentially used a well-defined model of the galaxy IC\,2531, varying its inclination angle and the distance to it (i.e. its spatial resolution). Two models for the dust component, an exponential disc and a \ser\ disc, are adopted to show if we can see a difference between the fitted dust emission profiles in these mock galaxies. IC\,2531 is a typical Sc galaxy. Sc is one of the most numerous types in our sample (see Table~\ref{tab:Hubble_distr}), and, therefore, by doing these simulations we strive to reproduce some characteristic view for the intermediate late-type galaxies in our sample. Of course, these mock galaxies cannot be completely compared to the selected sample which exhibit a large variety of morphological types, masses, environment, and other properties. Nevertheless, the behaviour of the fit parameters from the simulations can be easily compared to those for the real galaxies, from which we can conclude if our models are consistent with what we see in practice. On top of that, as shown below, the fitted parameters barely depend on galaxy morphology.

\subsection{Distributions of the parameters}
\label{sec:distributions}

\begin{figure*}
\centering
\includegraphics[width=12cm, angle=0, clip=]{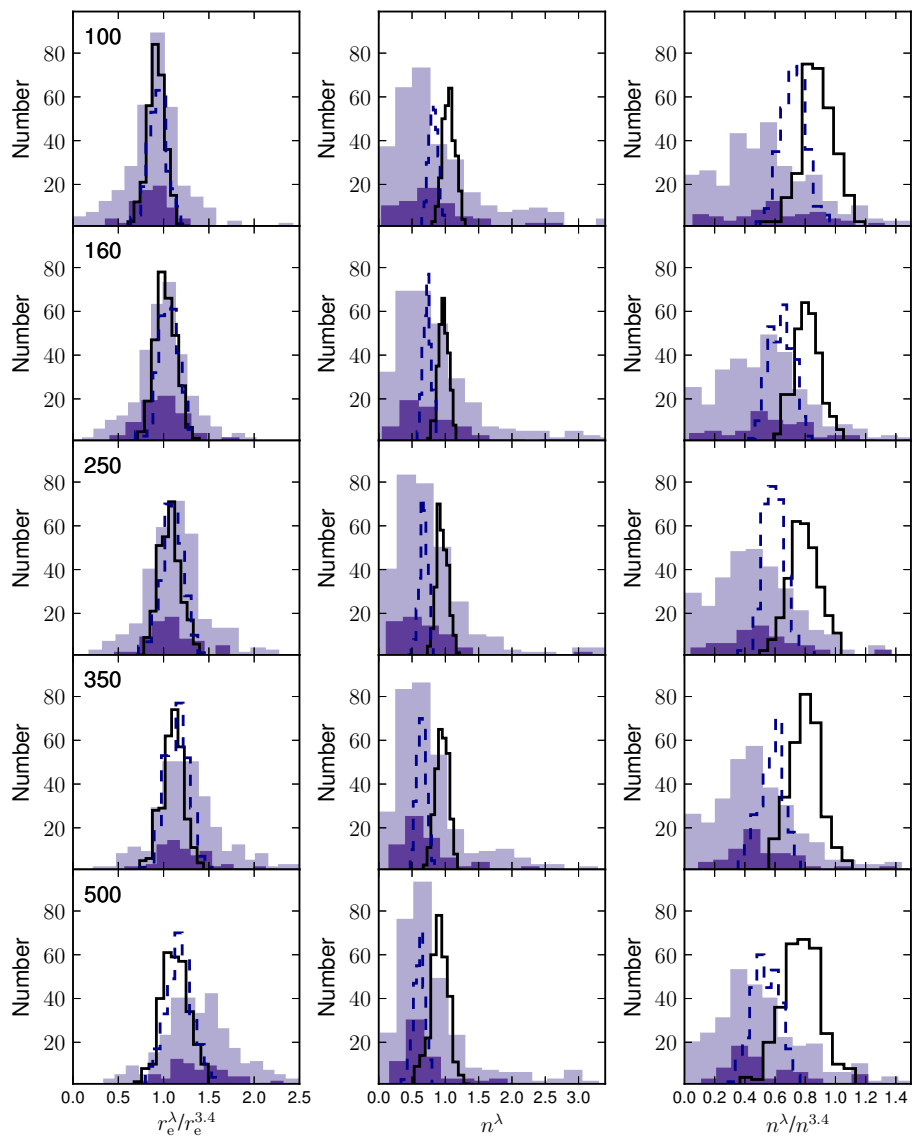}
\caption{Distributions of different parameters over wavelength: left panel -- the ratio of the effective radius in each \textit{Herschel} band to the effective radius in the \textit{WISE\,W1} band; middle panel -- the distribution of the \ser\ index in each \textit{Herschel} band; right panel -- the ratio of the \ser\ index in each \textit{Herschel} band to the \ser\ index in the \textit{WISE\,W1} band. By the lilac colour we show the subsample of 320 galaxies with reliable fitting results, whereas the dark lilac corresponds to the reference sample of 71 galaxies. The black solid and dark blue dashed lines refer to the simulations from the Appendix~\ref{sec:mock} with an exponential and \ser\ disc, respectively.}
\label{cors_IVm}
\end{figure*}

In Fig.~\ref{cors_IVm} we show different distributions of the retrieved structural parameters. We also show distributions of the fitted parameters for the mock galaxies from the Appendix~\ref{sec:mock}: for the exponential dust disc (the black solid line) and the \ser\ dust disc with $n_\mathrm{d}=0.6$ (the dark blue dashed line). 
 
Table~\ref{tab:Results_distrs} lists the mean and median values with the standard deviations of the retrieved parameters. We can see that, for example, the average \ser\ index at 250$\,\mu$m is $0.67\pm0.37$. 187 galaxies in our sample are fitted with $n^{250}\leq0.75$ ($\langle n^{250} \rangle=0.44\pm0.19$), 87 galaxies have $0.75<n^{250}\leq1.25$ ($\langle n^{250} \rangle=0.96\pm0.14$), and 46 -- with $n^{250}>1.25$ ($\langle n^{250} \rangle=2.17\pm0.79$). Here we use the border values 0.75 and 1.25 as the typical error for determining the \ser\ index of 1.0 in this band is 0.25 (25\%). From these numbers we can see that more than half of the sample galaxies exhibit significantly non-exponential discs. This fact will be discussed in Sect.~\ref{sec:discussion}.

\begin{figure*}
\centering
\includegraphics[width=5.5cm, height=5.5cm, angle=0, clip=]{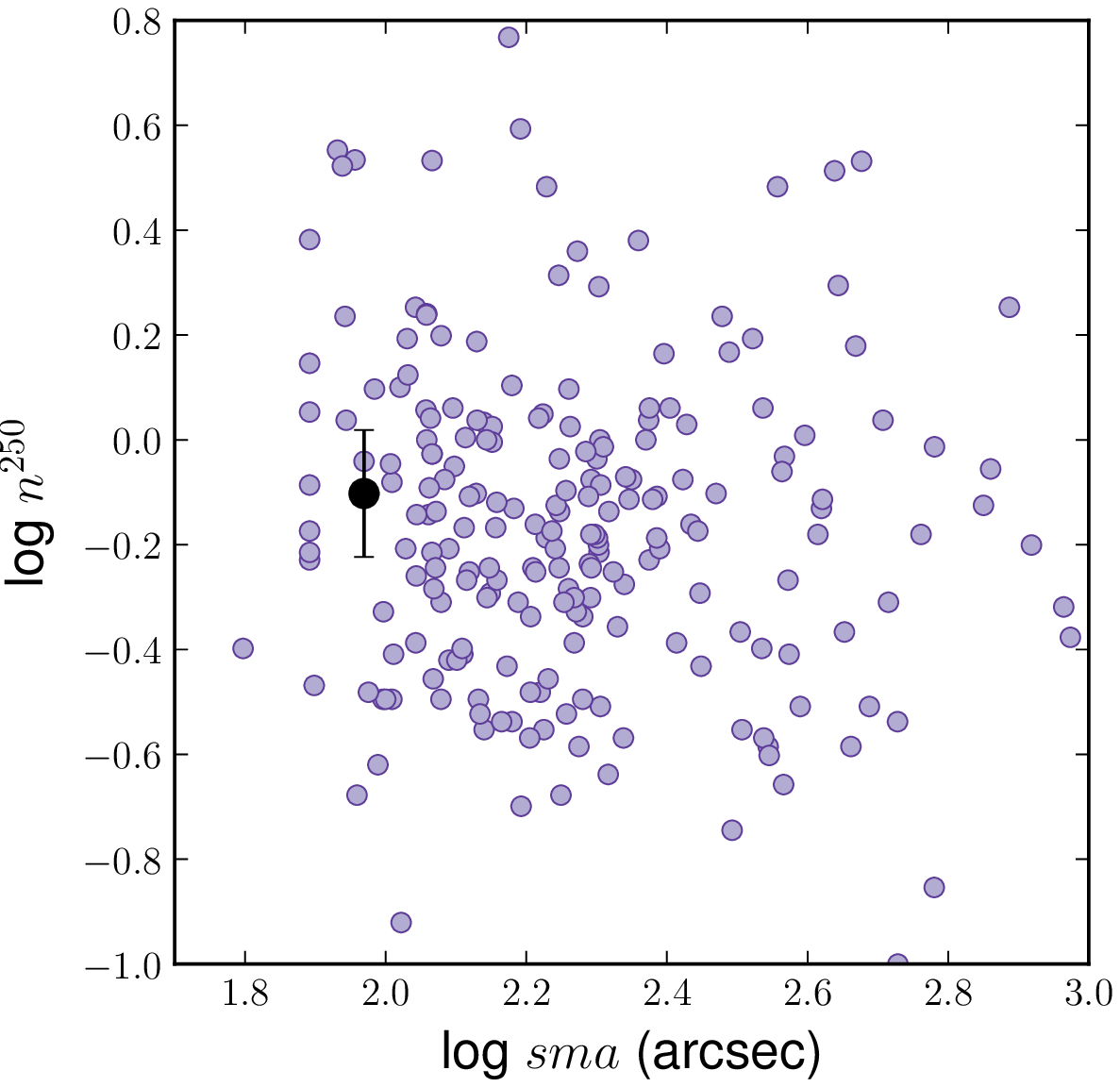}
\includegraphics[width=5.5cm, height=5.5cm, angle=0, clip=]{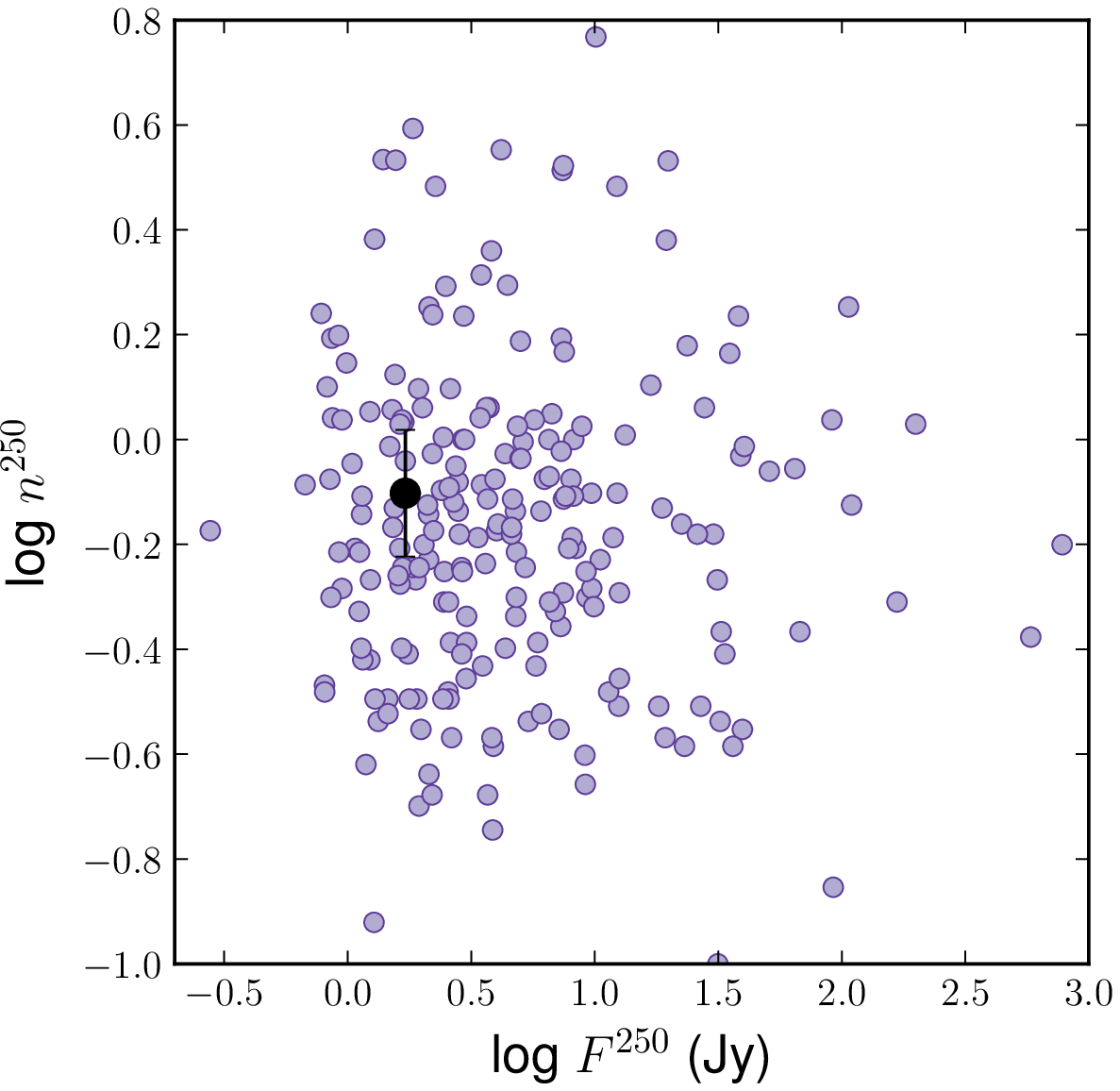}
\includegraphics[width=5.5cm, height=5.5cm, angle=0, clip=]{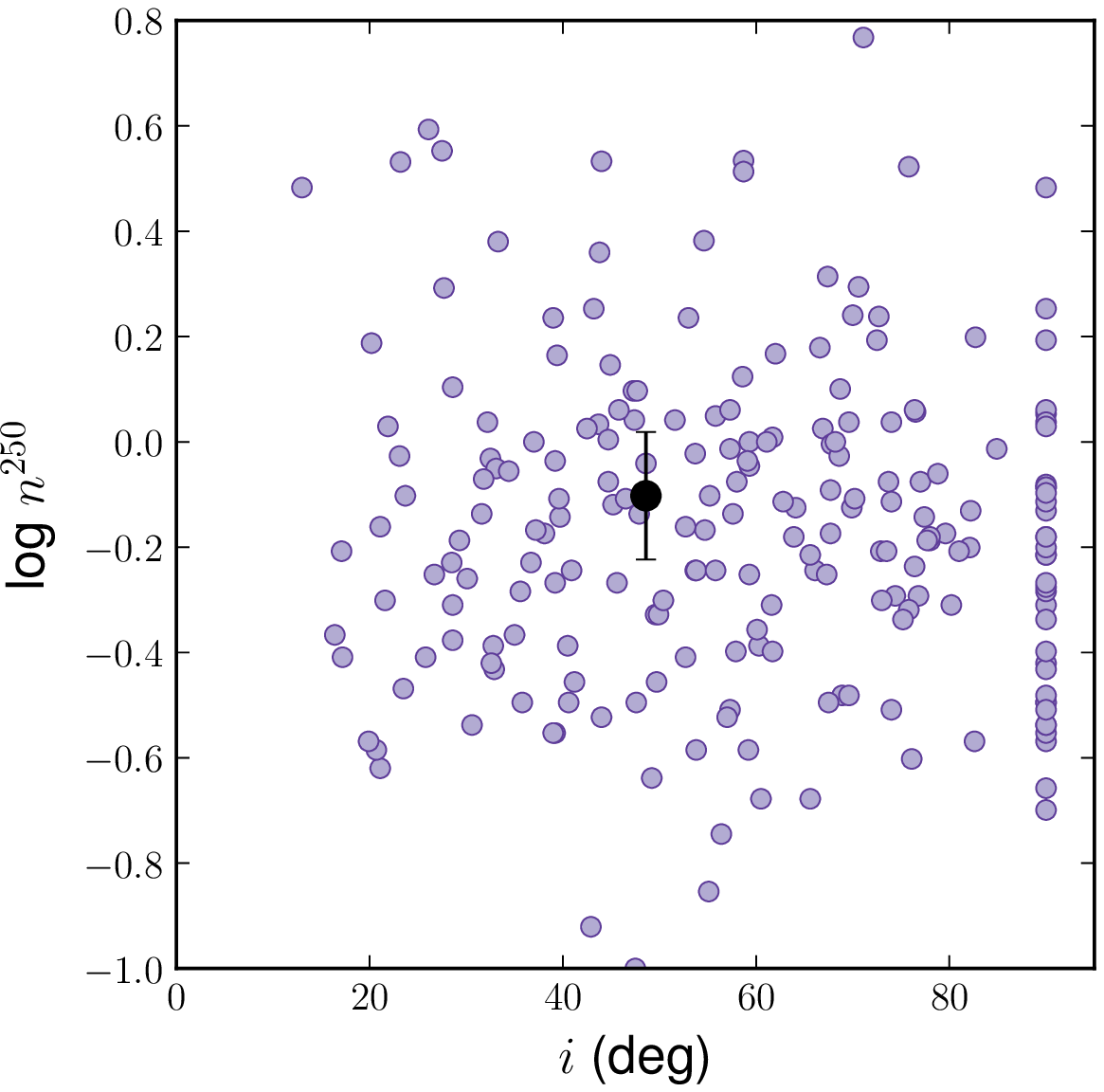}
\caption{Dependences of the \ser\ index on the radius (left panel), flux density (middle panel), and galaxy inclination computed in the Appendix~\ref{sec:add_analysis} (right panel). The black filled circle with the bar corresponds to NGC\,3622 (see text).}
\label{n250_vs_daim}
\end{figure*}

We decided to verify that our results are not affected by the effects of inclination and angular resolution. For all \textit{Herschel} bands we analysed if the \ser\ index correlates with the galaxy radius $sma$, galaxy flux density (both taken from \citetalias{2018A&A...609A..37C}), and galaxy inclination (computed in the Appendix~\ref{sec:add_analysis}). As one can see for SPIRE\,250 (all other bands show similar results) in Fig.~\ref{n250_vs_daim}, no correlations between these quantities are found. Therefore, we are confident that our results are not influenced by any of these effects.

\subsection{Dependence of the parameters on wavelength}
\label{sec:dependence_on_wavelength}

\begin{figure}
\centering
\includegraphics[width=8.5cm, angle=0, clip=]{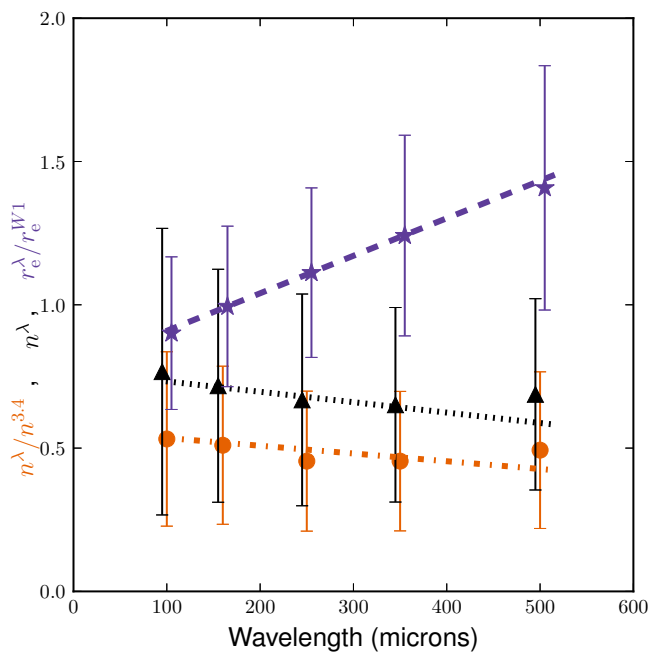}
\caption{Dependences of three different parameters on wavelength: the orange filled circles -- for the ratio of the \ser\ index in each \textit{Herschel} band to the \ser\ index in the \textit{WISE\,W1}, the black filled triangles -- for the \ser\ index, the lilac filled stars -- for the  effective radius in each \textit{Herschel} band to the effective radius in the \textit{WISE\,W1} band.}
\label{pars_wave}
\end{figure}

The dependence of the retrieved parameters on wavelength is shown in Fig.~\ref{pars_wave}.

Here we notice that one should always keep in mind that we do not expect to recover the effective radius of the dust density profile because there is always an IR color gradient (i.e. dust is warmer in the center), and, hence, at any IR wavelength the monochromatic light distribution will be more centrally concentrated (i.e. will have a smaller effective radius) than the actual dust distribution. As to the \ser\ index, our simulations indicate that the dust emission profiles in the SPIRE bands follow the dust mass density profiles very well, i.e. they should have the same \ser\ index.  

\begin{table*}
  \centering
  \caption{Distributions of the fit parameters (mean value, median value (inside the parentheses), and standard deviation).} 
  \label{tab:Results_distrs}
  \begin{tabular}{lccc}
    \hline \hline\\[-2ex]
    Band & $r_\mathrm{e}^{\lambda}/r_\mathrm{e}^{3.4}$ & $n^{\lambda}$ & $n^{\lambda}/n^{3.4}$ \\
    \hline\\[-1.5ex]
\textit{WISE\,W1} & 1.0 & $1.47(1.32) \pm 0.65$ & 1.0 \\     
PACS\,100 & $0.90(0.92) \pm 0.27$ & $0.77(0.65) \pm 0.50$ & $0.53(0.52) \pm 0.30$ \\ 
PACS\,160 & $0.99(1.00) \pm 0.28$ & $0.72(0.65) \pm 0.41$ & $0.51(0.51) \pm 0.28$ \\ 
SPIRE\,250 & $1.11(1.11) \pm 0.30$ & $0.67(0.62) \pm 0.37$ & $0.45(0.45) \pm 0.24$ \\ 
SPIRE\,350 & $1.24(1.22) \pm 0.35$ & $0.65(0.61) \pm 0.34$ & $0.45(0.44) \pm 0.24$ \\ 
SPIRE\,500 & $1.41(1.37) \pm 0.43$ & $0.69(0.65) \pm 0.33$ & $0.49(0.46) \pm 0.27$ \\ 
    \hline\\
  \end{tabular}
\end{table*} 

In the left panel of Fig.~\ref{cors_IVm} we presented the ratio of the effective radius in each \textit{Herschel} band to the effective radius in the \textit{WISE\,W1} band. For the sample galaxies, we can see a gradual increase of this ratio with wavelength, from $0.90\pm0.27$ in PACS\,100 to $1.41\pm0.43$ in SPIRE\,500. The view of the distributions for the mock galaxies is similar for both the exponential and \ser\ discs. From the comparison of the distributions for the real and mock galaxies one fact is obvious: the increase of the effective radius with wavelength is steeper for the real galaxies than for the simulated galaxies. Though the ratio $r_\mathrm{e}^{100}/r_\mathrm{e}^{3.4}=0.90$ for the observations is close to the simulations (0.91--0.94), the ratio $r_\mathrm{e}^{500}/r_\mathrm{e}^{3.4}=1.41$ is underestimated in our simulations by 20--25\%. This can be explained by several reasons. First, the dust temperature gradient of the model for the prototype edge-on galaxy IC\,2531 may differ from what we observe in the other galaxies, with different inclination angles. Second, the dust energy balance problem, observed in nearby edge-on galaxies (see \citealt{2000A&A...362..138P, 2001A&A...372..775M, 2004A&A...425..109A, 2005A&A...437..447D,2010A&A...518L..39B, 2011A&A...527A.109P,2012MNRAS.419..895D, 2015MNRAS.451.1728D,2016A&A...592A..71M} and recent study by \citealt{2018A&A...616A.120M}), may be responsible for the observed steeper increase of the effective radius in the real galaxies: in the FIR/submm, the dust emission derived from RT calculations underestimates the real observations by a factor 1.5--4. The dust heating mechanisms \citep{2011ApJ...738..124L,2011AJ....142..111B,2012MNRAS.419.1833B,2015MNRAS.448..135B,2015A&A...576A..33H} and the complex large- and small-scale dust structure \citep{2000A&A...362..138P, 2001A&A...372..775M, 2008A&A...490..461B, 2011A&A...531L..11B, 2011ApJ...741....6M, 2012MNRAS.419..895D, 2012A&A...541L...5H,2018A&A...616A.120M}, observed in galaxies, are among the possible explanations for this problem.

In the middle panel of Fig.~\ref{cors_IVm} we showed the distribution of the \ser\ index in each band. As we can see, the distributions look similar in all bands, though we can definitely distinguish an asymmetry of the $n^{\lambda}$-histograms: a vast majority of galaxies are concentrated near $n^{\lambda}\sim0.5$, whereas a flat distribution of galaxies with $n^{\lambda}>1$ is also present. The similar distributions by $n^{\lambda}$ in all \textit{Herschel} bands (this is seen well in Fig.~\ref{pars_wave}) implies that the dust emission has an essentially identical profile in the whole FIR/submm domain. However, according to our simulations, the fitted profile of the dust emission has a systematically lower \ser\ index with increasing wavelength. From this comparison we can conclude that the \ser\ model for the dust density distribution with a \ser\ index of about 0.5--0.7 fits the observations much better than the model with an exponential profile, which lies significantly off the peaks for the observations in all \textit{Herschel} bands. This is also seen well when we consider the ratio $n^{\lambda}/n^{3.4}$ (see the right panel of Fig.~\ref{cors_IVm}). We can see that the observed distributions in all five bands are rather broad ($\sigma=0.3$), with a mean value of approximately 0.5. However, we should notice that this ratio is overestimated (by 10--20\%) in our simulations as compared to the real galaxies. This can be explained by the fact that our simulations include only one prototype galaxy with a fixed bulge-to-total ratio, whereas galaxies in the selected sample may have a different bulge contribution, i.e. have a different morphological type and, consequently, different general \ser\ index in the \textit{WISE\, W1} band (see Fig.~\ref{n_Type} in Sect.~\ref{n_Type}). 

\begin{table*}
  \centering
  \caption{The numbers of galaxies in each bin ($T-0.5$, $T+0.5$] for a subsample of 320 galaxies with reliable fitting results.} 
  \label{tab:Hubble_distr}
  \begin{tabular}{lccccccccccccccc}
    \hline \hline\\[-2ex]
	$T$ & -4 & -3 & -2 & -1 & 0 & 1 & 2 & 3 & 4 & 5 & 6 & 7 & 8 & 9 & 10 \\
     Type & E & S0$^{-}$ & S0$^{\circ}$ & S0$^{+}$ & S0a & Sa & Sab & Sb & Sbc & Sc & Scd & Sd & Sdm & Sm & Ir \\
    \hline\\[-1.5ex]
	 Number & 1 & 3 & 2 & 13 & 4 & 21 & 27 & 40 & 47 & 52 & 54 & 25 & 14 & 6 & 11\\
    \hline\\
  \end{tabular}
\end{table*} 

\subsection{Dependence of the parameters on morphology}
\label{sec:dependence_on_morphology}

We also studied the dependence of the fitted parameters on morphological type. The distributions are shown in Table~\ref{tab:Hubble_distr}. As the number of early-type galaxies in the selected sample of 320 galaxies is not enough to consider them in a bin of 1 numerical Hubble stage $T$ separately, for them we use the broad bin [-4,0), i.e. we consider them as a single type of ``lenticulars''. As to Sdm ($T=8$) and Sm ($T=9$) galaxies, we combine them in one type Sdm for the same reason.

\begin{figure*}
\centering
\includegraphics[width=5.5cm, angle=0, clip=]{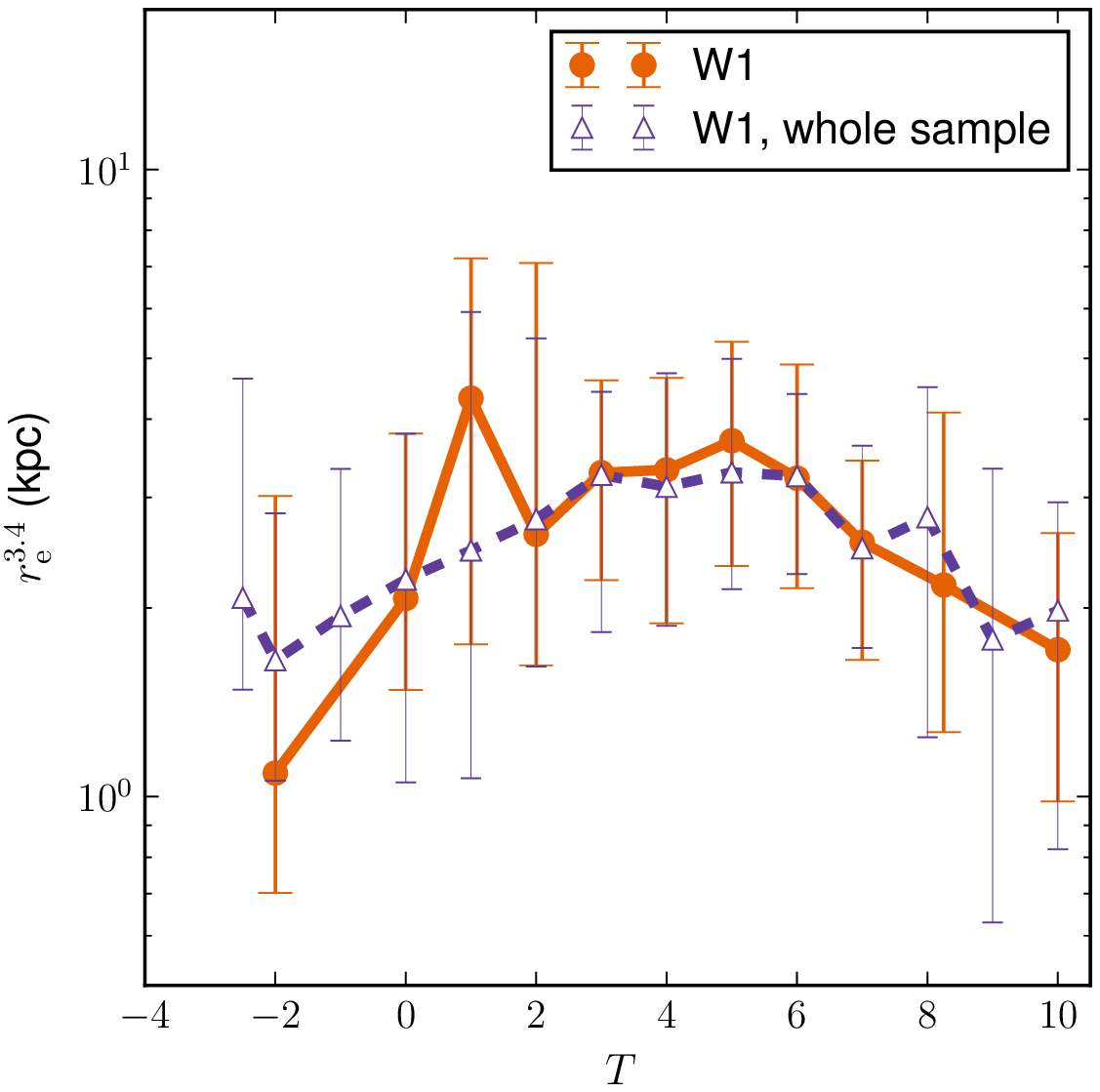}
\includegraphics[width=5.5cm, angle=0, clip=]{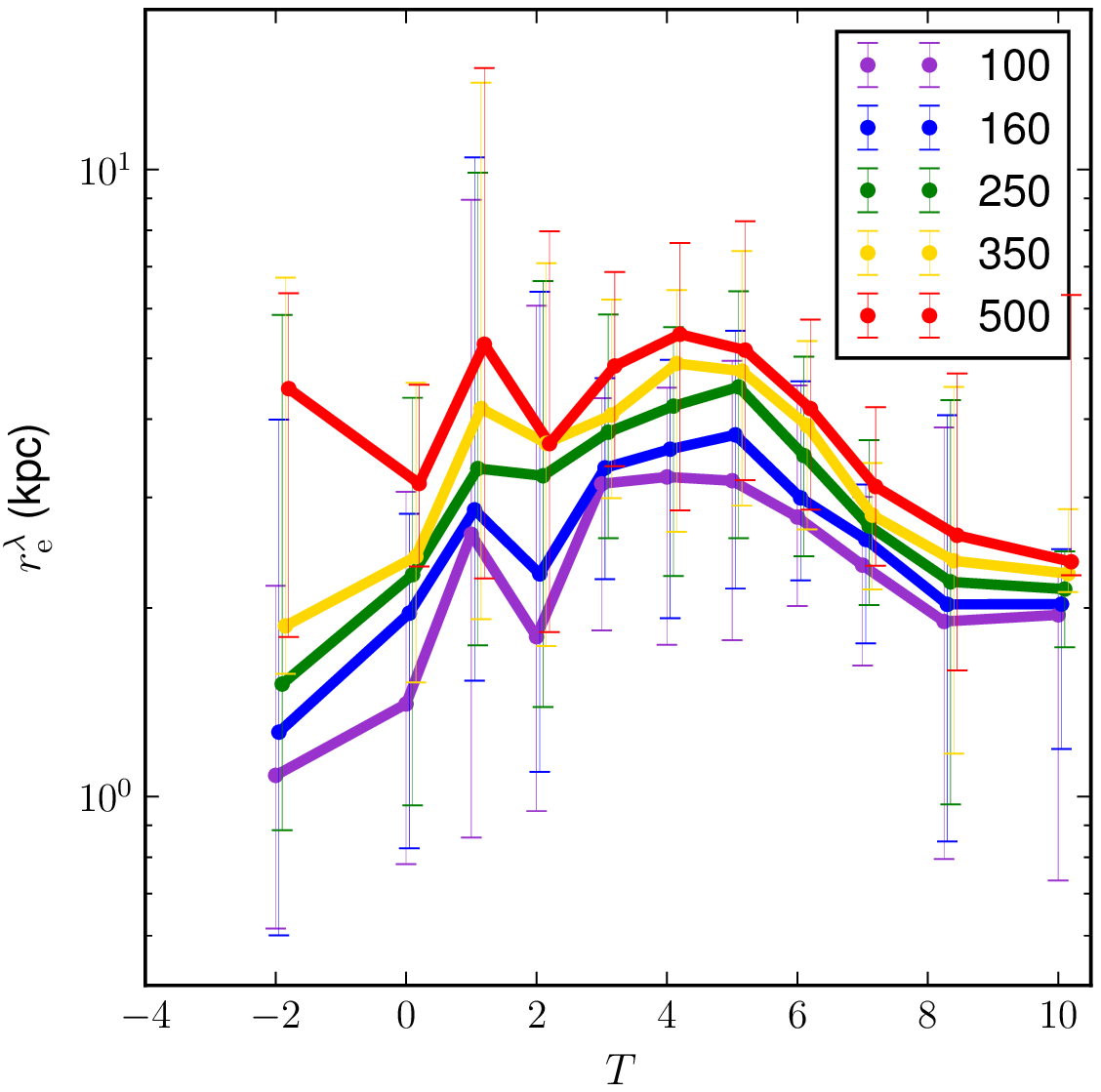}
\includegraphics[width=5.5cm, angle=0, clip=]{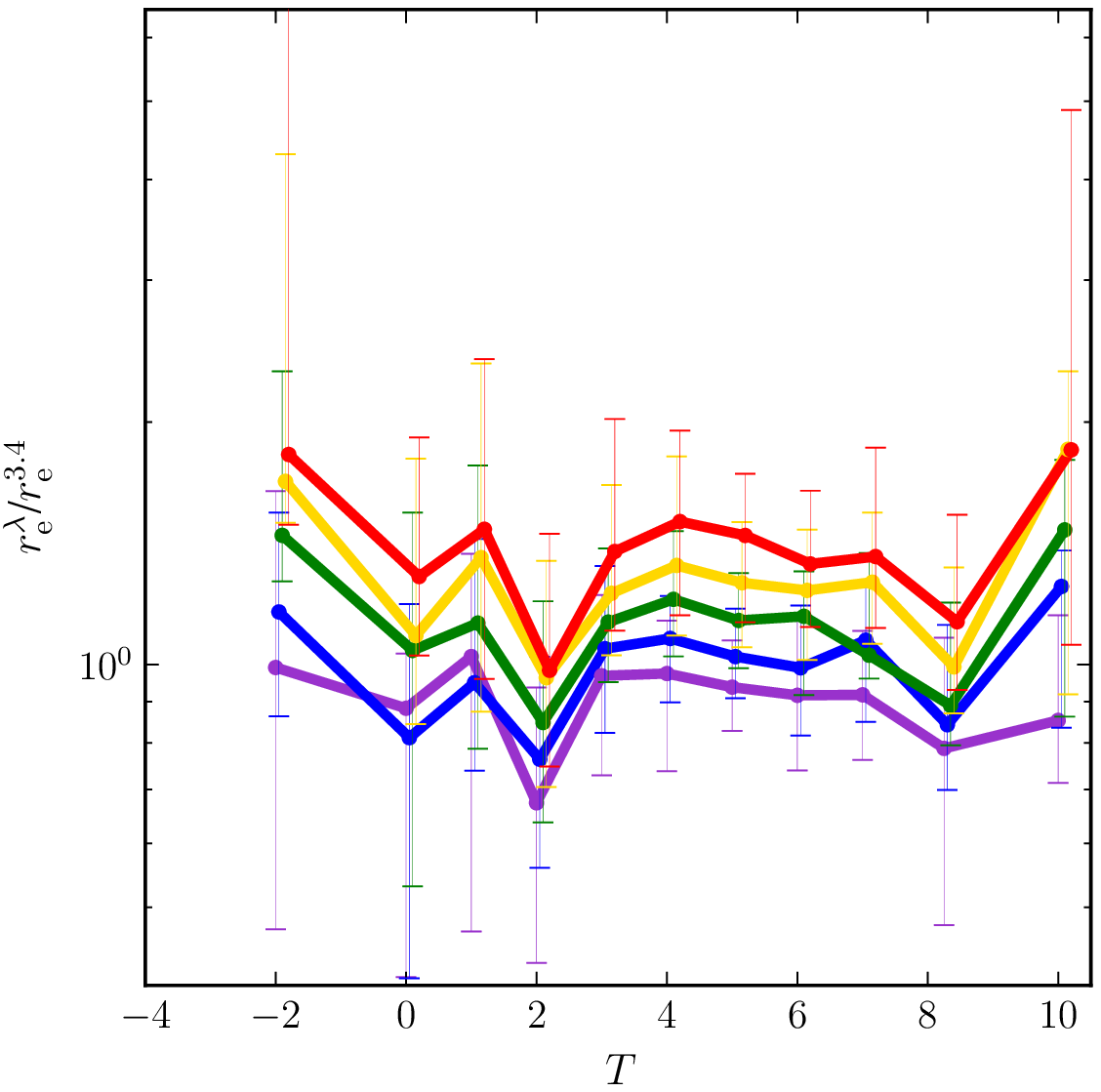}
\caption{
The dependence of the effective radius measured in the \textit{WISE\,W1} band on the Hubble stage (left panel), the dependences of the effective radius measured in the five \textit{Herschel} bands on the Hubble stage (middle panel), and the dependences of the ratio of the effective radius in each \textit{Herschel} band to the effective radius in the \textit{WISE\,W1} band on the Hubble stage (right panel). Each dependence is plotted for the averaged data for the subsample of 320 galaxies (except for the lilac line in the left plot which refers to the sample of 865 DustPedia galaxies) within a bin of 1 $T$, with 15 and 85 percentiles shown. All lenticular galaxies ($T\leq0$) are combined together and shown by $T=-2$. Sm and Sdm galaxies are joined together in one type $T=8$.}
\label{re_distrs}
\end{figure*}

The dependence of the effective radius as a function of morphological type is shown in Fig.~\ref{re_distrs}, the left and middle panels. In general, the scatter is large and the apparent trend in each band is feeble, within the scatter. In all bands, we can see an increase of the effective radius with wavelength up to $T\approx4-5$: on average, the most extended dust components, as well as stellar discs, are found in galaxies of these morphological types. For galaxies with $T>5$, we observe a decrease of the effective radius to the smallest discs in Sm galaxies (note that this decrease becomes sharper with increasing wavelength). However, for our selected sample a clear bump is seen for galaxies with $T=1$. A closer look at these galaxies revealed that many of them are viewed edge-on and possess prominent stellar haloes (e.g. NGC\,4235, NGC\,4260, NGC\,4300, NGC\,4424, NGC\,4506), similar to the Sombrero galaxy. Thus, the large effective radius of the single \ser\ model for these galaxies can be reasonably explained by their extended haloes. By the dashed line in the left plot we also show the distribution for the whole sample and can see a similar behaviour as for our subsample of 320 galaxies but without a bump at $T=1$. This means that the bump at $T=1$ for the subsample is likely a selection effect. 

The right panel of Fig.~\ref{re_distrs} shows the ratio $r_\mathrm{e}^{\lambda}/r_\mathrm{e}^{3.4}$. We can see that this ratio is almost constant with type in all \textit{Herschel} bands, with a tendency of decreasing $r_\mathrm{e}^{\lambda}/r_\mathrm{e}^{3.4}$ for late-type spirals. In the SPIRE bands, lenticular galaxies exhibit slightly more extended cold dust discs, as compared to their stellar discs, than spiral galaxies (see also \citealt{2012ApJ...748..123S}). The same tendency is observed in irregular galaxies which often show intensive star formation over the whole galaxy.

\begin{figure}
\centering
\includegraphics[width=8.2cm, height=8.3cm, angle=0, clip=]{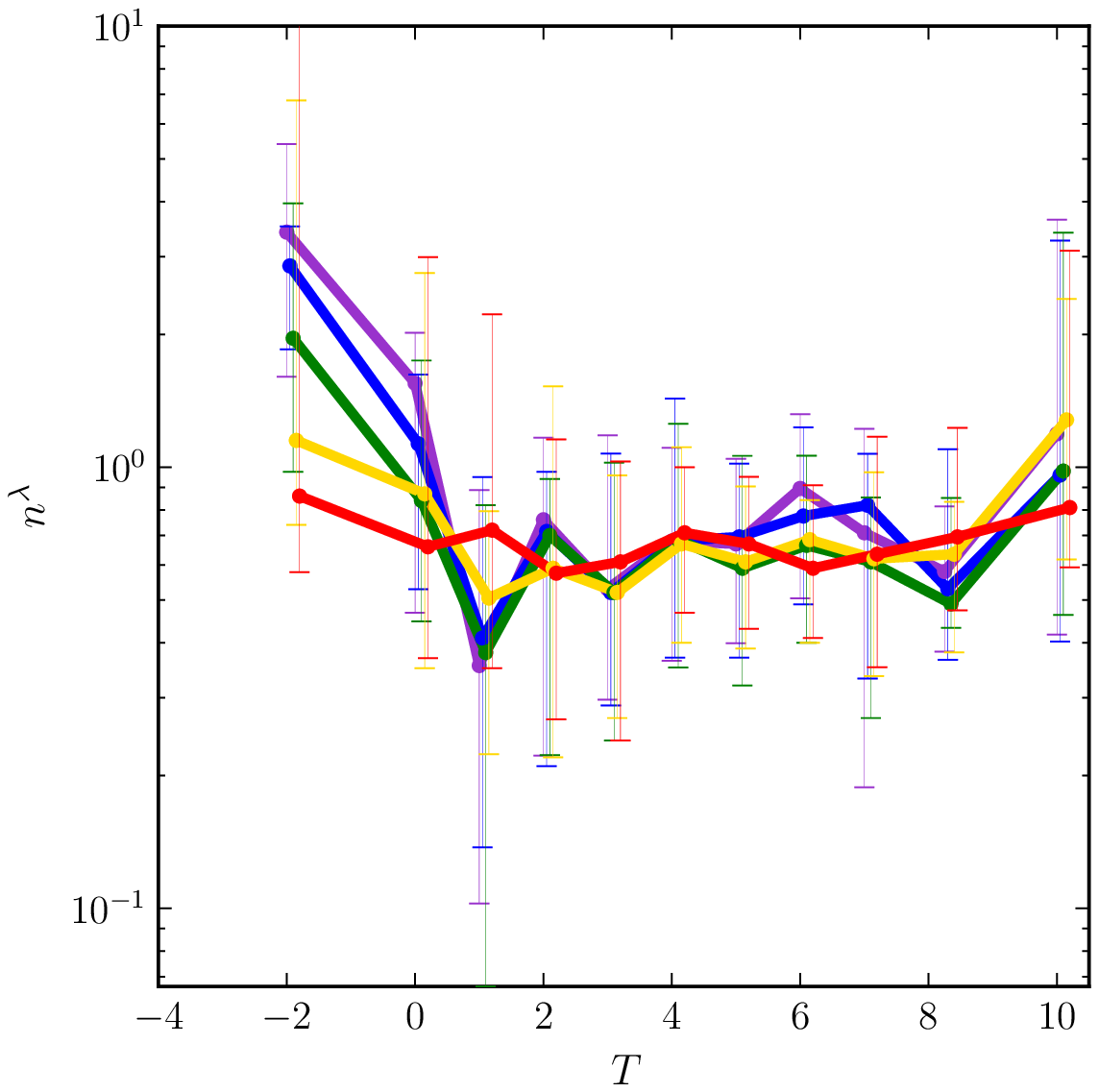}
\includegraphics[width=8.1cm, height=8.3cm, angle=0, clip=]{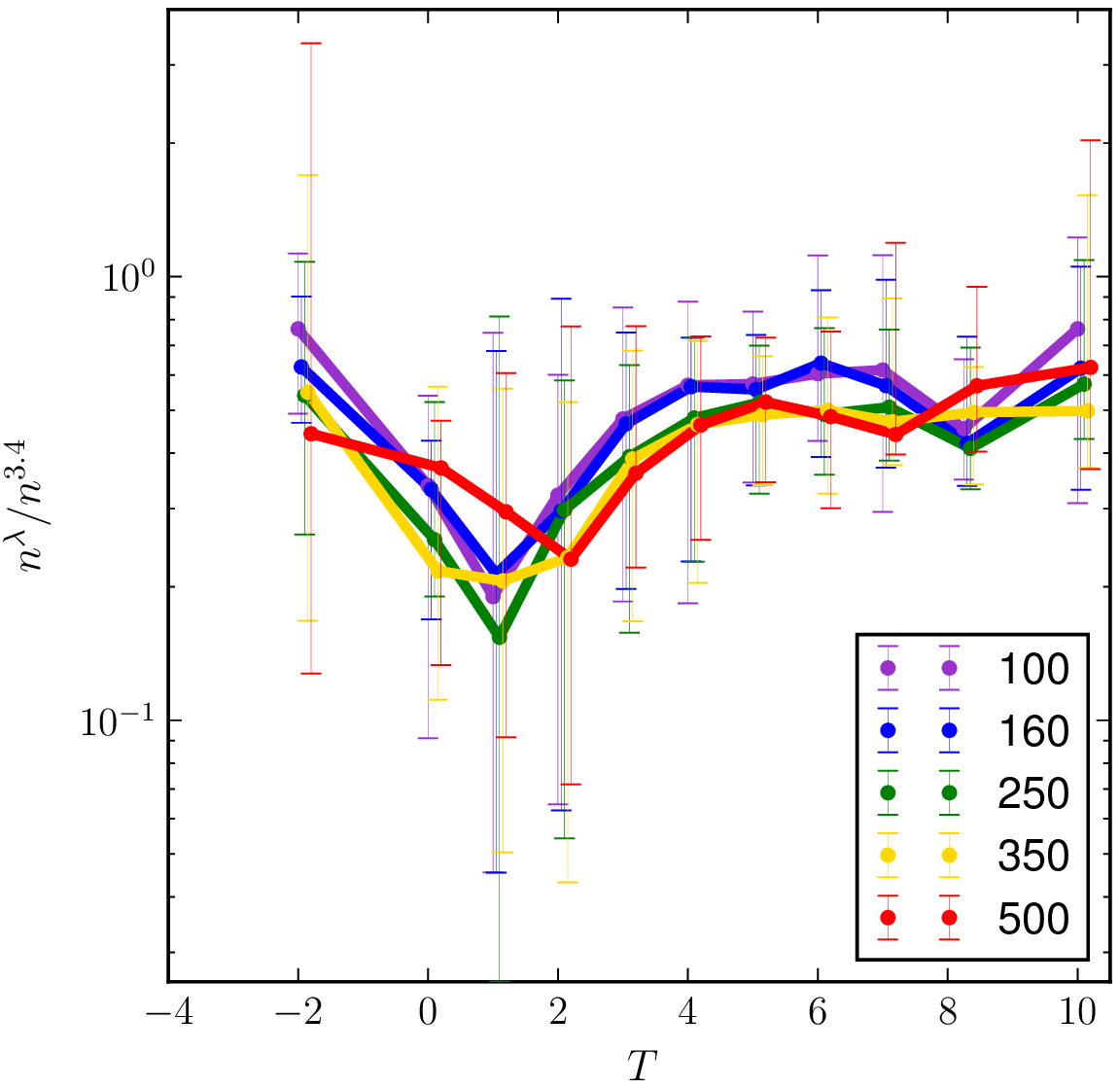}
\caption{The dependences of the \ser\ index on the type (top panel) and the ratio of the \ser\ index in each \textit{Herschel} band to the \ser\ index in the \textit{WISE\,W1} band on the type (bottom panel). The dependences are plotted for the averaged data for the subsample of 320 galaxies in each bin of 1 $T$ with 15 and 85 percentiles shown. All lenticular galaxies ($T\leq0$) are combined together and shown by $T=-2$. Sm and Sdm galaxies are joined together in one type $T=8$.}
\label{n_distrs}
\end{figure}

\begin{figure}
\centering
\includegraphics[width=8.5cm, angle=0, clip=]{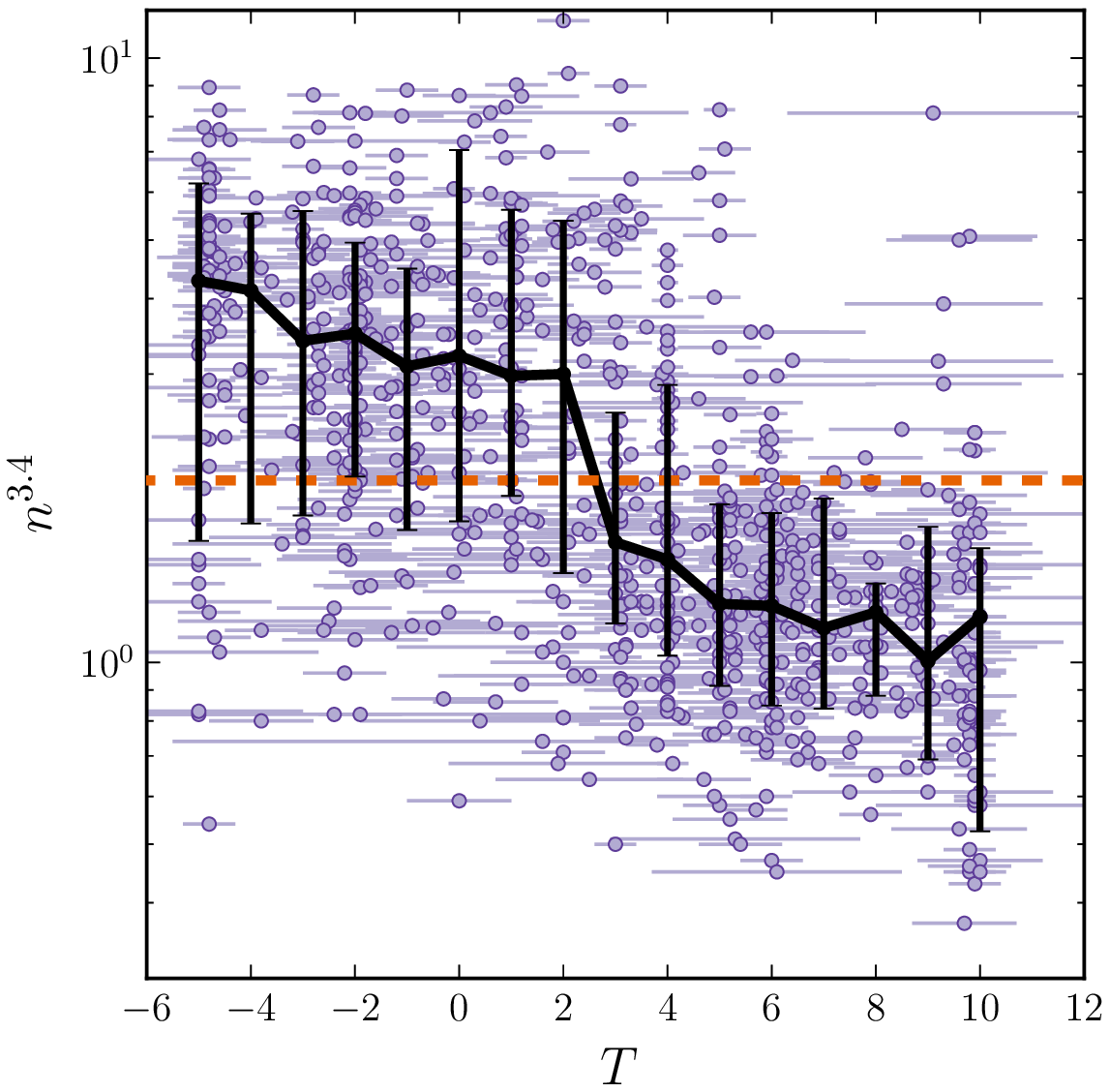}
\caption{The relationship between the \ser\ index in the \textit{WISE\,W1} band and the Hubble stage for the sample of 865 DustPedia galaxies. The median average and the 15 and 85 percentiles are shown by the black colour for each bin of 1 $T$. The orange dashed line refers to $n^{3.4}=2$, the border value to split galaxies into bulgeless ($n^{3.4}<2$) and bulge-dominated galaxies ($n^{3.4}>2$).}
\label{n_Type}
\end{figure}

In Fig.~\ref{n_distrs} (top panel) and Fig.~\ref{n_Type}, we present the distribution of the \ser\ index over Hubble stage. For the \textit{Herschel} data, we do not see a statistical difference of $n^{\lambda}$ for all morphological types, except that lenticular galaxies exhibit a large difference between the \ser\ index in all five bands, reaching values larger than 3 in PACS\,100 and going down to approximately 0.9 in SPIRE\,500.
We interpret this steep change of the \ser\ index by a possible bias in their fitting. The selected lenticular galaxies are rather small and often exhibit a complex morphology, with rings (e.g. NGC\,4324 and NGC\,5701), lenses (e.g. NGC\,3626), and different peculiarities (e.g. ESO\,495-021 and NGC\,2685). Two lenticular galaxies (IC\,691 and NGC\,7465) are classified as blue compact galaxies \citep{1981ApJ...247..823T}. Going from PACS\,100 to SPIRE\,500, these features can be washed out because of the increasing beam size. Nevertheless, we can discern that typical profiles of the selected lenticular galaxies in SPIRE\,250 often have extended wings (NGC\,5145), some galaxies show rather abrupt profiles with $n^{250}<0.5$ (NGC\,4531).

For the \textit{WISE\,W1} band (Fig.~\ref{n_Type}), we can see that late-type galaxies ($T>0$) have $n^{3.4}<2$. Notice also that bulgeless galaxies with $T>5$ have a \ser\ index close to 1, which means that, indeed, an exponential profile is a good approximation for describing SB distribution in stellar discs. On the contrary, early-type galaxies with a spheroidal dominance have more extended profiles with $n^{3.4}>2$.

We found that 218 galaxies with a morphological type between 3 and 7 (late-type spirals), which constituent the majority of the sample galaxies, have the mean value $\langle n^{250} \rangle = 0.67\pm0.34$.

The bottom panel in Fig.~\ref{n_distrs} depicts the ratio of the \ser\ index in each \textit{Herschel} band to that in the \textit{WISE\,W1} band. Galaxies with $T<0$ and $T\geq3$ demonstrate an essentially constant ratio of approximately $0.56\pm0.11$, whereas for the remaining galaxies of S0a--Sab types this ratio drops to $0.25\pm0.06$. A closer look at these galaxies revealed that many of them have a bright ring (for example, NGC\,1808 and NGC\,4772) or a grand-design structure (for example, NGC\,4260 and NGC\,5691) which are less apparent in the NIR and very prominent in the FIR. 28 out of 56 galaxies with $0<T<3$ have bright ring structures. This leads to a small \ser\ index ($n^{\lambda}\leq0.3$) in the FIR/submm and a potentially lower ratio $n^{\lambda}/n^{3.4}$ than observed in later spiral galaxies without such bright features in the FIR/submm. In its turn, $n^{3.4}$ decreases with type, and since $n^{\lambda}$ is almost constant for late-type spirals, we observe a gradual increase of $n^{\lambda}/n^{3.4}$ with Hubble type.

\section{Discussion}
\label{sec:discussion}

In the previous section, we presented the main results from our structural analysis of the DustPedia galaxies. In discussing these results, we comment here on some of the more noteworthy issues regarding our study, and concentrate on interpreting the parameter distributions obtained in this work. 

\subsection{On our simple dust emission modelling}
\label{sec:complex_morphology}

The results of our fitting are obtained for a large sample of DustPedia galaxies, with a diverse morphology and structural features. For the reference sample of 71 galaxies, where each galaxy have a smooth, regular distribution of dust emission, the results are essentially the same as for the subsample of 320 galaxies. We compared the distributions of the structural parameters in Fig.~\ref{cors_IVm} for both samples and verified that they show similar statistical features (e.g. the mean, median and standard deviation values). According to the Kolmogorov-Smirnov test, we cannot reject the hypothesis that the distributions of the two samples are the same since the p-value is high (>0.3) for all pairs of distributions presented in Fig.~\ref{cors_IVm}. 
 
For the \textit{Herschel} bands, there are a few galaxies (about 30 in each band) with a very low or large \ser\ index ($n^{\lambda}<0.3$ or $n^{\lambda}>10$). This indicates that a contrast bar, a ring, or a highly asymmetric structure (e.g. spirals) is present. In the case of a ring, the formal \ser\ index appears to be very small (for example, for the ring galaxy NGC\,1436, $n^{\lambda}\approx0.05$). In this case the code is trying to fit the ring structure by a \ser\ function with such a low \ser\ index which corresponds to a flat profile in the central part  and abrupt at the edge of the ring.

\subsection{Change of the effective radius with wavelength}
\label{sec:effective_radius}

Since in our study we derived the geometrical scale $r_\mathrm{e}^{\lambda}$ of the dust parameter using the \ser\ profile for the first time, it is impossible to directly compare it with the literature where an exponential law was fitted to the observed galaxy profiles. However, if from our sample we select galaxies with $n^\lambda\approx1$ (e.g., in the SPIRE\,250$\,\mu$m band, approximately 50 galaxies demonstrate exponential profiles), we can compute the disc scalelength as $h^{\lambda}\approx r_\mathrm{e}^{\lambda}/1.678$. For these galaxies we find $h^{160}/h^{100}=1.16\pm0.05$, $h^{250}/h^{100}=1.33\pm0.07$, $h^{350}/h^{100}=1.51\pm0.15$, and $h^{500}/h^{100}=1.64\pm0.26$. These are well-consistent with the results from  \citet{2017A&A...605A..18C} for a sample of 18 face-on spiral galaxies from the DustPedia sample: $h^{160}/h^{100}=1.10\pm0.01$, $h^{250}/h^{100}=1.30\pm0.03$, $h^{350}/h^{100}=1.45\pm0.05$, and $h^{500}/h^{100}=1.60\pm0.01$. The comparison with \citet{2016MNRAS.462..331S} for $h^{500}/h^{250}$ gives 1.23 versus their 1.15 (and 1.25 from  \citealt{2017A&A...605A..18C}). 

As \citet{1998A&A...335..807A} and \citet{2017A&A...605A..18C}, we attribute the steady increase of the effective radius of the dust emission with radius to the cold-warm dust temperature gradient. The dust heating in turn is consistent with the FIR colour gradient observed in the disc. This suggests that part of the FIR emission arises from grains heated by the diffuse ISRF gradually decreasing with galactocentric distance.

\subsection{Non-exponentiality of dust discs: comparison with the literature}
\label{sec:non_exp_literature}

\citet{2009ApJ...701.1965M} studied dust surface density profiles for the SINGS galaxies and concluded that in most cases they show an exponential behaviour. They also pointed out that the dust profiles in S0/a and Sab spirals usually have central depressions. Our results confirm their conclusion regarding early-type spirals -- as was shown in Fig.~\ref{n_distrs}, many early-type spirals in our sample have a \ser\ index close to 0.4--0.5 in the FIR/submm domain. Because of the flatness of the SB profile in the central region in these galaxies, the \ser\ index appears to be typically lower than 0.5. We should notice, however, that if we carefully look at the de-projected dust mass surface density profiles in their fig.~8, we can find many profiles with a central depletion or, vice versa, an excess of dust density among galaxies of Sb-Sd types and irregular galaxies. Similarly, in our sample galaxies we observe a large scatter of the \ser\ index -- profiles with a deficit or excess in the centre are present. The average profiles, which they provide in the same figure, obviously wash out the details for each individual galaxy and do not reflect the diversity of dust mass density profiles in the galaxies under consideration. For example, by averaging two profiles with a deficit and excess of dust mass density in the centre, we might obtain a fairly exponential decline in both the inner and outer parts of the profile.

The same explanation might be applicable to the conclusion by \citet{2016MNRAS.462..331S} who reported that `the distribution of dust is consistent with an exponential at all radii'. This result was obtained for an average SB profile by combining the signal of 110 spiral galaxies from the \textit{Herschel} Reference Survey \citep{2010PASP..122..261B}. A closer look at their figs.~2 and 3, indeed, does not reveal a deficiency in the central regions of their SB profiles. On the contrary, we can see a little excess in the centre (within $r/r_{25}<0.1$, where $r_{25}$ is defined as the radius of the isophotal level 25 mag\,arcsec$^{-2}$ in the $B$ band) at 250\,$\mu$m and 350\,$\mu$m.  Using their subsample of 45 large galaxies ($r_{25}>1.5\arcmin$), we found 27 common galaxies with our sample. This number is almost two times smaller than the number of the selected HRS galaxies because in our subsample only galaxies with all five PACS\,100--SPIRE\,500 observations were selected. Some galaxies (for example, NGC\,4298 and NGC\,5363) have the `C', `R' or `N' flags in the DustPedia database, and, therefore, we did not include them in our subsample. For the 27 common galaxies we computed the following average \ser\ indices: $0.67 \pm 0.37$ (SPIRE\,250), $0.62 \pm 0.29$ (SPIRE\,350), and $0.58 \pm 0.26$ (SPIRE\,500). Thus, for these common galaxies the results of our \ser\ fitting confirm the conclusion drawn for the larger sample -- the FIR dust emission has a distribution which is closer to a Gaussian rather than to an exponent.

In another work, \citet{2017A&A...605A..18C} made use of the DustPedia data to analyse SB profiles in the FIR and to accurately create dust mass density profiles for 18 face-on spiral galaxies. In their fig.~A.2 we can see examples of galaxies where surface density profiles for the mass of dust have an obvious depletion in the inner galaxy region (NGC\,300, NGC\,628, NGC\,3031, NGC\,3521, NGC\,3621, NGC\,4725, NGC\,4736, and NGC\,7793). Most of these galaxies are present in our subsample, and for them the mean \ser\ index in the SPIRE\,250$\,\mu$m band is $0.62\pm0.35$. 

We should point here that if we do not take into account the depression in the inner galaxy region and the truncation in the galaxy outskirts, which are often observed in galaxies, the dust emission and dust mass density profiles look more or less exponential (see further).

\subsection{Non-exponentiality of dust discs: possible reasons}
\label{sec:non_exp_reasons}

One of the most striking results of the present work is a non-exponential profile of the dust distribution in galaxies. Unlike the stellar distribution, the dust emission profile for many galaxies in our sample is close to Gaussian. Although our sample is incomplete in any sense due to selection restrictions, the representativity of the DustPedia sample, covering a wide range of morphological types and physical conditions within the interstellar medium of galaxies, should not affect the main finding of this study -- non-exponentiality of dust discs. The presence of the substantial number of galaxies with Gaussian dust discs, whereas their stellar distribution has an exponential decline or close to it, is an interesting fact by itself and needs an interpretation.

In Fig.~\ref{profile_examples} we show some typical examples of galaxies with different types of SB profiles in the SPIRE\,250$\,\mu$m band which are discussed below. 

\begin{figure*}
\centering
\includegraphics[width=6.8cm, angle=0, clip=]{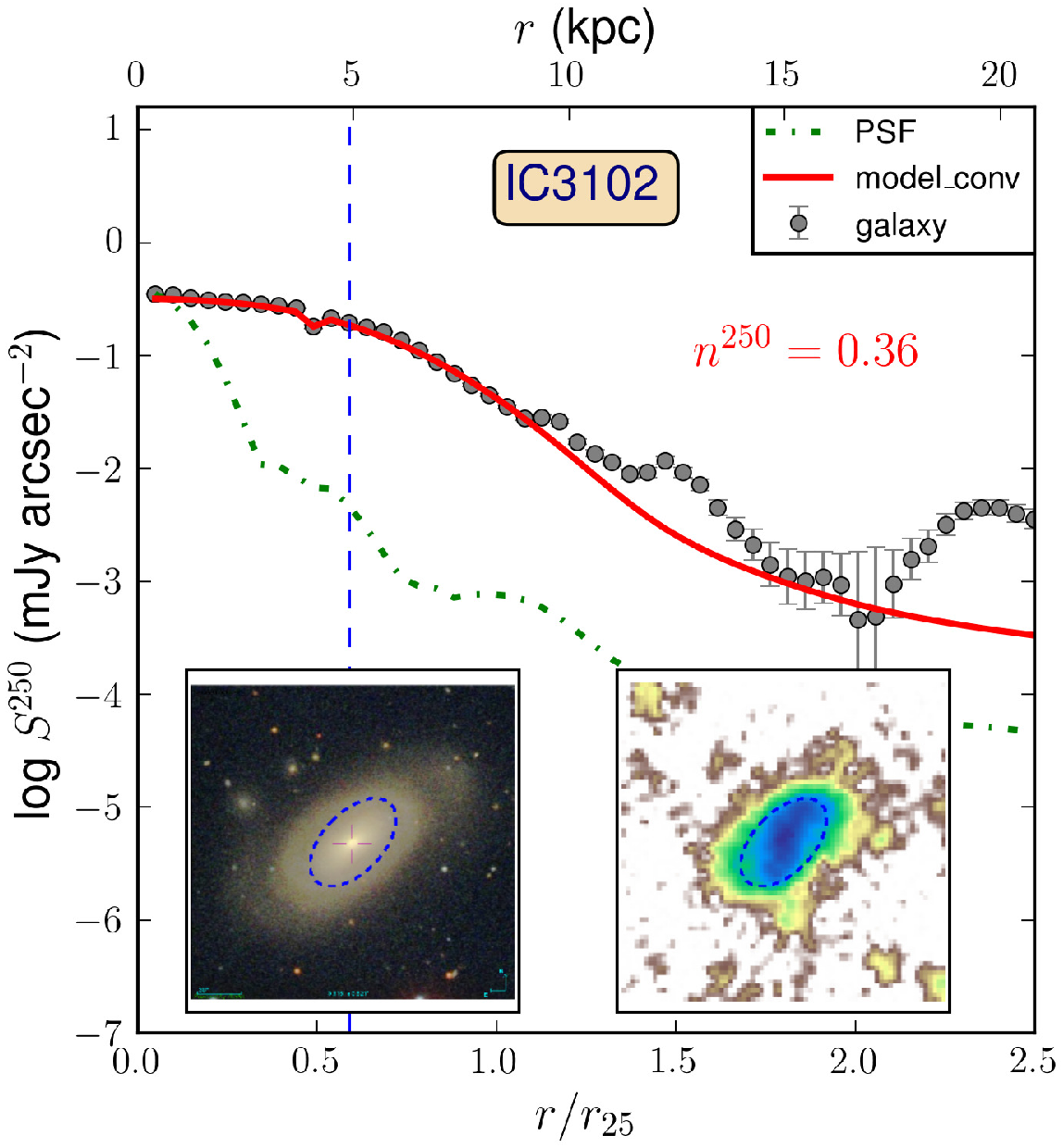}
\includegraphics[width=6.8cm, angle=0, clip=]{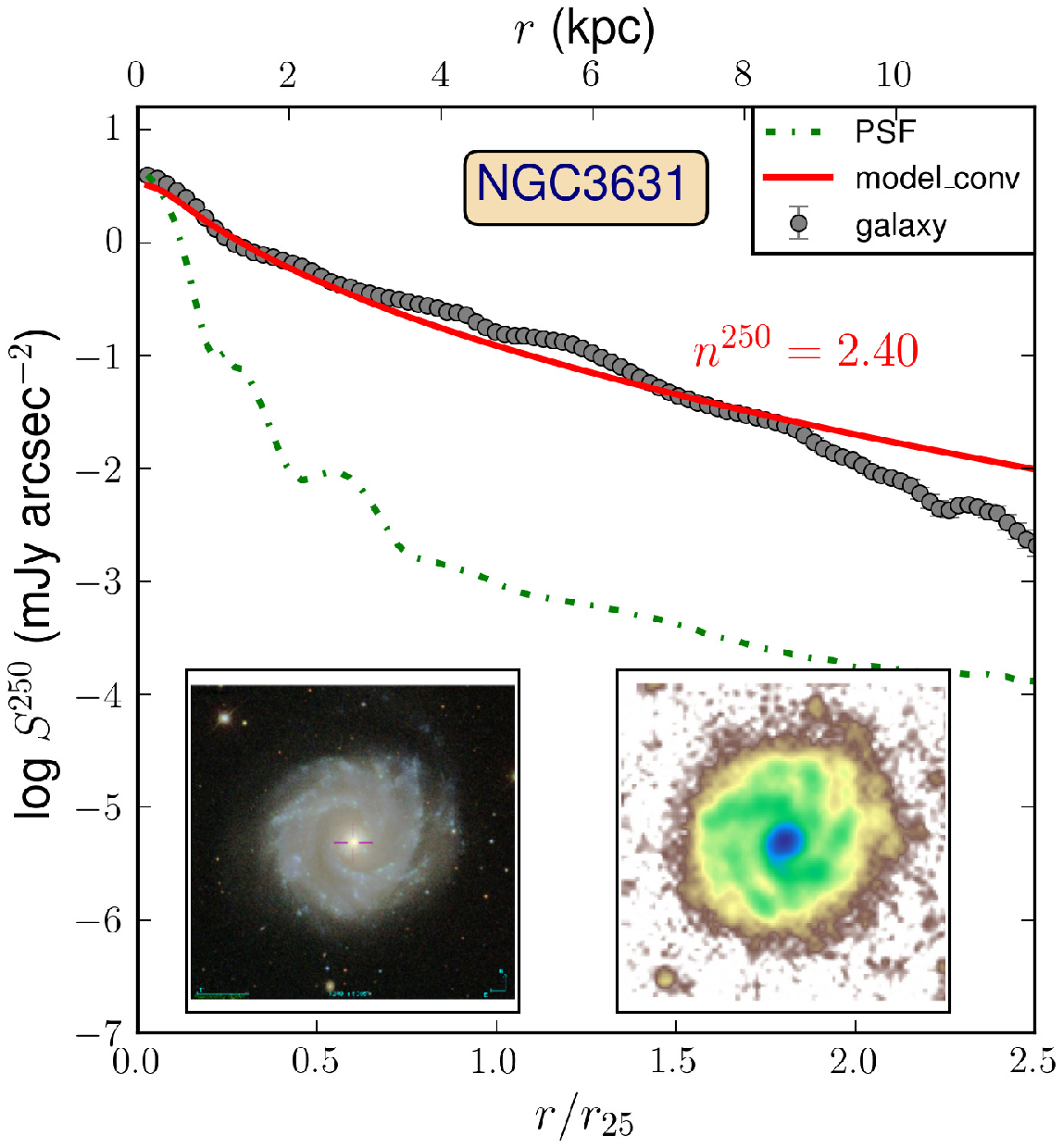}
\includegraphics[width=6.8cm, angle=0, clip=]{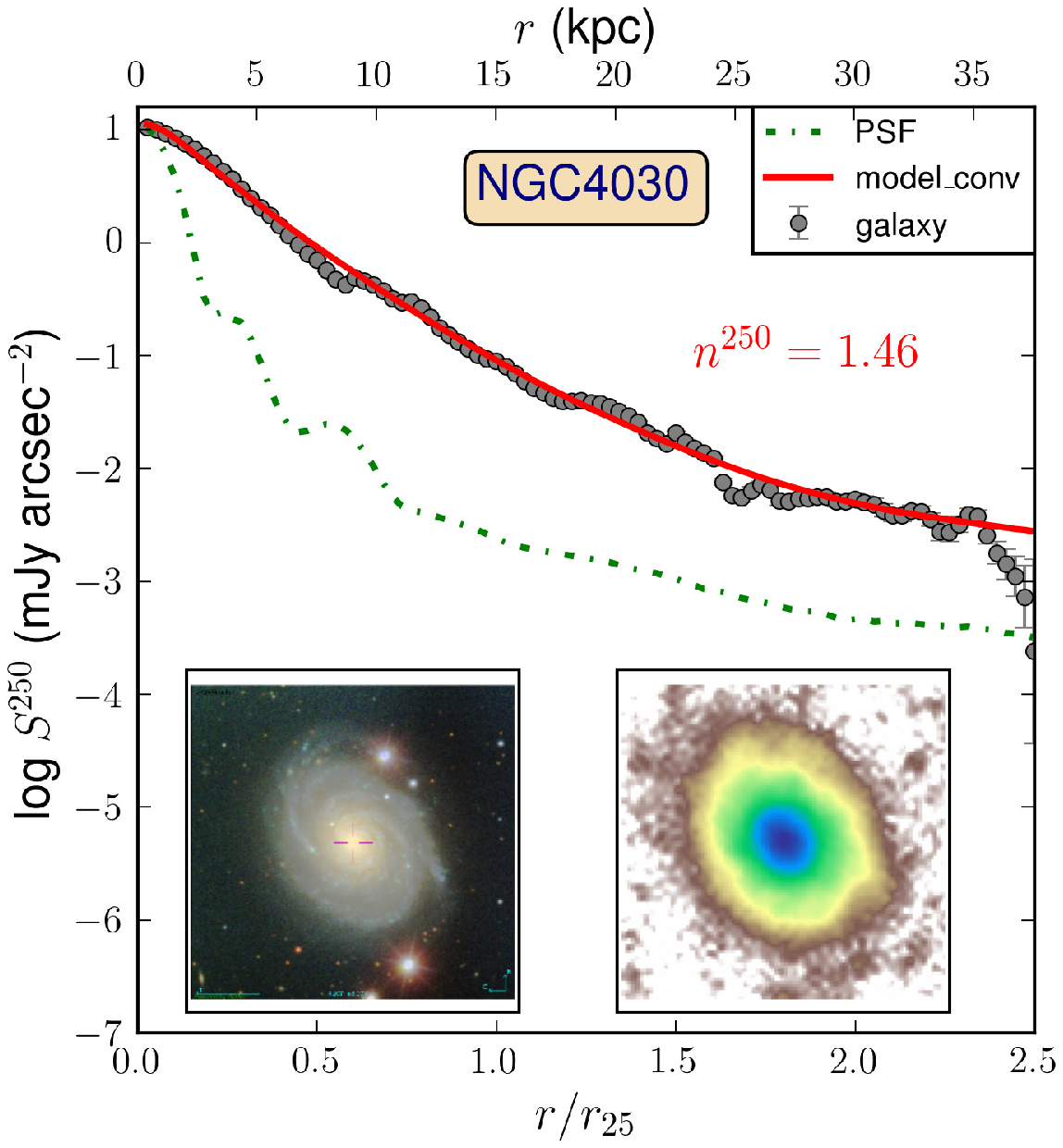}
\includegraphics[width=6.8cm, angle=0, clip=]{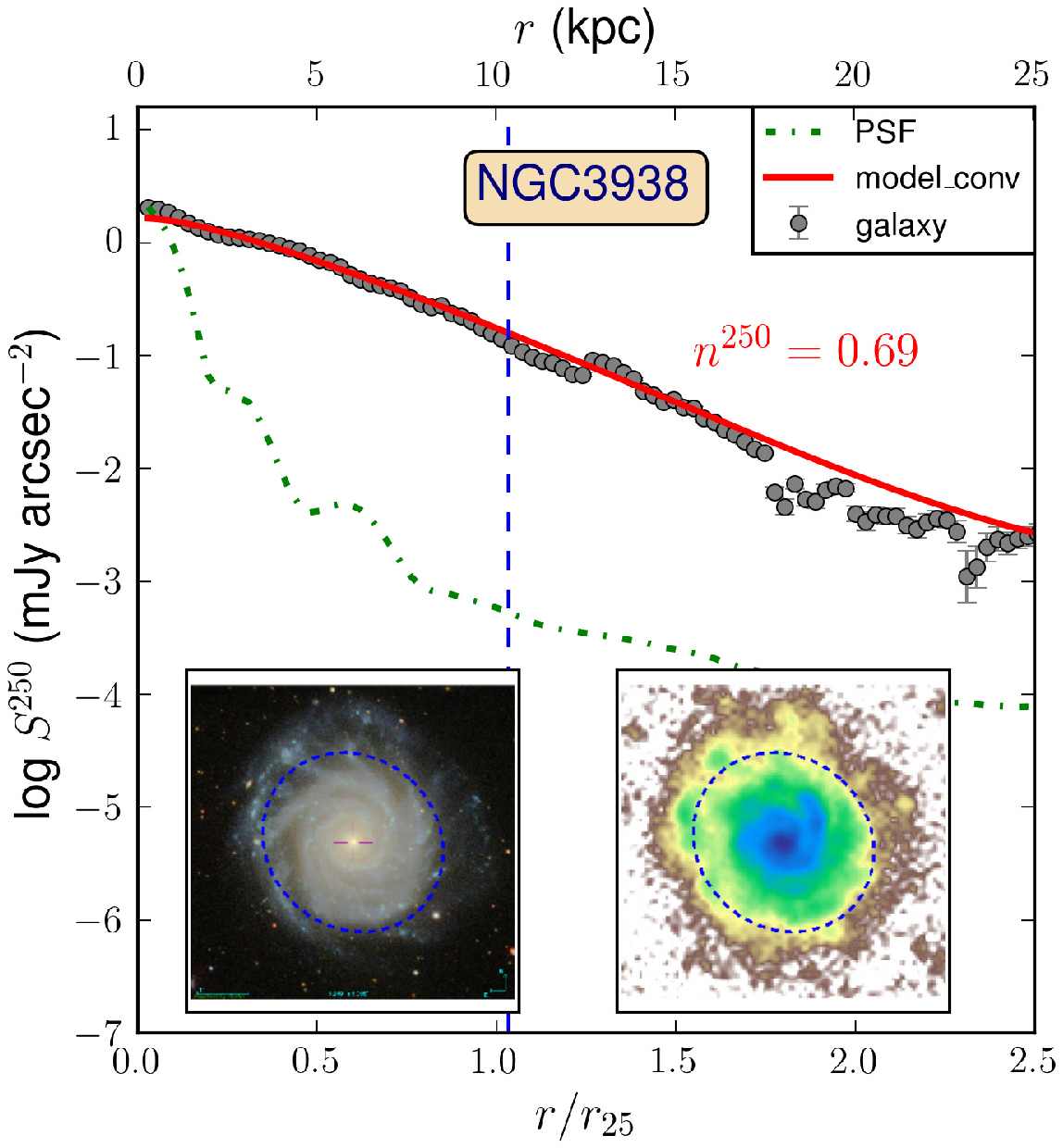}
\includegraphics[width=6.8cm, angle=0, clip=]{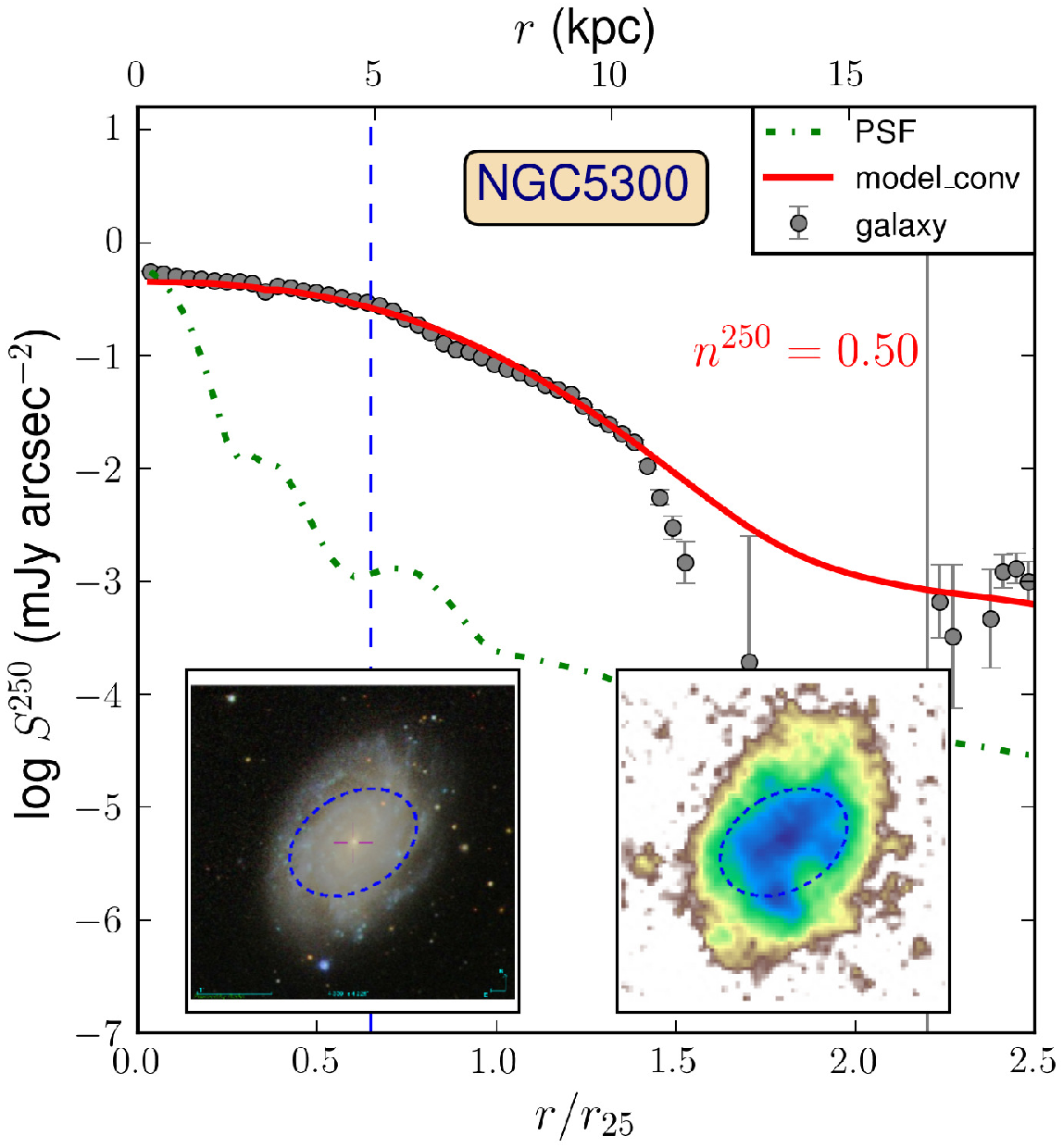}
\includegraphics[width=6.8cm, angle=0, clip=]{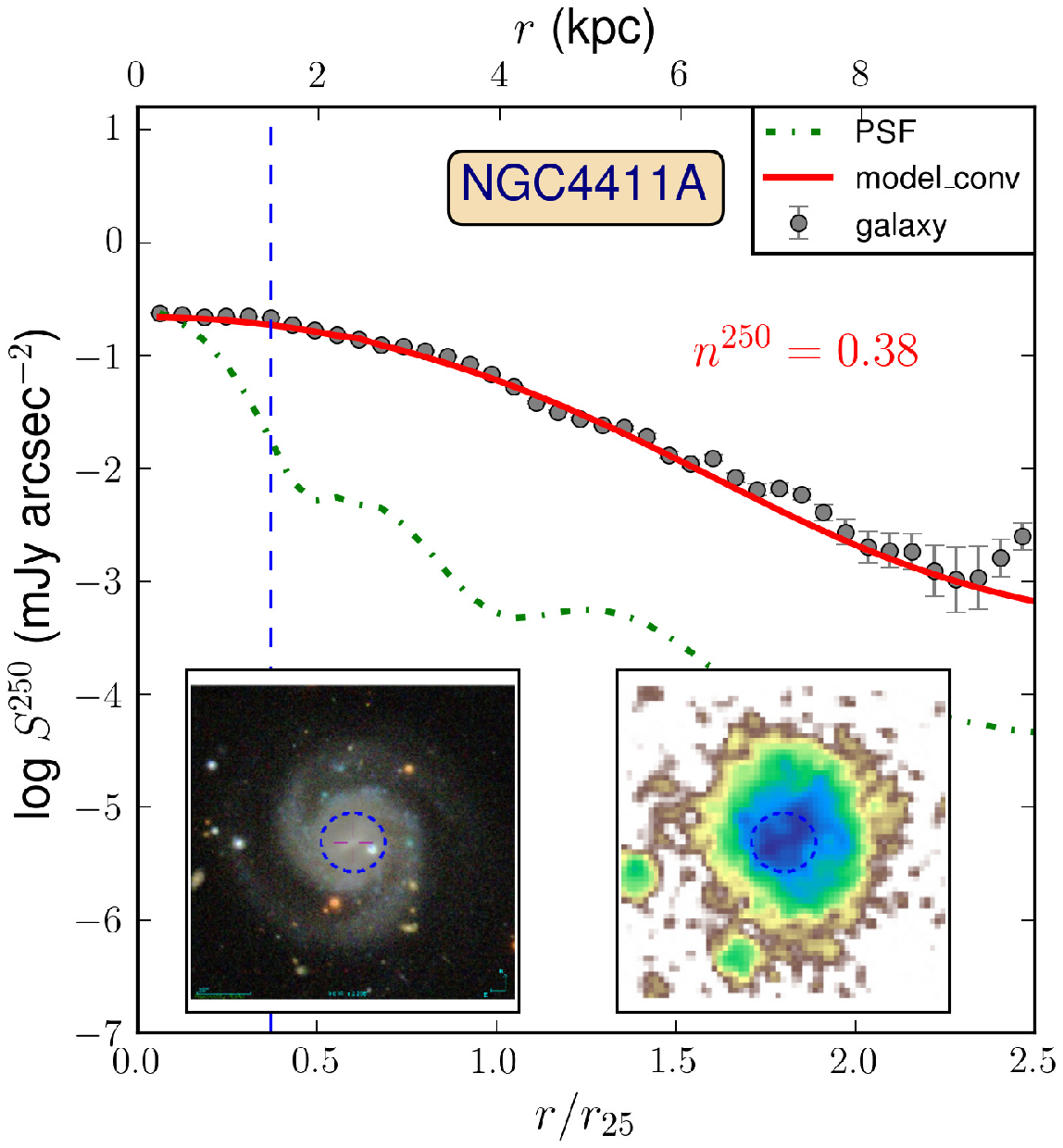}
\caption{Some typical galaxies in our sample with different features in their SB profiles (no correction for the galaxy inclination has been done). The grey circles represent the azimuthally averaged galaxy profile in the SPIRE\,250\,$\mu$m band. The model (convolved with the respecting PSF which is shown by the green dotdash line) is represented by the red solid line. The blue dashed line shows a break (if applicable) in the profile at some characteristic radius $r_\mathrm{c}$ (see text). The two images in each plot are a coloured SDSS snapshot (left, created with the Aladin sky atlas\protect\footnote{http://aladin.u-strasbg.fr/}, \citealt{2000A&AS..143...33B,2014ASPC..485..277B}) and a SPIRE\,250\,$\mu$m band image (right). Both images have the same scale and orientation.}
\label{profile_examples}
\end{figure*}

\addtocounter{figure}{-1}
\begin{figure*}
\centering
\includegraphics[width=6.8cm, angle=0, clip=]{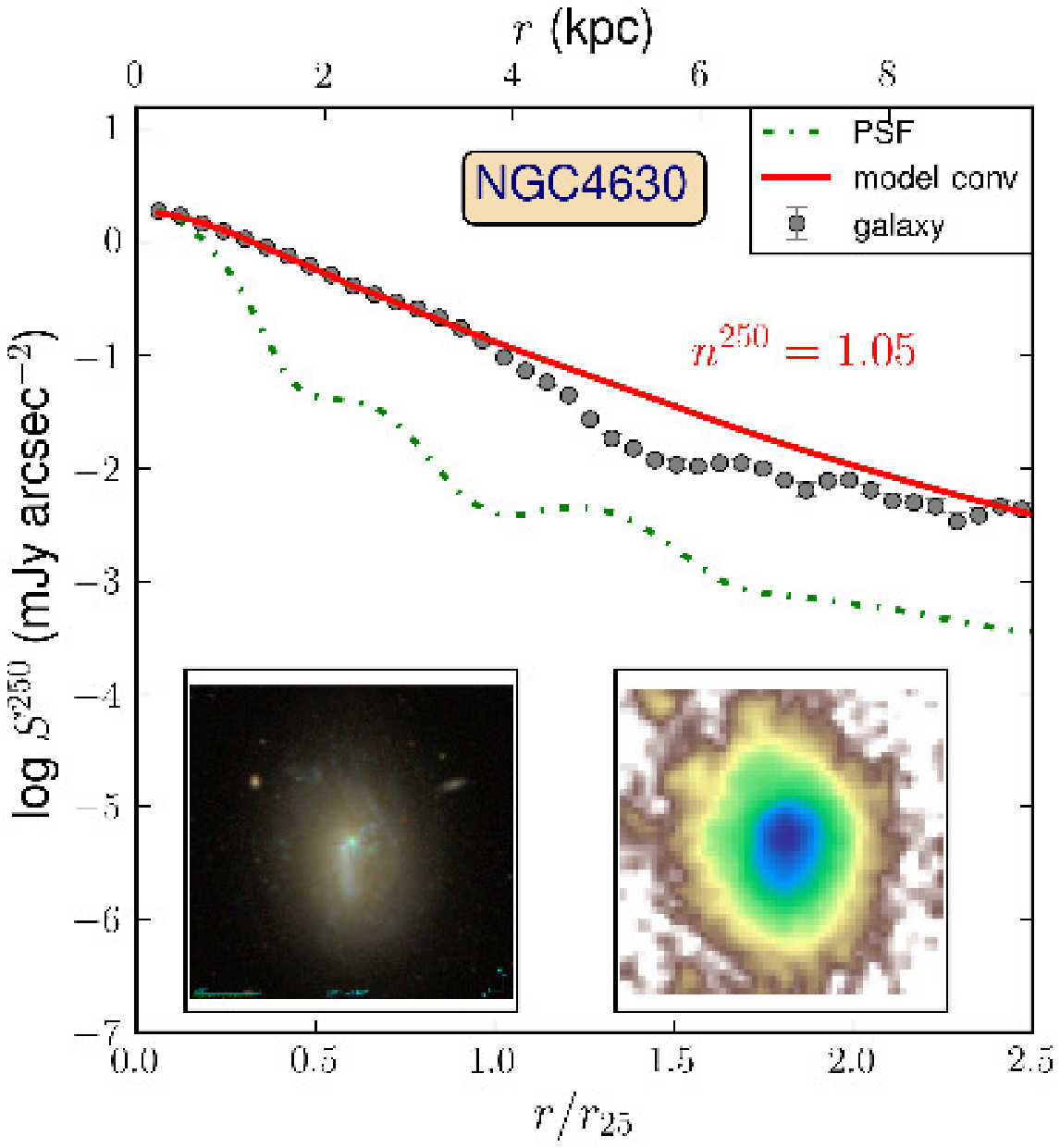}
\includegraphics[width=6.8cm, angle=0, clip=]{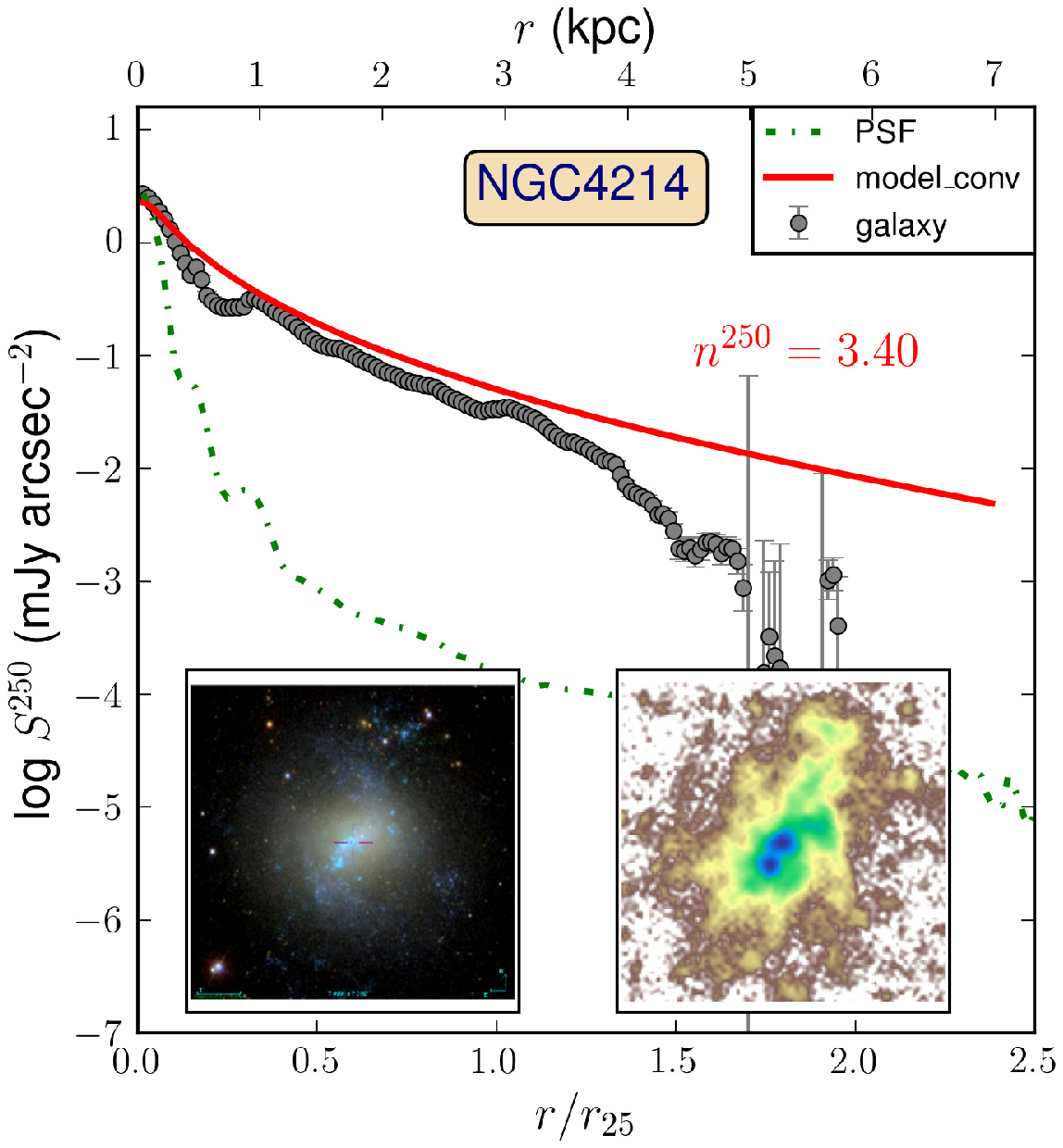}
\caption{(continued)}
\end{figure*}

\subsubsection{The presence of spiral arms}
\label{sec:non_exp_spirals}

\citet{2010A&A...518L..64S} pointed out regarding the FIR SB profiles for the three galaxies M\,81, M\,99, M\,100:  the result that `some of the disc profiles prefer a Gaussian to an exponential must not be overinterpreted because the presence of prominent spiral arms can create a distinctive bump in the disc part of the profile that is represented by a Gaussian better than by an exponential'. We agree that the spiral structure is indeed a very bright feature in many galaxies in our sample, and might potentially influence the fitting results. However, as has been shown in Sect.~\ref{sec:results}, galaxies of different morphology (i.e. from bright to very dim flocculent spirals to S0 galaxies without a spiral structure at all) may have a Gaussian profile. Moreover, a clear, rather extended dust depletion in the centre (sometimes, a plato or even a decrease of dust emission while going closer to the centre) with a more or less exponential profile in the periphery (see below) suggests that this cannot be related to the spiral structure, or, at least, that this cannot be regarded as the main reason for the observed overall Gaussian profile in galaxies.

\subsubsection{Depression of the H{\sc i} surface density in the galaxy central region}
\label{sec:non_exp_depression}

\begin{figure}
\centering
\includegraphics[width=8.3cm, angle=0, clip=]{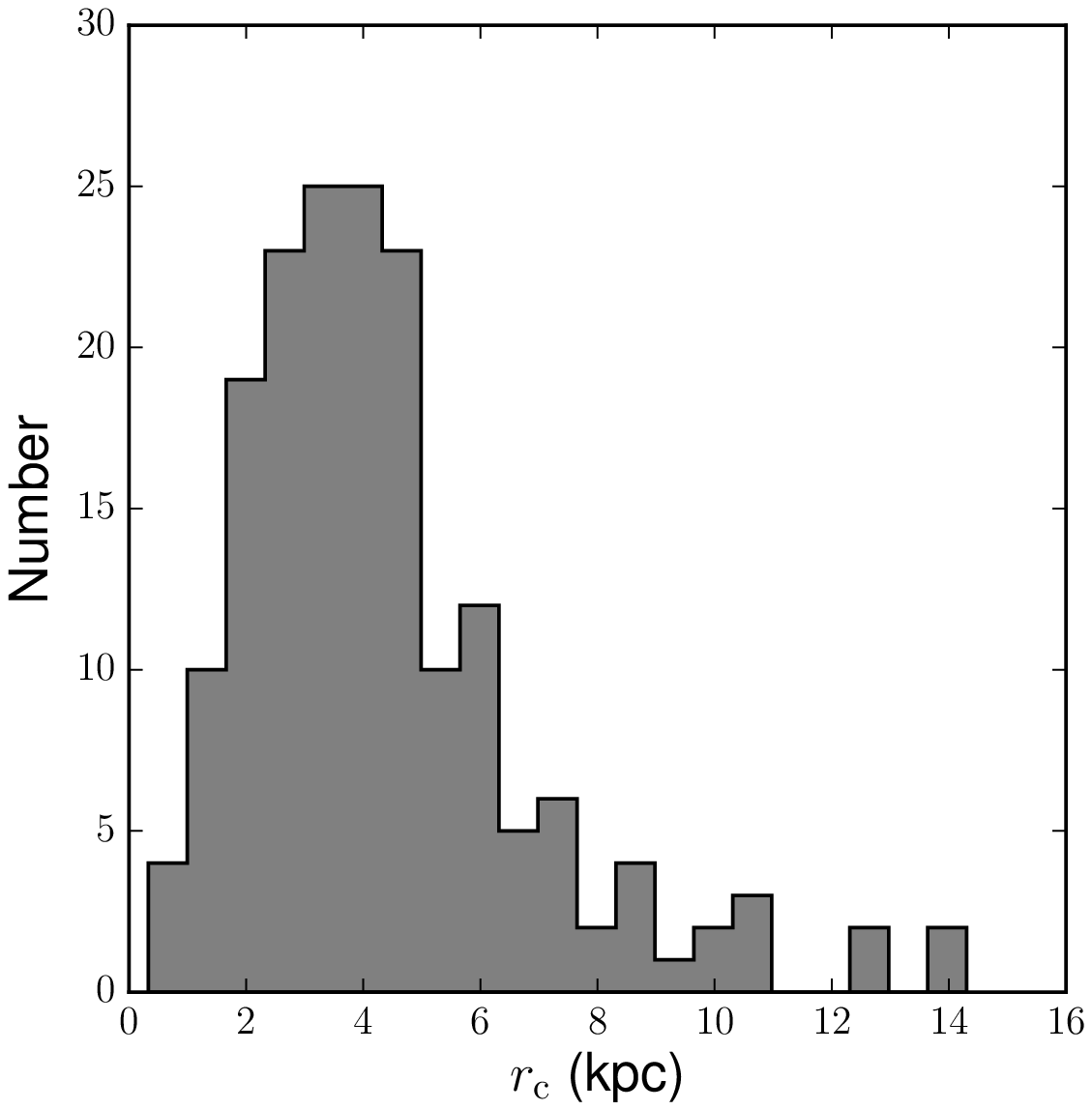}
\includegraphics[width=8.3cm, angle=0, clip=]{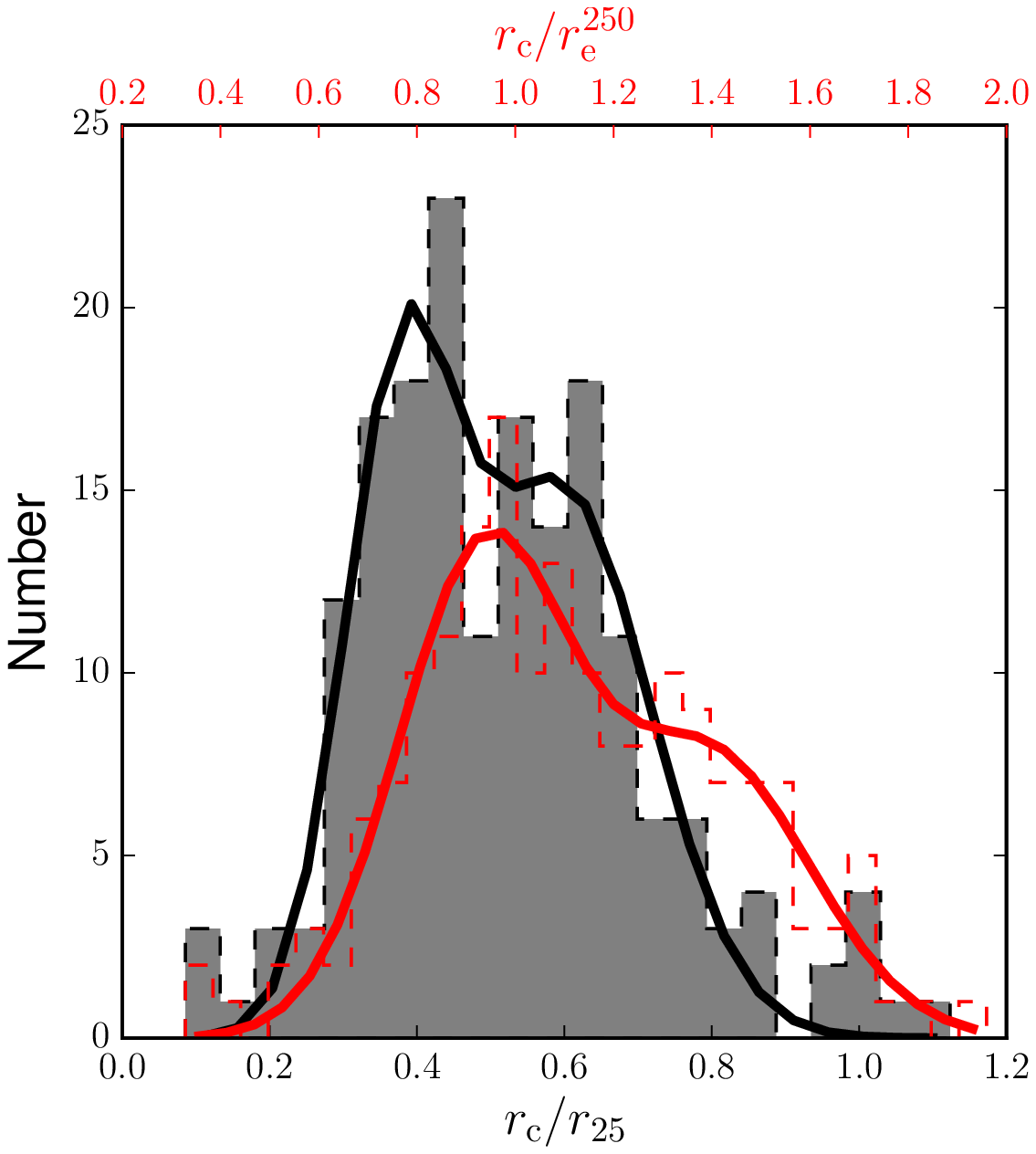}
\caption{Distributions by the characteristic radius expressed in kpc (top panel), and in units $r_{25}$ (grey colour) and $r_\mathrm{e}^{250}$ (red colour) (bottom panel). In the bottom panel we also show two bimodal distributions (as a sum of two Gaussians) fitted to the respecting histograms: for the $r_\mathrm{c}/r_{25}$ (black thick solid line) and $r_\mathrm{c}/r_{e}^{250}$ (red thick solid line). }
\label{char_radius_distr}
\end{figure}

Since most SB profiles of the galaxies under study show downbending (in Fig.~\ref{profile_examples} see panels for the galaxies IC\,3102, NGC\,3938, NGC\,5300, and NGC\,4411A), we estimated a radius where a prominent break takes place and can be unmistakably detected by the following approach. Using the IRAF task \textit{ellipse}, we created azimuthally averaged profiles for the sample galaxies in the SPIRE\,250$\,\mu$m band. Then we fitted a piecewise two-side linear function to each of the created profiles. Using this function, we can fit the SB profile in the inner and outer parts of a galaxy with two linear functions simultaneously, which intersect at a characteristic radius $r_\mathrm{c}$. We use the term ``characteristic radius'' following \citet{2014MNRAS.441.2159W}, who described the H{\sc i} galaxy profiles with a depression at the centre where the profile transitions to the inner flat region. In total, we fitted 178 galaxy profiles where we observe a distinct inner break. We show the distributions by the $r_\mathrm{c}$ in Fig.~\ref{bulge_bar_fracs} expressed in different units: kpc, optical radius $r_{25}$, and effective radius $r_\mathrm{e}^{250}$. One can see that the galaxies have a mean characteristic radius of $3.89\pm1.79$~kpc, and the distribution is asymmetric in shape. If we normalise the characteristic radius by the galaxy optical radius (taken from HyperLeda) or the effective radius found for the same SPIRE\,250$\,\mu$m band, we can see that the created distributions are broad (the mean values are $0.52\pm0.19\,r_{25}$ and $1.11\pm0.31\,r_\mathrm{e}^{250}$, respectively) and obviously have two peaks. This points to a bimodal distribution of the characteristic radius. We observe galaxies where a break may appear at rather long distances from the galaxy centre, as well as galaxies with shorter characteristic radii. Therefore, we fitted two Gaussian functions to these two distributions and found the following parameters of the observed dichotomy. For the optical radius: $r_\mathrm{c}=0.38\pm0.08\,r_{25}$ and $r_\mathrm{c}=0.60\pm0.12\,r_{25}$. For the effective radius: $r_\mathrm{c}=0.94\pm0.17\,r_\mathrm{e}^{250}$ and $r_\mathrm{c}=1.40\pm0.21\,r_\mathrm{e}^{250}$. If we compute the ratios between the short and long characteristic radii expressed in the units of optical and effective radius, we will receive 0.64 and 0.67, respectively, which are very close (taking into account that the effective radius was derived from the fitting whereas the optical radius was taken from HyperLeda). This bimodality suggests that the physical nature of these two types of characteristic radii is different. Below, we show that the short characteristic radius is related to the characteristic radius of the gas, whereas the long characteristic radius is, in fact, a break in the SB profile in the optical and NIR, which manifests a change in the stellar population.

As has been shown in many studies \citep[see e.g.][]{1984A&A...140..125W, 2002A&A...390..829S, 2014MNRAS.441.2159W}, in nearby spiral galaxies their H{\sc i} radial profiles decline exponentially in the outer regions and flatten or even decline near the galaxy centre. This is in contrast to the stellar surface density, which is peaked in the centre of the galaxy and drops steeply with radius. \citet{2012ApJ...756..183B} found for a sample of 32 nearby spiral galaxies that the combined H{\sc i} and H$_2$ gas profiles exhibit a universal exponentially declining radial distribution if the radius is scaled to $r_{25}$. To describe the shape of a two-component H{\sc i} radial profile and to obtain the deconvolved shape of it, \citet{2014MNRAS.441.2159W} developed a model which is an exponential function of radius in the outer regions and shows a depression towards the centre.

\citet{2014MNRAS.441.2159W} considered semi-analytic models to form a disc galaxy in a $\Lambda$ Cold Dark Matter universe with proper treating the conversion of atomic into molecular gas. They concluded that the depression in the inner H{\sc i} surface density profile arises because the gas reaches high enough surface densities in this region to be transformed into molecular gas. Because the level of star formation activity in the galaxy depends on its molecular gas content \citep{1998ApJ...498..541K}, correlations between star formation rate (and, hence, dust mass density) and H{\sc i} surface density arise consequently.

The second type of breaks, which is mostly detected at longer distances with respect to the galaxy optical radius (or the effective radius of the dust component), is related to the breaks in the SB profiles observed in galaxies in the optical. We discuss this below.

\subsubsection{Discs with different stellar populations}
\label{sec:non_exp_inner_region}

\citet{2010A&A...518L..72P} studied radial distribution of gas and dust in the two spiral galaxies M\,99 and  M\,100 and concluded that the dust shows the same breaks in the radial profiles as seen in the optical. Both of these galaxies exhibit a radial break in the profile at approximately $0.6\times r_{25}$. This is very close to the second peak of the distribution for the characteristic radius found above in this section. 

In the optical, such a type of breaks in the SB profiles is frequent. The scheme for classifying breaks in galaxy profiles was proposed in several studies \citep{2005ApJ...626L..81E,2006A&A...454..759P,2008AJ....135...20E}, which are grouped according to the specific behaviour of the declining brightness in the outer disc. The Type I profile is associated with a single exponentially declining disc with no signs of truncation within the optical radius. The Type II profile has a truncation with a steeper outer exponential (downbending profile). According to this classification, the breaks in the SB profile, which we observe in some galaxies of our sample, are of Type II. As was shown by \citet{2006A&A...454..759P}, \citet{2008AJ....135...20E} and \citet{2011AJ....142..145G}, these breaks are often found at a radius of approximately 2.5 disc scalelengths, or 0.5--0.6$\times r_{25}$ (again, this is very close to the longer characteristic radius found above). The frequency of this type is estimated to be 50--60\%. The Type III profile shows untitrancation in the periphery of the galaxy disc, i.e. has a shallower outer profile (upbending profile). 

\begin{figure}
\centering
\includegraphics[width=6.3cm, angle=0, clip=]{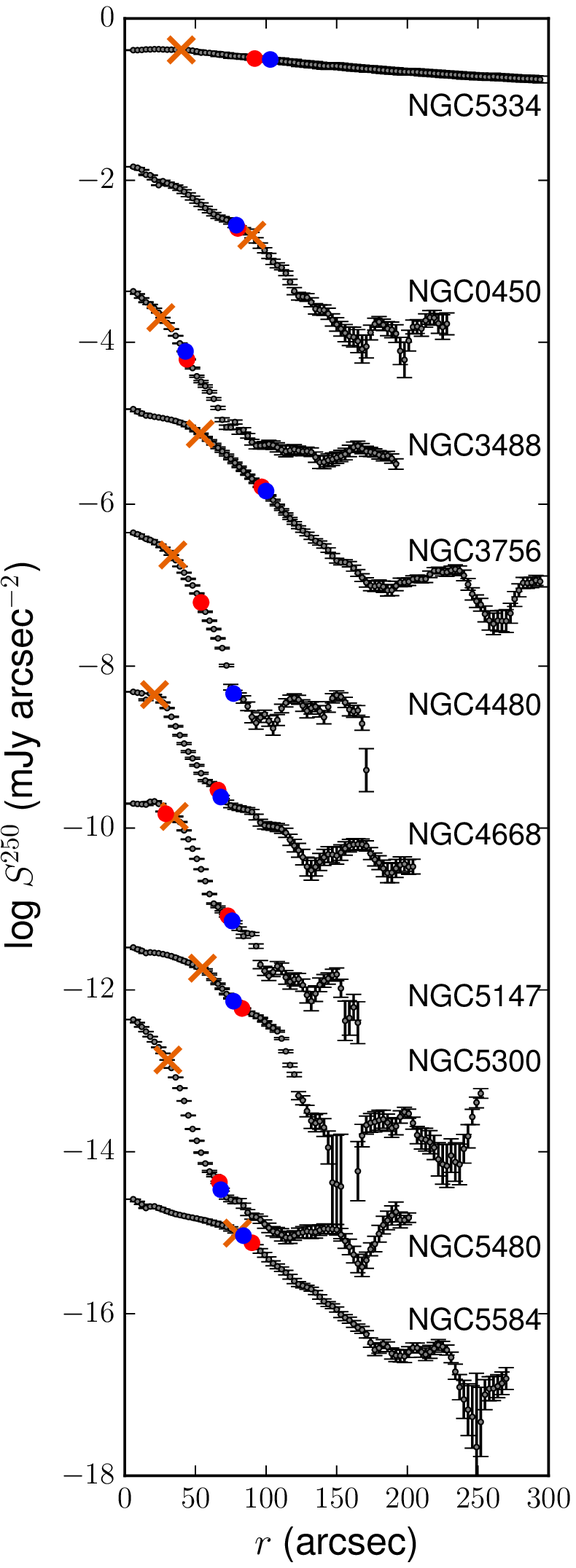}
\caption{SPIRE\,250 profiles for the 10 common galaxies with \citet{2006A&A...454..759P}. No correction for the galaxy inclination has been done. The orange cross is related to the characteristic radius found in this paper, the blue and red circles represent the break radii from \citet{2006A&A...454..759P} in the $g$ and $r$ bands, respectively. Each profile is shifted down relative to the previous one by 1.6 mJy\,arcsec$^{-2}$. }
\label{pohlen_compar}
\end{figure}

We found 10 common galaxies with the sample from \citet{2006A&A...454..759P} and compared our profiles plotted for these galaxies in the SPIRE\,250$\,\mu$m band with the ones presented in their study in the optical (see their Appendix~A). In Fig.~\ref{pohlen_compar} we show SPIRE\,250 profiles for these galaxies, along with the marked characteristic radii measured in this study and the break radii taken from \citet{2006A&A...454..759P} for both the $g$ and $r$ bands. For two galaxies, NGC\,450 and NGC\,5584, the break and characteristic radii are close (for NGC\,450, 79$\arcsec$ versus 90$\arcsec$ and for NGC\,5584, 84$\arcsec$ versus 80$\arcsec$, respectively). For the remaining galaxies, our characteristic radius is, on average, two-times smaller than their break radius ($\langle r_\mathrm{c} \rangle=4.60\pm1.50$~kpc and $\langle r_\mathrm{br} \rangle=9.65\pm3.67$~kpc). Comparing the profiles in the optical and FIR for these galaxies, we found a depletion in the centre of the dust emission profiles (which is absent in the optical profiles). Obviously, what we measure in the FIR is the short characteristic radius.

For the galaxies NGC\,3488 and NGC\,3756, there are no apparent breaks of Type II at longer distances from the galaxy centre which are detected in the optical profiles.

The three galaxies NGC\,4668, NGC\,5147, and NGC\,5480 exhibit Type III profiles: the up-bending profiles start at approximately the same break radius both in the optical and FIR. NGC\,5300 and NGC\,5334 have, apparently, both short and long characteristic radii -- the short radius is measured in the FIR, while the long radius (the break) can be detected both in the optical and FIR. To summarise, in general, the dust shows the same breaks not only of Type II, but of Type III in the radial profiles as seen in the optical. The depletion in the centre is only seen in  dust emission.

\begin{figure}
\centering
\includegraphics[width=8.3cm, angle=0, clip=]{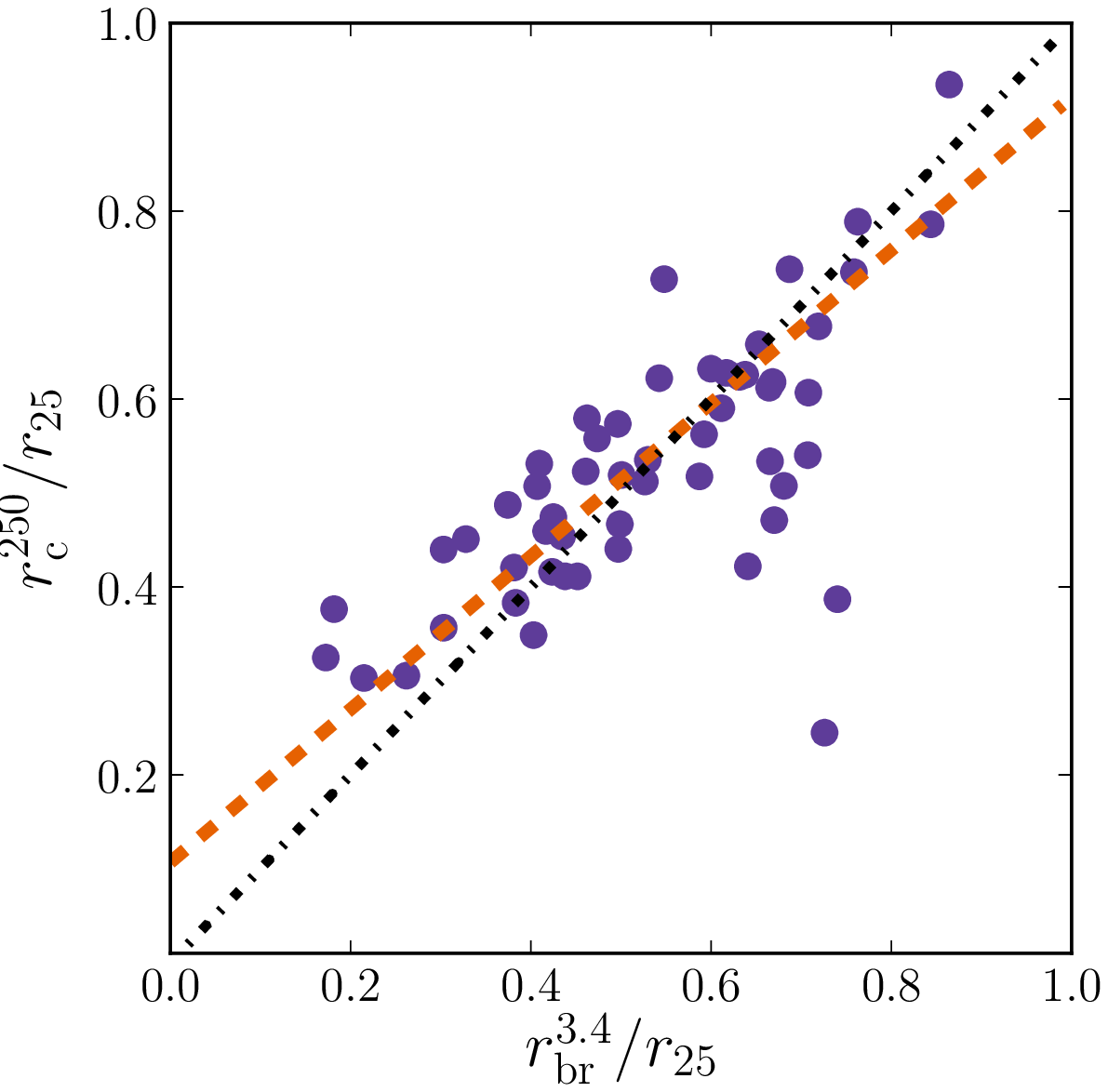}
\caption{The comparison between the break radius measured in the \textit{WISE\,W1} band and the characteristic radius in the SPIRE\,250$\,\mu$m band. The black dotdash line shows the one-to-one relation, whereas the orange line represents the regression line with $k=0.82$, $b=0.11$, and $\rho=0.83$.}
\label{WISE_compar}
\end{figure}

We also performed the same analysis of measuring the break radius for the \textit{WISE\,W1} azimuthally averaged profiles. We did this only for those galaxies which have a downbending profile with a prominent break in the SPIRE\,250$\,\mu$m band profile. The total number of galaxies for which both the characteristic and break radii were measured is 55. We show the comparison for both radii in Fig.~\ref{WISE_compar}. As one can see, they compare very well. The mean value of the characteristic radius in this sample is $0.54\pm0.18\,r_{25}$ which signifies that this is the long characteristic radius associated with the Type II break in the NIR.

\citet{2008ApJ...684.1026A} and \citet{2008ApJ...683L.103B} studied Type II breaks and found that they usually show a minimum in their color profile (the so-called U-shaped colour profile) at the break, or very near to it: the colour gets bluer out to the break radius, as one would expect from an inside-out formation scenario, and then becomes redder past the break radius. As shown by \citet{2012ApJ...753..138R}, the strength of the break decreases towards longer wavelengths (i.e. for older stellar populations). These and other studies (e.g. \citealt{2008ApJ...683L.103B,2010ApJ...716L...4Y,2012ApJ...752...97Y}) suggest that Type II breaks are attributed to changes on the stellar populations at the break position rather than to drops in the number density of stars. Several mechanisms have been proposed to explain this fact: star formation thresholds \citep{1989ApJ...344..685K}; stellar migration \citep{2007ApJ...667L..49D,2011ApJS..195...18R,2010ApJ...716L...4Y,2012ApJ...752...97Y}, confirmed by numerical simulations (see \citealt{2008ApJ...675L..65R,2009MNRAS.398..591S}); or stellar migration induced by resonances produced by strong bars \citep{2006ApJ...645..209D,2008MNRAS.386.1821F,2013ApJ...771...59M}.

Antitruncations of stellar discs, which are also observed in the corresponding dust emission profiles, are connected to activity in the outer galaxy parts which can be related to the disc (interactions with other galaxies triggering outer star formation, see \citealt{2006ApJ...636..712E}) or the galaxy spheroid (minor merging/smooth gas accretion, see \citealt{2007ApJ...670..269Y,2012A&A...548A.126M}).

Obviously, the observed breaks in the dust discs should have the same origin as proposed for explaining the breaks in the stellar discs.

\subsubsection{Triggered star formation in the galaxy inner region}
\label{sec:non_exp_inner_region}

Interestingly, in some galaxies of the same morphological type the \ser\ index of the dust emission profile is Gaussian, whereas in others we can see a profile close to exponential or even with a \ser\ index much larger than 1 (see Fig.~\ref{profile_examples} for the galaxies NGC\,3631, NGC\,4030, and NGC\,4214). A careful investigation of this fact showed that when a triggered star formation in the central region takes place, in the FIR we do not see a depletion in the centre of the SB profile, and often, on the contrary, we observe  a ``bump'' of dust emission in the centre. This star formation activity in the inner region can be caused by several reasons.

\citet{2010A&A...518L..64S} studied a bright central component in three nearby spiral galaxies. They noticed that the rising profile in the inner part of M\,100 is related to the more prominent bulge, bar, or inner-disc component, whereas in M\,81 a small AGN boosts the emission from transiently heated dust grains.

Using the NED database, we selected galaxies which are classified as `Sy' or `LINER', meaning that with these galaxies some activity in the centre is associated. The total number of such galaxies in the DustPedia sample is 90, from which 34 galaxies have determined \ser\ models. For these galaxies, we found their average \ser\ index in the SPIRE\,250$\,\mu$m band to be $1.29\pm0.47$. The large scatter can be explained by the fact that the activity (power) of the AGN is different in these galaxies. In some, the AGN does not contribute much emission in the FIR (NGC\,613), whereas in others it is the most luminous component (for example, NGC\,7465).

For seven NED-classified starburst galaxies out of 320 galaxies in our sample, we found that their profiles show particularly high \ser\ indices which decrease with wavelength from PACS\,100 to SPIRE\,500: $2.93\pm1.97$, $2.64\pm1.60$, $2.22\pm1.79$, $1.73\pm1.38$, and $1.20\pm0.62$. 
Six starburst galaxies (NGC\,3049, NGC\,3504, NGC\,4123, NGC\,4416, NGC\,4779, and NGC\,4904) have an apparent bar structure and a relatively bright nucleus. As was shown in \citet{1994ApJ...425L..13M}, \citet{2007A&A...468...61D}, and \citet{2008A&A...492...31D}, through the bar instability, which may be triggered by interaction with other galaxies or a minor merging event, gas can be effectively funneled towards the galactic nucleus. This ignites bursts of star formation near the galaxy centre, which, in its turn, leads to efficient dust producing. The star formation burst heats the dust in the inner galaxy region and increases the fraction of warm dust emission in the FIR domain of the galaxy SED. This leads to increasing the \ser\ index in these galaxies, with its dependence on wavelength, as the contribution of the warm dust emission to the total emission decreases with wavelength.

\begin{figure*}
\centering
\includegraphics[width=8.0cm, height=8.0cm, angle=0, clip=]{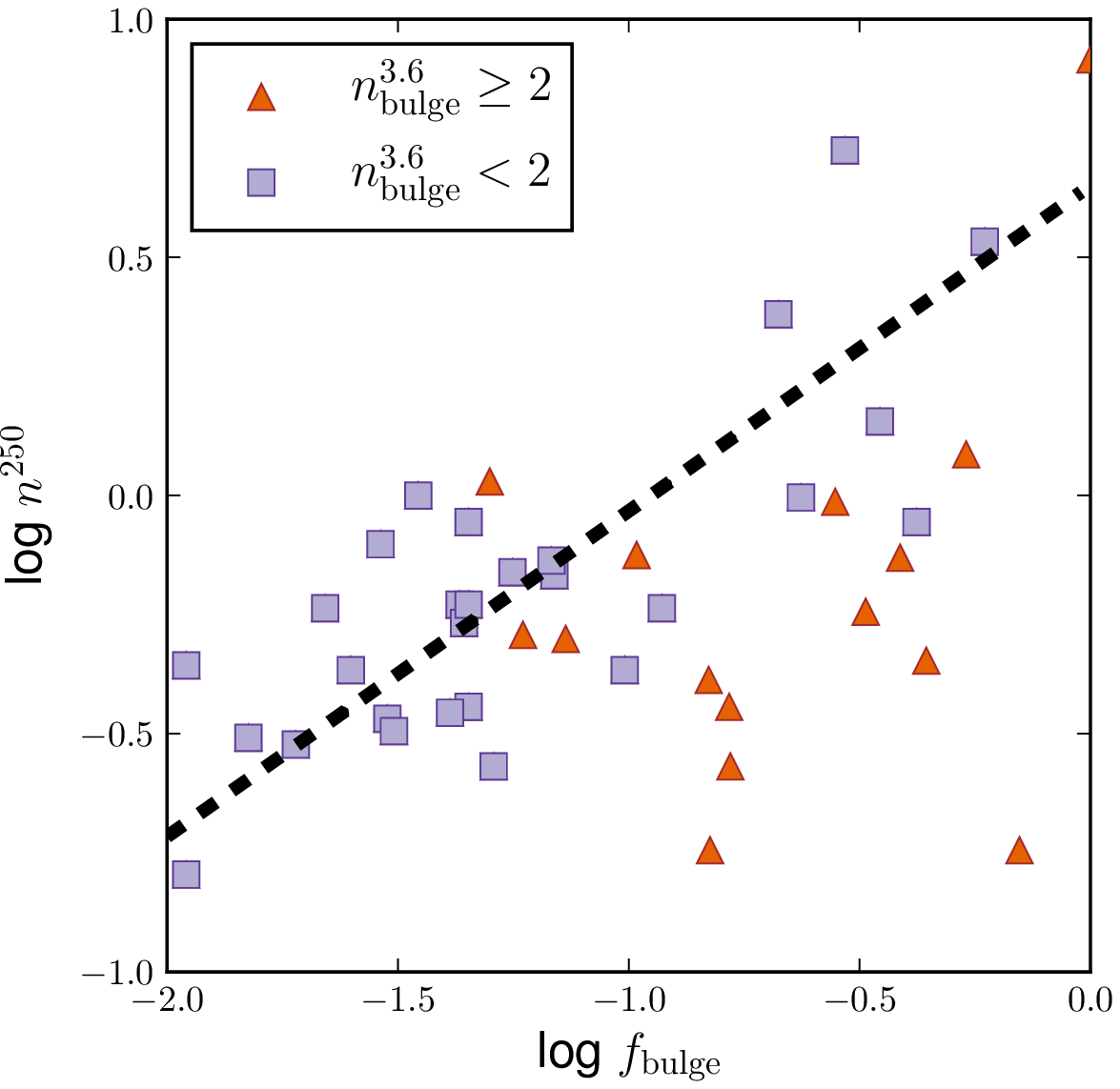}
\includegraphics[width=8.0cm, height=8.0cm, angle=0, clip=]{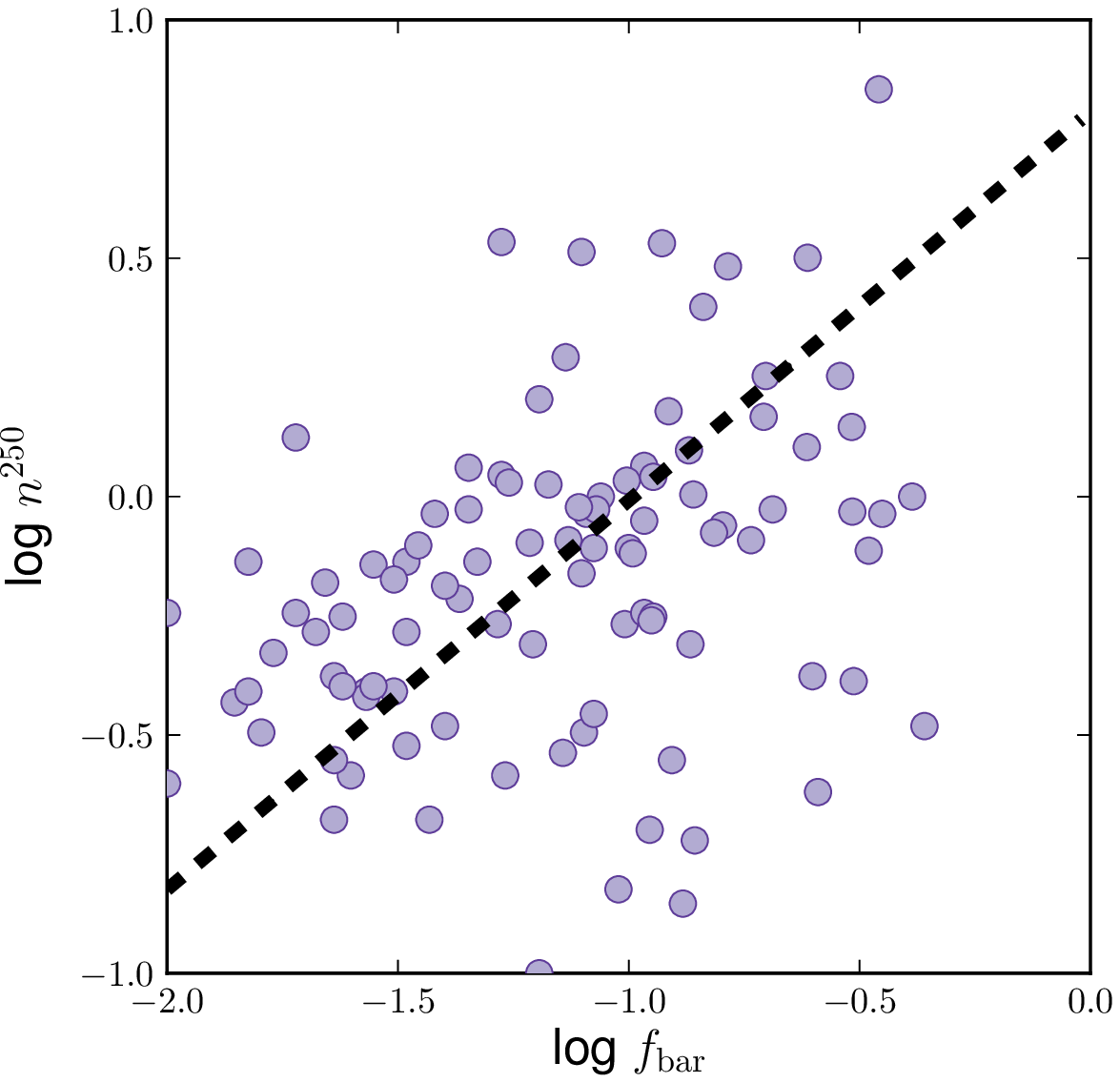}
\caption{Correlations between the bulge (left plot) and bar (right plot) fraction to the total galaxy model, based on the S$^4$G decomposition, and the \ser\ index in the SPIRE\,250$\,\mu$m band. The black dashed lines in each plot represent the regression lines: $\log\,n^{250}=0.68\,f_\mathrm{bulge}+0.65$ ($\rho=0.78$) and $\log\,n^{250}=0.82\,f_\mathrm{bar}+0.81$ ($\rho=0.60$).}
\label{bulge_bar_fracs}
\end{figure*}

Another reason which explains the steeper profile in the inner region in some galaxies is the presence of a pseudobulge. There exist at least two families of bulges in galaxies (see reviews by \citealt{2004ARA&A..42..603K} and \citealt{2013seg..book....1K}): pseudobulges (bulges made through secular evolution) and classical bulges (bulges made through hierarchical mergers). Pseudobulges can be identified by the presence of a disc-like structure in the centre of the galaxy (nuclear spirals, nuclear bars, and high ellipticity in bulge) and classical bulges with smooth isophotes that are round compared to the outer disc and show no discy structure in their bulge. \citet{2006ApJ...642L..17F} showed that pseudobulges have significantly higher specific SFRs than classical bulges. Also, the PAH emission profiles of pseudobulges are brighter and more centrally concentrated without a transition in star formation properties from the outer disc to the central pseudobulge. On the contrary, classical bulges exhibit remarkably regular star formation profiles in the inner region with a sharp change (decline) in star formation rate toward the galaxy centre. In addition, it was shown that galaxies with bars or ovals, on average, have brighter central PAH emission than galaxies without strong drivers of secular evolution (see also \citealt{1999ApJ...525..691S,2005ApJ...632..217S}).

To verify that the bulge type and the bar strength have impact on dust emission, we selected galaxies with available multicomponent decomposition in the S$^4$G database.

\citet{2009ApJ...697..630F} showed that more massive pseudobulges have higher SFR density. In Fig.~\ref{bulge_bar_fracs}, left plot, we show the correlation between the bulge fraction to the total galaxy luminosity (mass) in the 3.6$~\mu$m band and the \ser\ index in SPIRE\,250 (which is related to the central SFR density), for classical bulges and pseudobulges (we identify them by the \ser\ index $n_\mathrm{bulge}^{3.6}>2$ for classical bulges and $n_\mathrm{bulge}^{3.6}<2$ for pseudobulges, see \citealt{2010ApJ...716..942F}). One can see that for the pseudobulges there exists a correlation between these two quantities, whereas for the classical bulges without a prominent star formation this correlation does not exist. This correlation indicates that the more luminous/massive the pseudobulge is as compared to the total galaxy luminosity/mass, the steeper the dust emission profile in the inner region.

We also studied a similar correlation but for the bar fraction to the total galaxy luminosity (Fig.~\ref{bulge_bar_fracs}, right plot). This correlation is weaker but still we can see that galaxies with stronger bars  (defined as having a higher fraction to the total galaxy luminosity at 3.6$~\mu$m) have dust emission profiles with a larger \ser\ index. However, the situation here is complicated by the fact that bars may have some inner structures with a triggered star formation: from pseudobulge to AGN galaxy-scale \citep[see also][]{2016A&A...595A..63V}.

In our sample we have a few irregular galaxies ($T=9-10$) which do not have a distinct well-organised structure, though most of them feature a patchy disc with some trace of a bar and/or spiral arms. Among them we can find a large variety of structural features which are typical for irregular galaxies: a starburst nucleus (NGC\,4194), a pseudobulge structure (NGC\,4214, \citealt{2011ApJ...733L..47F}), an elongated bar (NGC\,3846A), a ring (NGC\,1427A), and a faint spiral pattern (NGC\,4032). UGC\,5720 is also classified as a blue compact dwarf galaxy \citep{1981ApJ...247..823T,2003ApJS..147...29G}. In the optical, in almost all of these galaxies we observe multiple blue ``beads'' of starforming regions, which in the FIR are associated with bright dust emission features. As shown in Fig.~\ref{n_distrs}, the mean \ser\ index for irregular galaxies is close to 1, however there are Irr galaxies with a very small \ser\ index (e.g. for NGC\,3846A $n^{250}=0.37$), as well as with a large \ser\ index (for NGC\,4214 $n^{250}=3.40$). Galaxies with a large \ser\ index demonstrate intensive star formation in the central region which naturally results in a higher \ser\ index. The inner region in irregular galaxies with a Gaussian profile does not show a high star formation and this explains their small \ser\ index.

\subsubsection{Dust heating}
\label{sec:non_exp_heating}

The continuum emission from dust depends both on the column-density and temperature of the dust. The dust is heated by starlight, coming from both sites of the hot young stars in star-formation regions and from more evolved (old) stellar populations, as well as from the presence of AGN \citep[e.g.][]{2007ApJ...657..810D}. FIR/submm wavelengths give information on the dominant heating mechanism of dust, though no clear picture about the exact fraction of dust heating contributed by each stellar population has been created \citep[e.g.,][]{2011ApJ...738..124L,2011AJ....142..111B,2012MNRAS.419.1833B,2015MNRAS.448..135B,2015A&A...576A..33H}: it depends on the intrinsic SED of the stellar populations, the dust mass and optical properties, as well as the relative dust-star geometry. For example, the heating of dust in the galaxy centre can be initiated by the old stellar population of the luminous bulge in S0-Sb galaxies and AGN galaxies (dust torus?). This might be the reason for the higher \ser\ index in the FIR wavebands observed in these galaxies. 

In this study we do not analyse the dust-density but only dust emission profiles of galaxies. Therefore, formally, our conclusions relate to the dust emission profiles and cannot be extended to the dust density profiles automatically. To create the dust mass density distribution in a galaxy some additional modelling should be done (see e.g. \citealt{2009ApJ...701.1965M,2016MNRAS.462..331S,2017A&A...605A..18C}), which is beyond the scope of this paper and will be done in a subsequent DustPedia study. However, the comparison with the other works (see above), where such dust mass density distributions have been analysed, and for our simulations in the Appendix~\ref{sec:mock} showed that the dust emission in the FIR/submm region generally follows the dust mass density distribution. Therefore, this suggests that the conclusions of this study should work for the mass density distribution of dust in galaxies as well.

\section{Conclusions}
\label{sec:conclusions}

We have studied the two-dimensional distribution of stellar and dust tracers in galaxies selected from the DustPedia sample. For this purpose we used \textit{WISE\,W1} and \textit{Herschel} (from PACS\,100 to SPIRE\,500) observations gathered in the DustPedia database. All galaxy images have been homogeneously treated and prepared for fitting. Using a new version of the Python wrapper \textsc{deca}, we fitted the 2D SB distribution of the galaxies with a single \ser\ profile and determined their best-fit structural parameters. For the galaxy images in the \textit{WISE\,W1} band, we used the \textsc{galfit} code, whereas for the \textit{Herschel} observations, both \textsc{galfit} and \textsc{galfitm} were applied. Using \textsc{deca}, we were able to properly fit even those galaxies which reside in a close pair or interact, and most of the galaxies where a bright foreground star contaminates the light from the target galaxy.

Our main results can be summarised as follows:
\begin{enumerate}
\item We derived structural parameters for the whole DustPedia sample based on their \textit{WISE\,W1} images. These parameters are the overall \ser\ index (related to galaxy morphology), the effective radius (which is an indication of the galaxy's size), and its total luminosity, upon which the stellar mass was computed. 
\item For 320 DustPedia galaxies we obtained reliable 2D \ser\ models based on the available \textit{Herschel} images in all five FIR bands.
\item The obtained set of the parameters puts constrains on the structural properties of the dust component. We found that there is no significant change in the \ser\ index at 100--500$\,\mu$m. This indicates that the dust emission profile is essentially the same in the fitted \textit{Herschel} bands. The average value at 250$\,\mu$m is $0.67\pm0.37$ (187 galaxies are fitted with $n^{250}\leq0.75$, 87 galaxies have $0.75<n^{250}\leq1.25$, and 46 -- with $n^{250}>1.25$). This signifies that, formally, cold dust in galaxies is distributed by a non-exponential law. Plenty of galaxies in our sample demonstrate a Gaussian dust emission profile -- with a depression in the central region up to the characteristic radius $r_\mathrm{c}\approx 0.4\,r_{25}$ and a rather exponential decline in the outer region. We suggest that this is related to the depletion in the inner region of the H{\sc i} surface density profiles. In this region the atomic gas reaches high enough surface densities to be transformed into molecular form.
When a triggered star formation in the central region takes place (e.g. because of a bar instability, formation of a pseudobulge, AGN activity etc.), we observe a SB profile of the cold dust emission with a steeper decline in this region (no depression in the centre). We confirm the result by \citet{2010A&A...518L..72P} that the dust mostly shows the same breaks of Type II in the radial profiles as seen in the optical and NIR at $r_\mathrm{br}\approx 0.6\,r_{25}$. The Type III breaks are seen in the FIR and optical as well.
\item The effective radius of the dust SB profile steadily increases with wavelength, from $0.90\pm0.27$ in PACS\,100 to $1.41\pm0.43$ in SPIRE\,500, most likely because of the cold-dust temperature gradient with galactocentric distance. 
\end{enumerate}

The fit results for the \textit{WISE\,W1} and \textit{Herschel} data are available on the DustPedia database.

Our future follow-up studies will be focused on a detailed decomposition of well-resolved DustPedia galaxies, by taking into account several dust components which are associated with different galaxy substructures.

\begin{acknowledgements}

A.V.M. expresses gratitude for the grant of the Russian Foundation for Basic Researches number mol\_a 18-32-00194.

IDL gratefully acknowledges the support of the Flemish Fund for Scientific Research (FWO-Vlaanderen).

The DustPedia project is funded by the European Union, as a European Research Council (ERC) 7\textsuperscript{th} Framework Program (FP7) project (PI Jon Davies, proposal 606824).

JF acknowledges the financial support from UNAM-DGAPA-PAPIIT IA104015 grant, M\'exico.

This research made use of \textsc{montage} (\url{http://montage.ipac.caltech.edu/ }), which is funded by the National Science Foundation under Grant Number ACI-1440620, and was previously funded by the National Aeronautics and Space Administration's Earth Science Technology Office, Computation Technologies Project, under Cooperative Agreement Number NCC5-626 between NASA and the California Institute of Technology.

This research has made use of the NASA/IPAC Infrared Science Archive (IRSA; \url{http://irsa.ipac.caltech.edu/frontpage/}), and the NASA/IPAC Extragalactic Database (NED; \url{https://ned.ipac.caltech.edu/}), both of which are operated by the Jet Propulsion Laboratory, California Institute of Technology, under contract with the National Aeronautics and Space Administration. This research has made use of the HyperLEDA database (\url{http://leda.univ-lyon1.fr/}; \citealp{2014A&A...570A..13M}).

This research makes use of data products from the Wide-field Infrared Survey Explorer, which is a joint project of the University of California, Los Angeles, and the Jet Propulsion Laboratory/California Institute of Technology, and NEOWISE, which is a project of the Jet Propulsion Laboratory/California Institute of Technology. WISE and NEOWISE are funded by the National Aeronautics and Space Administration.

This work is based in part on observations made with the {\it Spitzer} Space Telescope, which is operated by the Jet Propulsion Laboratory, California Institute of Technology under a contract with NASA.

\textit{Herschel} is an ESA space observatory with science instruments provided by European-led Principal Investigator consortia and with important participation from NASA. The \textit{Herschel} spacecraft was designed, built, tested, and launched under a contract to ESA managed by the \textit{Herschel}/\textit{Planck} Project team by an industrial consortium under the overall responsibility of the prime contractor Thales Alenia Space (Cannes), and including Astrium (Friedrichshafen) responsible for the payload module and for system testing at spacecraft level, Thales Alenia Space (Turin) responsible for the service module, and Astrium (Toulouse) responsible for the telescope, with in excess of a hundred subcontractors.

This research has made use of "Aladin sky atlas" developed at CDS, Strasbourg Observatory, France. 
\end{acknowledgements}

\bibliographystyle{aa} 
\bibliography{decomp}



\appendix

\section{Data preparation}
\label{sec:data_preparation}

We exploited the aperture photometry analysis which had been carried out in \citetalias{2018A&A...609A..37C}. For each galaxy, \citetalias{2018A&A...609A..37C} determined a photometric master aperture ellipse \footnote{See the photometry files at \url{http://dustpedia.astro.noa.gr/Photometry}}, which is described by the semimajor axis ($sma$) and ellipticity ($\epsilon$). Also, carefully calculated total fluxes and their uncertainties over a large set of wavelengths are provided.

First, for each galaxy, we rebinned its \textit{Herschel} images to the PACS\,100~$\mu$m band image which has the best resolution among all chosen PACS and SPIRE bands. The corresponding sigma- and PSF-images were rebinned as well, to match their rebinned flux density maps. 

To properly estimate background emission, we masked out all objects on the images which lie outside of the aperture ellipse. The masking was performed using a combined mask created by {\sc SExtractor} \citep{1996A&AS..117..393B} and the Astropy \citep{2013A&A...558A..33A} Python library\footnote{\url{https://github.com/astropy/photutils}}, the routine \textit{Findpeaks}.

After masking background and foreground sources, a 2D 5th-order polynomial sky model was fitted to remove large-scale background emission. It was estimated within an annulus with the inner semimajor axis $1.250\,sma$, the outer semimajor axis $1.601\,sma$, and the ellipticity $\epsilon$ (see \citetalias{2018A&A...609A..37C}). For galaxies without enough empty space for a robust determination of the background emission in this way, we fitted the background simultaneously with the galaxy (see Sect.~\ref{sec:decomp}). Also, the same approach is applied to galaxies which have some bright glow in the \textit{WISE\,W1} band from close saturated stars.  

After removing background emission, we used the same routine, as previously, to mask objects overlapping with the primary galaxy, with the revision of the created masks and fixing incorrect detections if needed. In this step, foreground stars, bright clumpy details (e.g. bright starforming and H{\sc ii} regions), and small satellite galaxies, which overlap the body of the galaxy, were masked out. Obviously, all these features cannot be described by a single \ser\ function, see Sect.~\ref{sec:decomp}. However if the contamination of a non target source was too high (from a foreground star or an extended object), this source was not included in the final mask and, instead, was fitted simultaneously with the primary galaxy (see Sect.~\ref{sec:decomp}). 

To match the final \textit{WISE\,W1} galaxy image, the \textit{WISE\,W1} PSF image was rotated by some angle. This position angle was found using the following approach. First, in each retrieved \textit{WISE\,W1} map, we selected 5 point-like sources (unsaturated, non-overlapping stars with a high signal-to-noise ratio). Then we fitted the selected stars with a PSF function (available in \textsc{galfit}) multiple times, rotating each time the PSF image by an angle from $0^{\circ}$ to $360^{\circ}$, with a step of $1^{\circ}$. The PSF position angle with the smallest $\chi^2$ value was then chosen to be the correct angle. After that, we found an average angle for all five stars and rotated the original PSF image for each galaxy frame by the corresponding position angle. 

As to the FIR/submm observations, all \textit{Herschel} PSF images were rotated by an average angle (the average of the list with PSFs, which are the scanning directions for all obsIDs used in the image) between the scanning direction of the telescope and the North direction. 

The final step is to cut out the galaxy by cropping the galaxy images (if possible) to encompass the ellipse with a semimajor axis of 2\,$sma$ and an ellipticity $\epsilon$. We insure that all images have sufficient empty background space included. For the \textit{Herschel} images, if an original image, provided in the DustPedia database for a galaxy, is smaller than the size of the required cut-out image for this galaxy, than no cropping was done. 

\section{The fitting of mock galaxy images}
\label{sec:mock}

To verify the robustness of our fitting method (see Sect.~\ref{sec:decomp}) in its ability to recover the properties of the dust distribution in galaxies, we apply the following approach. \citet{2016A&A...592A..71M} performed a consistent fitting of the stellar and dust structural components and obtained a plausible complex model of one of the galaxies from the \textit{HER}OES (\textit{HER}schel Observations of Edge-on Spirals, \citealt{2013A&A...556A..54V}) sample, IC\,2531. Despite the fact that for this galaxy they found the dust energy balance problem, we adopted their results in the current study to create realistic galaxy simulations and then estimate the dust distribution parameters in the FIR/submm for these galaxy images using the technique described in Sect.~\ref{sec:decomp}. Potentially, the dust energy balance problem in IC\,2531 might affect the reliability of our simulations (as compared to real galaxies). However, as no commonly accepted solution to the dust energy problem has yet been proposed, we decided to consider this simplified galaxy model to merely test our fitting method and to study the dependence of the retrieved dust distribution parameters on wavelength. We should point out that this problem is unlikely to be related to the observed depression in the dust emission profile discovered in many DustPedia galaxies (see Sect.~\ref{sec:non_exp_reasons} for possible explanations of this fact). If we used a dust density profile with a depression in the centre (e.g. a \ser\ profile with $n\approx0.5-0.7$), then the dust energy problem would be even worse as less emission would be radiated in the central region as compared to the exponential profile.

The model of IC\,2531, adopted from \citet{2016A&A...592A..71M} consists of several stellar components (a superthin young disc, a thin disc, a thick disc, and a bulge) and a dust disc. The model parameters, as well as some other quantities, which we use in our panchromatic RT modelling, are listed in Table~\ref{tab:IC2531.tab}.

Using the model of IC\,2531, we created a set of RT simulations at the wavelengths 3.4~$\mu$m, 100~$\mu$m, 160~$\mu$m, 250~$\mu$m, 350~$\mu$m, and 500~$\mu$m, which correspond to the \textit{WISE\,W1} and \textit{Herschel} bands. We used the RT code {\sc skirt} to create the mock galaxy images. In our simulations, we varied the galaxy inclination angle $i$ and the distance to it $D$ using the normal distributions for the these parameters ($\langle i \rangle=57^\circ$, $\sigma_i=22^\circ$ and $\langle D \rangle=21.6$~Mpc, $\sigma_D=11.1$~Mpc -- these are the average values for the DustPedia sample). By doing so, we created a sample of 300 galaxies: for each galaxy we tried both an exponential and \ser\ disc to describe the dust component (see Table~\ref{tab:IC2531.tab}). The output mock galaxy images in each band were then rebinned and convolved (the details of the pixel size and the PSF in each band are provided in Sect.~\ref{sec:sample}) to match the corresponding \textit{WISE\,W1} and \textit{Herschel} bands. Also, the following Gaussian noise was added to the created frames, estimated from the real observations of IC\,2531:  0.00035 mJy/pix (in \textit{WISE\,W1}), 0.51 mJy/pix (in PACS\,100), 0.68 mJy/pix (in PACS\,160), 0.91 mJy/pix (in SPIRE\,250), 1.31 mJy/pix (in SPIRE\,350), and 1.47 mJy/pix (in SPIRE\,500).

\begin{table}
\caption{The structural parameters of the multicomponent decomposition model for IC\,2531 adopted from \citet{2016A&A...592A..71M}, with an originally used double-exponential dust disc or an alternative \ser\ dust disc (not used in \citealt{2016A&A...592A..71M}). The parameters of the \ser\ dust disc were chosen as follows: $r_\mathrm{e,d}=1.678\,h_\mathrm{R,d}$, $n_\mathrm{d}=0.6$ (similar to what we observe in real galaxies, see Sect.~\ref{sec:results}), and the intrinsic flattening $q_\mathrm{d}=h_\mathrm{R,d}/h_\mathrm{z,d}$. The broken exponential function includes the inner disc scalelength $h_\mathrm{R,inn}$, the outer disc scalelength $h_\mathrm{R,out}$, the break radius $R_\mathrm{b}$, the disc scaleheight $h_\mathrm{z}$, and the total disc luminosity $L$ in the given waveband (the 2MASS $H$ band for all stellar components except  the young stellar disc, for which the \textit{GALEX\,NUV} band was used).  The bulge is described by a \ser\ law.}
\label{tab:IC2531.tab}
\centering
\resizebox{\columnwidth}{!}{
    \begin{tabular}{cccc}
    \hline
    \hline\\[-1ex]    
    Component & Parameter &  Value & Units \\[0.5ex]
    \hline\\[-0.5ex]
    Superthin disc:&$h_\mathrm{R,inn}^\mathrm{st}$ & $8.0$ & kpc \\[+0.5ex]
    ({\it BrokenExponentialDisk3D})&$h_\mathrm{R,out}^\mathrm{st}$ & $3.33$& kpc \\[+0.5ex]
              & $h_\mathrm{z}^\mathrm{st}$ & $0.1$& kpc \\[+0.5ex]
              & $R_\mathrm{b}$ & $21.41$ & kpc  \\[+0.5ex] 
              &  $L_\mathrm{st,NUV}$ & $1.57$ & $10^{12}\,\mathrm{L}_{\odot}$ \\[+0.5ex]
              &  Age & 0.1 & Gyr \\[+0.5ex]    
    Thin disc: & $h_\mathrm{R,inn}^\mathrm{t}$ & $8.0$ & kpc \\[+0.5ex]
    ({\it BrokenExponentialDisk3D}) & $h_\mathrm{R,out}^\mathrm{t}$ & $3.33$& kpc \\[+0.5ex]
              & $h_\mathrm{z}^\mathrm{t}$ & $0.61$& kpc \\[+0.5ex]
              & $R_\mathrm{b}$ & $21.41$ & kpc  \\[+0.5ex] 
              &  $L_\mathrm{t,H}$ & $3.72$ & $10^{10}\,\mathrm{L}_{\odot}$ \\[+0.5ex]
              &  Age & 5.0 & Gyr \\[+0.5ex] 
    Thick disc:& $h_\mathrm{R}^\mathrm{T}$ & $24.87$& kpc \\[+0.5ex]
    ({\it ExponentialDisk3D})& $h_{z}^{T}$ & $1.57$ & kpc \\[+0.5ex]
              &  $L_\mathrm{T,H}$ & $2.12$ & $10^{10}\,\mathrm{L}_{\odot}$ \\[+0.5ex]
              &  Age & 5.0 & Gyr \\[+0.5ex] 
    Bulge:&$r_\mathrm{e,b}$ & $1.86$ & kpc \\[+0.5ex]
    (\ser) & $n_\mathrm{b}$ & $2.26$ & ---\\[+0.5ex]
              & $q_\mathrm{b}$ & $0.85$ & ---\\[+0.5ex]
              &  $L_\mathrm{b}$ & $1.97$ & $10^{10}\,\mathrm{L}_{\odot}$ \\[+0.5ex]
              &  Age & 8.0 & Gyr \\[+0.5ex]  
    \hline\\[-1ex]
    Dust disc (double-exponential):& $h_\mathrm{R,d}$ & $8.44$ & kpc \\[+0.5ex]
                      & $h_\mathrm{z,d}$ & $0.25$ & kpc \\[+0.5ex]
                      & $M_\mathrm{d}$   & $4.08$ & $10^7~M_{\odot}$ \\[+0.5ex]    
    Dust disc (\ser\ model):&$r_\mathrm{e,d}$ & $14.16$ & kpc \\[+0.5ex]
              & $n_\mathrm{d}$ & $0.6$ & ---\\[+0.5ex]
              & $q_\mathrm{d}$ & $0.03$ & ---\\[+0.5ex]
              & $M_\mathrm{d}$   & $4.08$ & $10^7~M_{\odot}$ \\[+0.5ex]    
    \hline\\[-0.5ex]
    \end{tabular}
}    
\end{table}

Using our code \textsc{deca} (see Sect.~\ref{sec:decomp}), we performed a single \ser\ fitting of the created mock galaxy images in the \textit{WISE\,W1} and \textit{Herschel} bands. The results of our fitting are summarised in Table~\ref{tab:IC2531fits.tab}. Here we concentrate on the distributions by the \ser\ index, whereas we comment on the other parameters (for example, the effective radius) in Sect.~\ref{sec:results}. As one can see, the \ser\ index of the dust emission profile, recovered by \textsc{deca}, is approximately 0.9--1.0 if the dust mass density distribution is given by an exponential law. If the dust density is given by a \ser\ law with $n_\mathrm{d}=0.6$, then for the PACS bands the retrieved \ser\ index appeared to be somewhat higher ($n_\mathrm{d}=0.7-0.8$), as compared to the \ser\ index of the dust density profile. However, for the SPIRE bands, it is approximately 0.6--0.7, i.e. the \ser\ index of the dust emission is almost the same as the \ser\ index of the dust mass density profile. Thus, we can conclude that the distribution of dust emission in the SPIRE bands should be very close to the real dust density distribution. It is no surprise since the 500$\,mu$m flux serves as a proxy of the dust mass. Also, the adopted fitting technique is able to recover the dust emission profiles fairly well. The histograms for these parameters are plotted in Figs.~\ref{cors_IVm}--\ref{n_distrs}.

\begin{table}
\caption{The results of the fitting for the mock galaxy images with a double-exponential and \ser\ dust disc. Only the average values for the whole sample of 300 galaxies are listed. The effective radius for an exponential disc is related to its scalelength by $r_\mathrm{e}^{\lambda}\approx1.678\,h_\mathrm{R}^{\lambda}$.}
\label{tab:IC2531fits.tab}
\centering
\resizebox{\columnwidth}{!}{
    \begin{tabular}{ccccc}
    \hline
    \hline\\[-1ex]
Band & $n^\lambda$ & $n^{\lambda}/n^{3.4}$ & $r_\mathrm{e}^{\lambda}$ &  $r_\mathrm{e}^{\lambda}/r_\mathrm{e}^{3.4}$ \\
     &             &                      &         (kpc)            &                                             \\
\hline 
\multicolumn{5}{c}{Exponential dust disc:}\\
\textit{WISE\,W1} & $1.19 \pm 0.10$ & $1.00 \pm 0.01$ & $9.14 \pm 0.73$ & $1.00 \pm 0.01$ \\ 
PACS\,100 & $1.05 \pm 0.09$ & $0.88 \pm 0.11$ & $8.32 \pm 0.68$ & $0.91 \pm 0.10$ \\ 
PACS\,160 & $0.97 \pm 0.08$ & $0.82 \pm 0.10$ & $9.35 \pm 0.76$ & $1.02 \pm 0.12$ \\ 
SPIRE\,250 & $0.95 \pm 0.08$ & $0.80 \pm 0.10$ & $9.71 \pm 0.79$ & $1.06 \pm 0.12$ \\ 
SPIRE\,350 & $0.95 \pm 0.10$ & $0.80 \pm 0.11$ & $10.09 \pm 0.91$ & $1.10 \pm 0.13$ \\ 
SPIRE\,500 & $0.92 \pm 0.14$ & $0.78 \pm 0.13$ & $10.37 \pm 1.06$ & $1.13 \pm 0.15$ \\[+0.5ex]
\multicolumn{5}{c}{\ser\ dust disc ($n_\mathrm{d}=0.6$):}\\
\textit{WISE\,W1} & $1.14 \pm 0.10$ & $1.00 \pm 0.01$ & $9.55 \pm 0.77$ & $1.00 \pm 0.01$ \\ 
PACS\,100 & $0.82 \pm 0.07$ & $0.72 \pm 0.09$ & $8.99 \pm 0.73$ & $0.94 \pm 0.11$ \\ 
PACS\,160 & $0.73 \pm 0.06$ & $0.64 \pm 0.08$ & $10.10 \pm 0.81$ & $1.06 \pm 0.12$ \\ 
SPIRE\,250 & $0.67 \pm 0.06$ & $0.59 \pm 0.07$ & $10.63 \pm 0.86$ & $1.11 \pm 0.13$ \\ 
SPIRE\,350 & $0.64 \pm 0.07$ & $0.57 \pm 0.08$ & $11.01 \pm 0.94$ & $1.15 \pm 0.13$ \\ 
SPIRE\,500 & $0.61 \pm 0.09$ & $0.54 \pm 0.09$ & $11.20 \pm 1.08$ & $1.17 \pm 0.15$ \\ 
 
    \hline\\[-0.5ex]
    \end{tabular}
}    
\end{table}

\section{Robustness of the results and comparison with the literature}
\label{sec:comparison}
To verify how robust our fitting is, we applied the following test. The border value $r_\mathrm{e}^{\lambda}$/FWHM$^{\lambda}=0.5$ (see e.g. \citealt{2010AJ....139.2097P} and comments to the {\sc galfit} code\footnote{\url{https://users.obs.carnegiescience.edu/peng/work/galfit/galfit.html}}) separates galaxies with a reliable \ser\ fitting (with a larger effective radius) and galaxies which are too small for recovering their structural parameters. Below we consider the \textit{WISE\,W1} and \textit{Herschel} bands separately.

\subsection{WISE W1}
\label{sec:WISE_W1}

\begin{figure*}
\centering
\includegraphics[width=5.8cm ,height=5.8cm, angle=0, clip=]{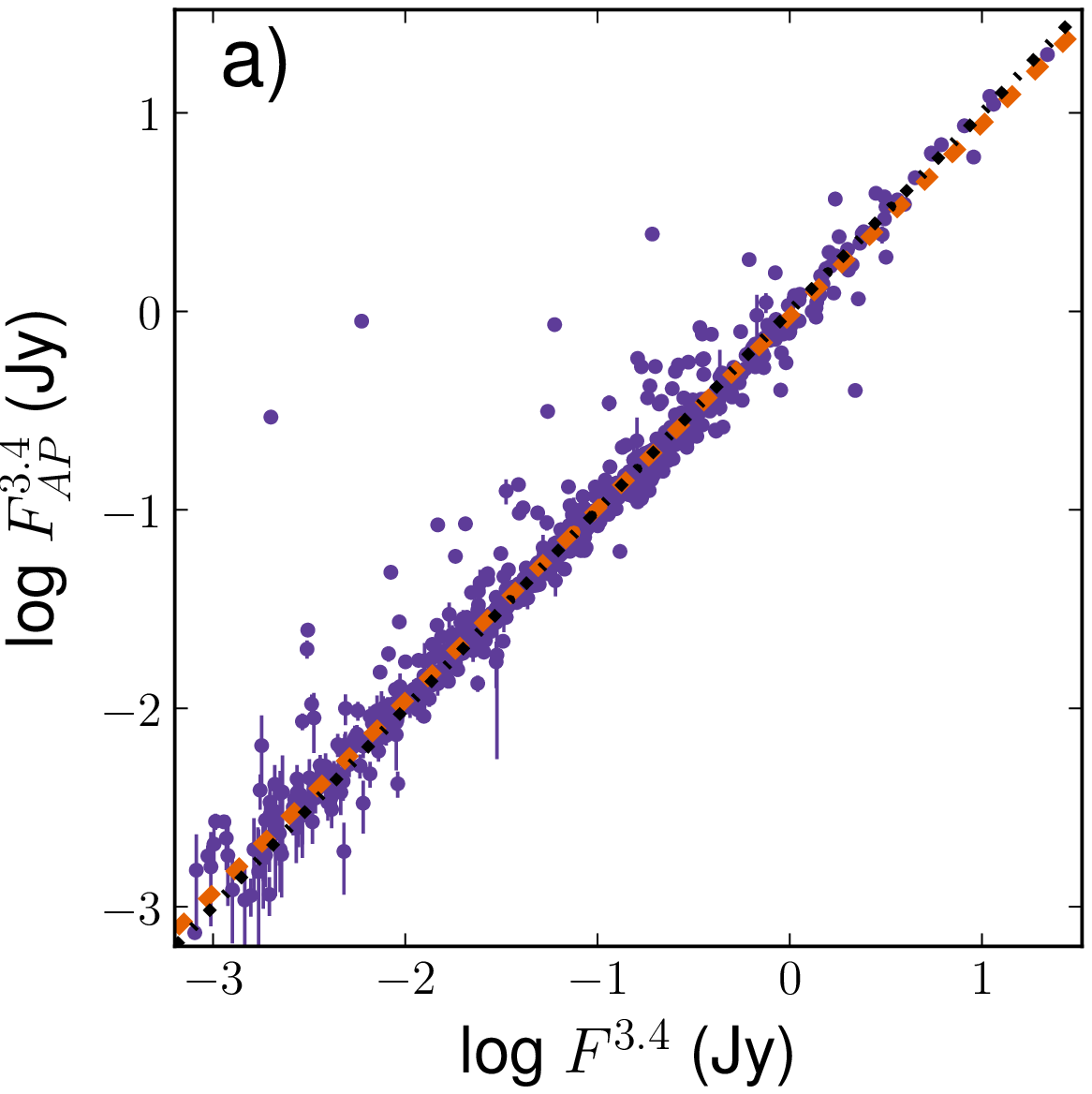}
\includegraphics[width=5.8cm ,height=5.8cm, angle=0, clip=]{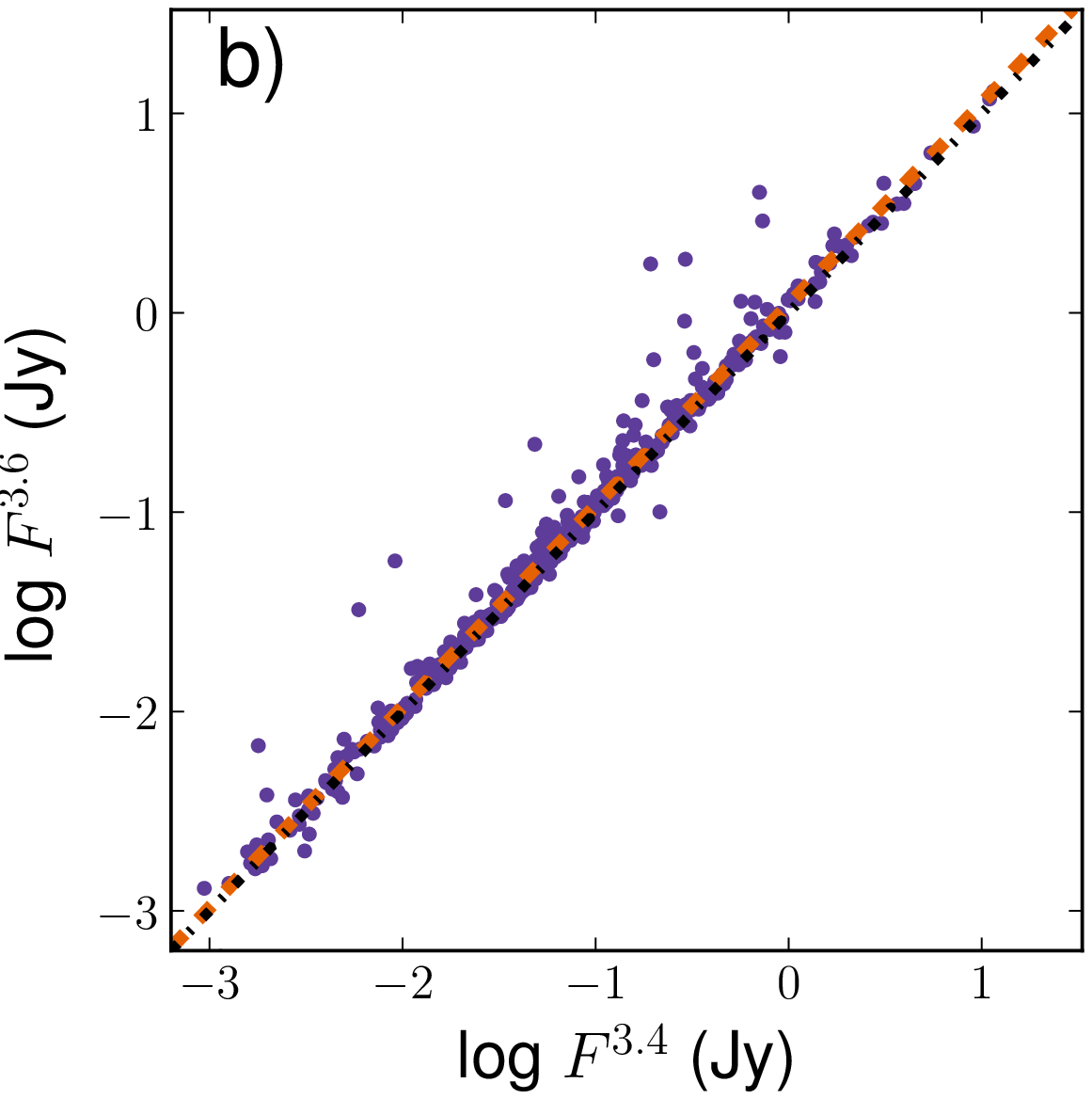}
\includegraphics[width=5.9cm ,height=5.8cm, angle=0, clip=]{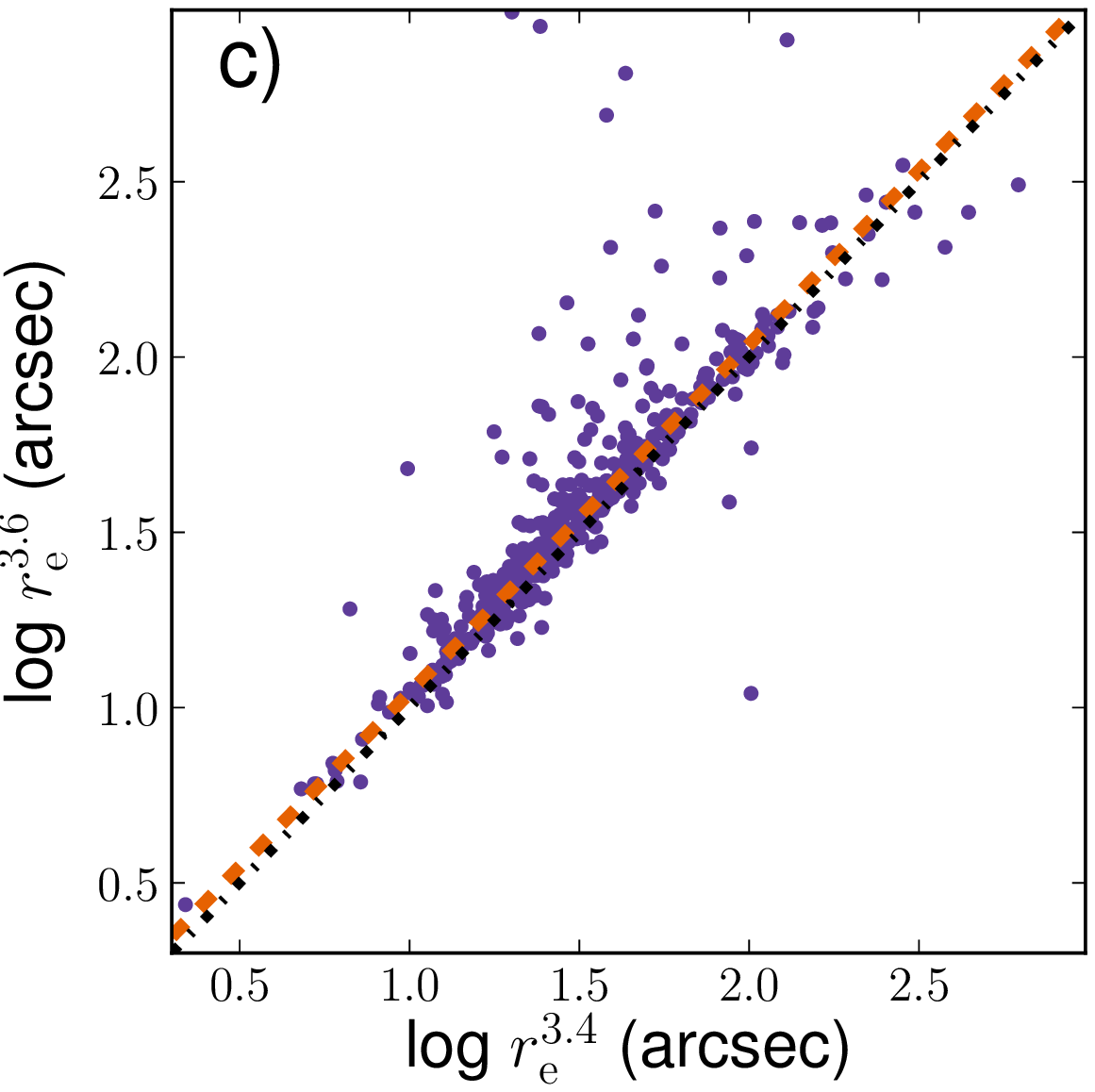}
\includegraphics[width=5.8cm ,height=5.8cm, angle=0, clip=]{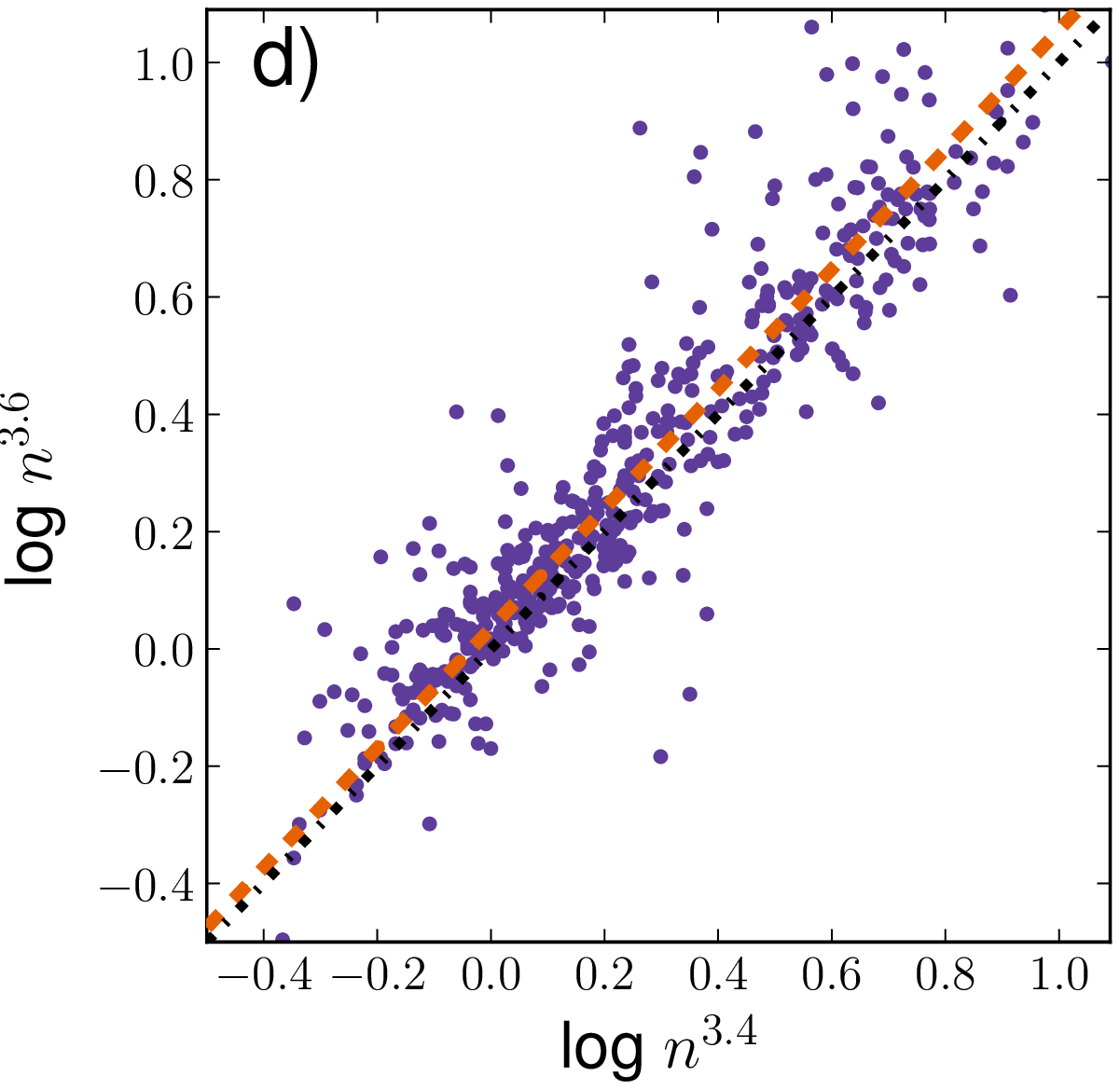}
\includegraphics[width=5.8cm ,height=5.8cm, angle=0, clip=]{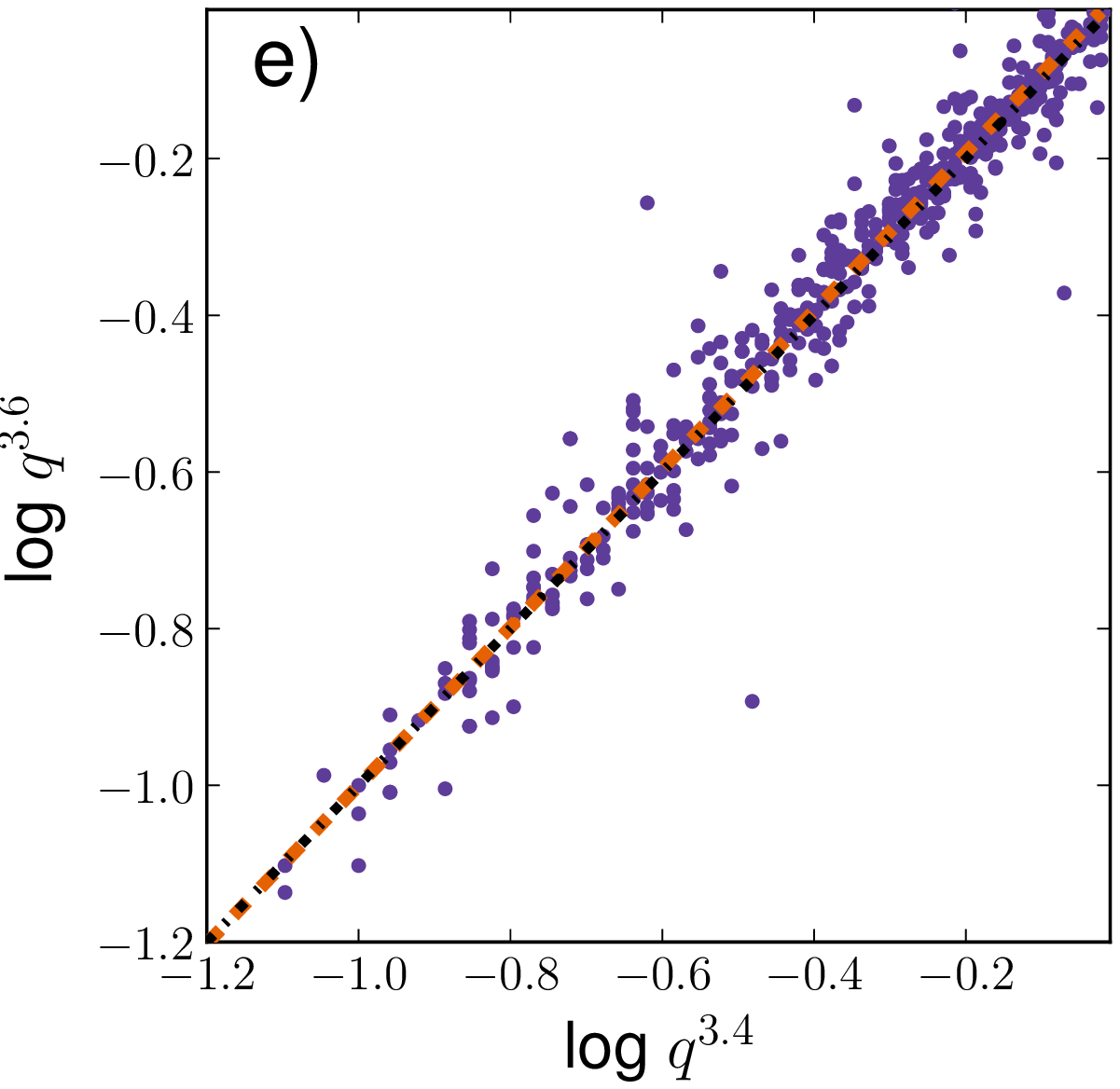}
\includegraphics[width=5.8cm ,height=5.8cm, angle=0, clip=]{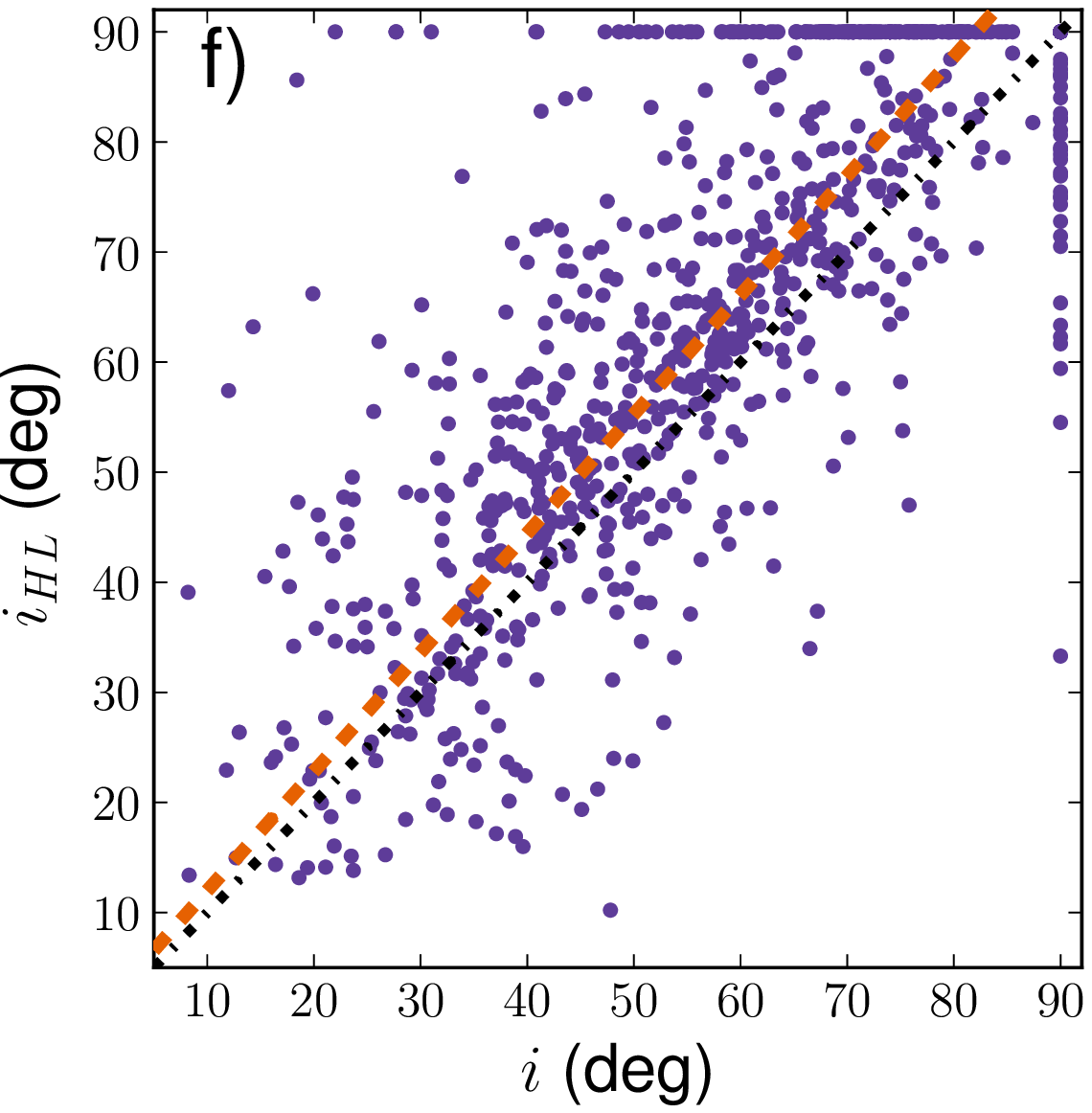}
\caption{The comparison between our results ($x$-axis) and the same parameter from the literature ($y$-axis): a) between the single \ser\ model and aperture flux from \citetalias{2018A&A...609A..37C}, b) between the model flux in \textit{WISE\,W1} and the one from the S$^4$G single \ser\ modelling, c) between the effective radius in \textit{WISE\,W1} and the one from the S$^4$G single \ser\ modelling, d) between the \ser\ index in \textit{WISE\,W1} and the one from the S$^4$G single \ser\ modelling, e) between the model flattening in \textit{WISE\,W1} and the one from the S$^4$G single \ser\ fitting, f) between our estimate of the galaxy inclination and the one taken from HyperLeda. The clustering dots at $90^{\circ}$ for both estimates are either galaxies for which an edge-on model in the S$^4$G modelling pipeline was adopted, or for which the apparent flattening was less then the intrinsic one (see the Appendix~\ref{sec:add_analysis}). The black dotdash lines show the one-to-one relationship and the orange dashed lines refer to the regression lines.}
\label{comparison_W1}
\end{figure*}
 
In the \textit{WISE\,W1} band, for only 5 galaxies (PGC\,2807061, ESO\,434-040, IC\,0691, NGC\,4194, and PGC\,40819) our results cannot be considered as reliable by any means since these galaxies are too small to be fitted. The \textit{WISE\,W1} images for another 5 galaxies (ESO\,116-012, UGC\,4684, PGC\,28759, UGC\,6566, and UGC\,7636) contain a very bright saturated star, therefore for these galaxies our analysis failed to compute realistic parameters of the \ser\ model, even when a simultaneous fitting of the star and the target galaxy was performed. 
Thus, in total, 865 galaxies (98.86\% of the whole DustPedia sample) have reliable single \ser\ fits in the sense that they are sufficiently resolved to be fitted. We should stress, however, that for a small fraction of these galaxies their \ser\ models are meaningless because of a quite complex geometry of their structure (see Sect.~\ref{sec:discussion}).  

We compared the retrieved model fluxes with the aperture fluxes from \citetalias{2018A&A...609A..37C} (see Fig.~\ref{comparison_W1}\textit{a}). One can see that the fluxes compare very well ($k=0.967$, $b=-0.010$, $\rho=0.97$), except for some outliers which appeared to be interacting galaxies (for example, NGC\,2480, IC\,2163, ESO\,491-021) or galaxies with bright foreground stars (for example, ESO\,411-013, NGC\,2974, ESO\,377-039).

539 galaxies in our sample are common with the sample from the \textit{Spitzer} Survey of Stellar Structure in Galaxies (S$^4$G, \citealt{2010PASP..122.1397S,2013ApJ...771...59M,2015ApJS..219....5Q}). The \textit{WISE\,W1} and the \textit{Spitzer} \citep{2004ApJS..154....1W} IRAC~3.6\,$\mu$m passbands \citep{2004ApJS..154...10F} are overlapping enough to make a reliable comparison. 
In the frame of the S$^4$G project, a single \ser\ fitting has already been performed \citep[see table~6 in][]{2015ApJS..219....4S} using the same fitting code {\sc galfit}.
Therefore, we compared the results of the fitting for the \textit{WISE\,W1} and S$^4$G IRAC 3.6~$\mu$m data (see Fig.~\ref{comparison_W1}, panels \textit{b}, \textit{c}, \textit{d}, and \textit{e}). As one can see, the comparison is very good ($\rho=0.92-1.0$ in all panels), and, despite the different resolution and deepness of the \textit{WISE\,W1} and IRAC~3.6\,$\mu$m observations, the results of the \ser\ modelling compare well. 

In Fig.~\ref{comparison_W1}\textit{f}, we show the comparison between the HyperLeda inclinations and those found in the Appendix~\ref{sec:add_analysis}. One can see a large scatter in this correlation which can be explained by a large scatter in the correlation between the apparent flattening $logr25$ from HyperLeda (this is the axis ratio of the galaxy isophote of 25 mag\,arcsec$^{-2}$ in the  $B$ band) and the apparent flattening fitted in this work. Also, we can clearly see a systematic overestimation of the inclination in HyperLeda as compared to the inclination based on our results ($\rho=0.75$, $k=1.10$) -- this is especially apparent for the highly-inclined galaxies. It can be explained by the fact that we used the intrinsic disc flattening $\langle q_0 \rangle\approx0.14$ (see Appendix~\ref{sec:add_analysis}), whereas in HyperLeda $\langle q_0 \rangle\approx0.23$.
According to the formula for calculating galaxy inclination (see eq.~\ref{Hubble_form}), the number of galaxies with $90^\circ$ inclinations is larger in the latter case than if we imply a lower intrinsic flattening.

\subsection{\textit{Herschel}}
\label{sec:Herschel} 
 
As for \textit{WISE\,W1}, we built the distribution of the ratio of the effective radius to the PSF FWHM in each \textit{Herschel} band. We found that 320 galaxies (94\% out of the subsample of 339 galaxies) are large enough to have a plausible \ser\ model, i.e. their $r_\mathrm{e}^{\lambda}$/FWHM$^{\lambda}>0.5$ in each band.

\begin{figure}
\centering
\includegraphics[width=6.0cm, angle=0, clip=]{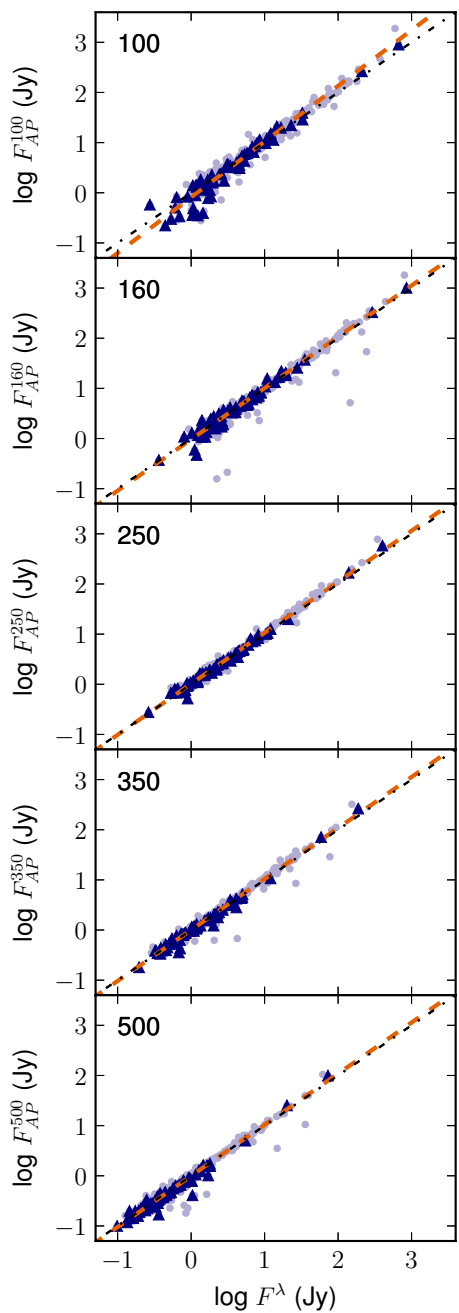}
\caption{The comparison between the single \ser\ model ($x$-axis) and the aperture ($y$-axis) taken from \citetalias{2018A&A...609A..37C}. The lilac filled circles refer to all galaxies with reliable models, while the dark-blue filled triangles correspond to the reference sample. The black dotdash line shows the one-to-one relationship and the orange dashed line depicts the fit for all galaxies shown.}
\label{compar_fluxes_her}
\end{figure}

In Fig.~\ref{compar_fluxes_her} we present the comparison between the aperture fluxes from \citetalias{2018A&A...609A..37C} and the \textsc{galfitm} model fluxes.
For PACS\,100 and PACS\,160, we can see that the observed and model fluxes are not completely consistent ($k=1.13$ and $b=-0.11$ for PACS\,100 and $k=1.04$ and $b=-0.04$ for PACS\,160). However, for SPIRE\,250--SPIRE\,500 the consistency is much better ($k\approx1.02$ and $b\approx-0.01$). This discrepancy may be explained, as above, by the better resolution in the PACS bands than in the SPIRE bands: bright emission details, which are seen in PACS, are smeared out in SPIRE. 
However, in general our models follow the observations fairly well, taking into account that our simple \ser\ model cannot adequately describe the complex geometry of the dust component in some cases (see discussion in Sect.~\ref{sec:discussion}).

\section{The inclinations and stellar masses}
\label{sec:add_analysis}

In addition to the single \ser\ modelling, which was done for all DustPedia galaxies in the \textit{WISE\,W1} band, we
used results of the multicomponent modelling for the 539 galaxies common with S$^4$G. The models were taken from the S$^4$G pipeline 4 \citep{2015ApJS..219....4S}\footnote{\url{http://www.oulu.fi/astronomy/S4G\_PIPELINE4/MAIN/}}. 

For galaxies with the bulge-to-total luminosity ratio $B/T$ less than 1 (i.e. for disc galaxies), we estimated the disc inclination angle $i$ toward the observer using the relation \citep{1926ApJ....64..321H}

\begin{equation}
\sin^2i = \frac{1-10^{\,2\,\log q}}{1-10^{\,2\,\log q_0}}\,,
\label{Hubble_form}
\end{equation}
in which $q$ is the ratio between the apparent minor and major axes of a galaxy outer disc (retrieved from our 2D fitting) and $q_0$ is the intrinsic disc flattening. 

To estimate the inner flattening $q_0$ of disc galaxies, we selected a sub-sample of edge-on galaxies from the S$^4$G sample. We used the multicomponent decomposition results from \citet{2015ApJS..219....4S}, where at least one component of an edge-on (isothermal) disc is present.

\begin{figure}
\centering
\includegraphics[width=8.5cm, angle=0, clip=]{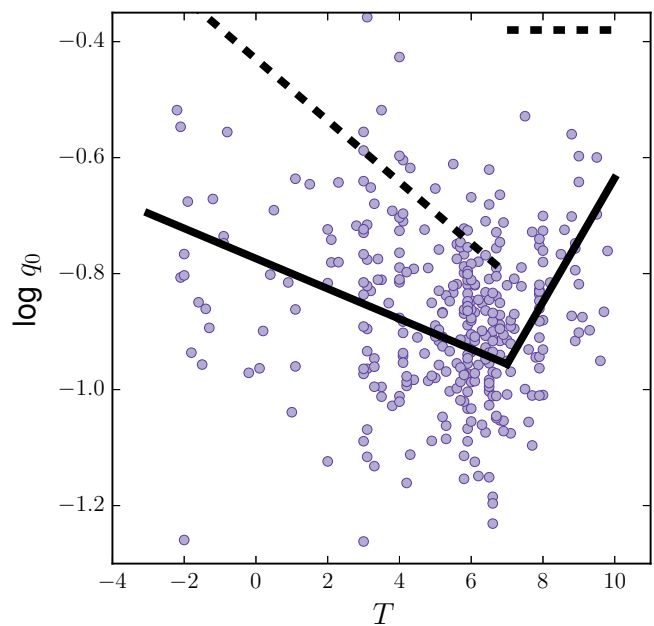}
\caption{Intrinsic disc flattening versus Hubble stage (see text). The black dashed line represents the HyperLeda piecewise formula from \url{http://leda.univ-lyon1.fr/leda/param/incl.html}, the black solid line is an approximation for the S$^4$G edge-on galaxies.}
\label{Type_q0}
\end{figure}

In order to calculate the intrinsic disc flattening $q_0=h_\mathrm{z}/h_\mathrm{R}$, we used the computed disc scaleheight $z_0\approx2\,h_\mathrm{z}$ (here $h_\mathrm{z}$ is the disc scaleheight of the exponential disc) and the disc scalelength $h_\mathrm{R}$ for the outer disc (if the model consists of two stellar discs) or the single disc. Both values $h_\mathrm{z}$ and $h_\mathrm{R}$ were taken from the S$^4$G decomposition. We estimated the dependence of the disc flattening on the Hubble type $T$ from HyperLeda (see Fig.~\ref{Type_q0}). As one can see, the dependence of $q_0$ on $T$ is very poor, however, we split up galaxies by their $T$ and found an average trend (the red dashed line) separately for $T\leq7$ ( $\log q_0 = -0.026\,T - 0.774$) and $T>7$ ($\log q_0 = 0.107\,T - 1.705$), similar to what is done in HyperLeda to compute their inner flattening of galaxies (for $T\leq7$, $\log q_0 = -0.43-0.053\,T$ and for $T>7$, $\log q_0 = -0.38$, both the trends are shown with the blue dashed line). As one can see, the estimated flattening is $\approx1.6$ times lower than from HyperLeda taken for the same $T$. According to our estimation, the average value for late-type galaxies is $\langle q_0 \rangle\approx0.14$, consistent with \citet{1994AJ....107.2036G} and \citet{2015MNRAS.451.2376M}.

Having the derived dependence of the intrinsic disc flattening $q_0$ on Hubble stage, we computed the galaxy inclination for all DustPedia galaxies using eq.~\ref{Hubble_form}. If a galaxy has an edge-on disc model in S$^4$G, we count it as viewed edge-on with $i\equiv90^{\circ}$. For the other galaxies common with S$^4$G, we used the fitted disc flattening $q$. For the remaining galaxies, we use the galaxy flattening $q$ taken from the single \ser\ fitting in \textit{WISE\,W1}. Although such inclination estimates are inhomogeneous since they are taken from the two different sources, this seems to be the best solution to measure the galaxy inclinations for the given sample as precisely as possible. The regression analysis shows that $i_{HL}=1.10\,i$ and $\rho=0.75$ (see Fig.~\ref{comparison_W1}\textit{f}), i.e. there is a systematic overestimation of the galaxy inclination in HyperLeda, mainly because their intrinsic flattening is 1.6 times larger than found here. Also notice that the scatter in this correlation is quite large, which might be explained by a large scatter in the correlation $q$ versus $r_{25}$ ($\rho=0.78$, not shown here), where $r_{25}$ is taken from HyperLeda.
Since $r_{25}$ is the galaxy (not disc) axis ratio and it was measured in the $B$ band (while we consider the \textit{WISE\,W1} band with a drastically reduced contamination of dust), our inclination estimates should be more precise.

To properly estimate \textit{WISE\,W1} total fluxes of the galaxies, we used the following method. For the frames with more than one fitted object, where apart from the target galaxy some other objects (interacting galaxies or background stars) are present, we subtracted the models of the non-target (contaminating) objects from the given galaxy image. For the obtained residual frames and for the frames of galaxies with only one fitted object (the galaxy), we filled the masked pixels with the model pixels and then measured the total flux within the master aperture from \citetalias{2018A&A...609A..37C}. By so doing, we were able to reliably reconstruct the direct flux coming from the target galaxy taking into account the light from the detected contaminants. The extracted fluxes were then converted to Vega magnitudes\footnote{See Table\,1, \url{http:/wise2.ipac.caltech.edu/docs/release/allwise/expsup/sec4_3a.html}} and then to AB magnitudes\footnote{See Table\,3, \url{http://wise2.ipac.caltech.edu/docs/release/allsky/expsup/sec4_4h.html}}. Extended-source correction needs to be applied when performing aperture photometry upon \textit{WISE} maps\footnote{See Table 5, \url{http://wise2.ipac.caltech.edu/docs/release/allsky/expsup/sec4_4c.html}}, to account for the fact that the calibration of \textit{WISE} data is based upon profile fitting of point sources. To convert the calculated AB-magnitudes into stellar masses, we computed the total galaxy luminosities using the `best' distances tabulated in \citetalias{2018A&A...609A..37C} and applied the mass-to-light luminosity ratio from \citet{2018MNRAS.473..776K} $M_*/L_{3.4}=0.65$, which was recently estimated in the \textit{WISE\,W1} band for non-star forming galaxies.

\end{document}